\documentclass[fleqn,usenatbib,useAMS]{mnras}
\usepackage{xcolor}

\usepackage{lipsum}
\usepackage{comment}
\usepackage{graphicx}
\usepackage{amsmath}
\usepackage{amssymb}
\usepackage[super]{nth}
\usepackage[T1]{fontenc}
\usepackage{ae,aecompl}
\usepackage{newtxtext}

\DeclareRobustCommand{\VAN}[3]{#2}
\let\VANthebibliography\thebibliography
\def\thebibliography{\DeclareRobustCommand{\VAN}[3]{##3}\VANthebibliography}

\title[SILCC VIII - FUV]{SILCC -- VIII: The impact of far-ultraviolet radiation on star formation and the interstellar medium}

\author[T.-E.
Rathjen et al.]{
 Tim-Eric Rathjen,$^{1,2}$\thanks{E-mail: rathjen@ph1.uni-koeln.de}
 Stefanie Walch,$^{1}$
 Thorsten Naab,$^{2}$
 Pierre N\"urnberger,$^{1}$
 Richard W\"unsch,$^{3}$
 \newauthor
 Daniel Seifried,$^{1}$
 \& Simon C. O. Glover$^{4}$
 \\
 $^{1}$I. Physikalisches Institut, Universit\"at zu K\"oln, Z\"ulpicher Str. 77, 50937 K\"oln, Germany\\
 $^{2}$Max Planck Institute for Astrophysics, Karl-Schwarzschild-Str. 1, 85748 Garching, Germany\\
 $^{3}$Astronomical Institute of the Czech Academy of Sciences, Bo\v{c}n\'{i} II 1401, 141 00 Prague, Czech Republic\\
 $^{4}$Zentrum f\"ur Astronomie, Institut f\"ur Theoretische Astrophysik, Universit\"at Heidelberg, Albert-Ueberle-Str. 2, 69120 Heidelberg, Germany
}
\date{Accepted XXX.
Received YYY; in original form ZZZ}
\pubyear{2024}
\definecolor{ltgray}{gray}{.89}

\begin{document}
\label{firstpage}
\pagerange{\pageref{firstpage}--\pageref{lastpage}}
\maketitle

\begin{abstract}
We present magnetohydrodynamic simulations of star formation in the multiphase interstellar medium to quantify the impact of non-ionising far-ultraviolet (FUV) radiation within the \textsc{Silcc Project} simulation framework.
Our study incorporates the radiative transfer of ionising radiation and self-consistent modelling of variable FUV radiation from star clusters, advancing beyond previous studies using static or simplified FUV fields.
This enables a more accurate capture of the dynamic interaction between radiation and the evolving ISM alongside other stellar feedback channels.
The interstellar radiation field (ISRF) near young star clusters can reach $G_0 \approx 10^4$ (in Habing units), far exceeding the solar neighbourhood value of $G_0 = 1.7$.
Despite these high intensities, FUV radiation minimally impacts the integrated star formation rate compared to ionising radiation, stellar winds, and supernovae.
A slight reduction in star formation burstiness is linked to increased photoelectric (PE) heating efficiency by the variable FUV field.
Dust near star-forming regions can be heated up to 60 K via the PE effect, with a broad temperature distribution.
PE heating rates in variable FUV models exhibit higher peaks but lower averages than static ISRF models.
Simulations under solar neighbourhood conditions without stellar winds or ionising radiation but with supernovae yield unexpectedly high star formation rates of $\sim0.1~\mathrm{M_\odot~yr^{-1}~kpc^{-2}}$.
Our analysis reveals increased cold neutral medium (CNM) volume-filling factors (VFF) outside stellar clusters, reduced thermally unstable gas, and sharper warm–cold gas separation.
The variable FUV field also promotes a cold diffuse gas phase with a molecular component, exhibiting a VFF of $\sim 5{-}10$ per cent.
\end{abstract}


\section{Introduction}\label{sec:intro}

The interstellar medium (ISM) hosts various gas phases that interact dynamically and chemically, playing a key role in star formation and galaxy evolution \citep[][and references therein]{Wolfire1995, Ferriere2001, Naab2017}.
Understanding these interactions is crucial for advancing our knowledge of how galaxies form and evolve.

Star formation occurs in the cold neutral medium (CNM), characterised by its high gas densities and efficient shielding from the interstellar radiation field (ISRF).
Stellar feedback mechanisms in the form of ionising radiation, stellar winds, and supernova (SN) explosions, have a significant impact on the ISM, driving turbulence, shaping its morphology, and regulating the star formation activity \citep[][and references therein]{Girichidis2020a, Schinnerer2024}.
However, the relative importance of various stellar feedback mechanisms and their implications for star formation are not yet conclusively understood.
Ionising radiation and winds from massive O- and B-type stars are thought to regulate or even halt further gas accretion onto newly formed stellar associations and modify the properties of the ambient gas before the onset of SNe. This early stellar feedback might take precedence over the aforementioned cataclysmic events at the end of a massive star's lifetime \citep[see e.g.][]{Walch2012, Gatto2017, Haid2019, Rathjen2021}.

Among the early feedback processes, ionising ultraviolet radiation (EUV, $E_\gamma > 13.6,\mathrm{eV}$) was suggested to be the main agent for dispersing natal molecular clouds \citep{Kannan2020, Rathjen2021, Chevance2022}.
This process regulates the local star formation rate (SFR) by drastically reducing gas accretion onto protostars and star clusters while creating HII regions \citep[see e.g.][]{McKee1989, Ali2019, Rathjen2021}.
The ionised gas within HII regions is mostly optically thin for the non-ionising far-ultraviolet (FUV, $E_\gamma = 5.6-13.6\,\mathrm{eV}$) radiation emitted by star clusters.
This radiation is notable for its ability to dissociate molecular hydrogen (Lyman-Werner band photons) and directly heats the ambient gas via photoelectric (PE) heating on dust grains.
These processes influence the chemical composition and physical properties of the surrounding gas, especially in photodissociation regions (PDRs) \citep[see e.g.][]{Draine1978, Wolfire1995, Hollenbach1999}.
It has been suggested that PE heating could explain the thermal pressure in the diffuse gas and set the global SFR of galactic discs in dynamical and thermal equilibrium \citep[see e.g.][]{Ostriker2010, Kim2011}.
In regions sufficiently shielded from EUV and FUV light, low-energy cosmic rays (CRs) are the main agents of $\mathrm{H}_2$ ionisation and the resulting heating \citep{Bakes1994}.
Together, the strength of the UV radiation field and the low-energy CR ionisation rate, $\zeta_\mathrm{CR}$, control the thermal and chemical state of the warm neutral medium (WNM) and the CNM \citep{Wolfire2003}.
Stars are sources of the UV radiation field, which therefore varies in space and time.
Hence, an assumed constant ISRF based on estimates and local observations \citep[e.g.][]{Draine1978} might be insufficient to describe the state of the ISM properly.

Observations \citep{Calzetti2000, Leroy2008, Ossenkopf2013} and theoretical models \citep{Rollig2007, Bisbas2021, Pound2023} of nearby star-forming regions have provided valuable insights into the effects of FUV radiation on the ISM.
These studies have demonstrated the presence and structure of PDRs, where FUV radiation from young stellar clusters interacts with surrounding molecular clouds, influencing the chemical and physical properties of neutral gas layers \citep[see e.g.][for a review]{Hollenbach1999}.
Additionally, observations have revealed specific FUV-driven processes such as photodissociation of molecules, heating of neutral gas, and the formation of particular molecular species characteristic of PDRs.
Theoretical models and numerical simulations at various scales have supplemented these observations and have further studied the influence of FUV radiation on the ISM and its star-forming properties \citep[e.g.][]{Kim2017, Hill2018, Bialy2020}.
On the star cluster scale, \citet{Ali2019} have studied how the varying FUV radiation fields in simulated massive star clusters affect the photoevaporation of protoplanetary discs near massive stars.
ISM simulations of galactic patches have included the relevant physical processes to study the impact of stellar feedback on the cold gas phase \citep[e.g.][]{Butler2017, Kannan2020, Rathjen2021, Hu2022, Rathjen2023, Kim2023a, McCallum2024}.
Recently, numerical advances have made it possible to simulate (dwarf) galaxies with adequate resolution to capture the major physical processes and explicit star formation and feedback as either isolated systems or in cosmological context \citep[e.g.][]{Forbes2016, Hu2017, Emerick2018, Lahen2019, Agertz2020, Tress2020, Smith2021, Andersson2024, Deng2024, Fotopoulou2024}.
However, individual studies incorporate different physical processes and combinations of stellar feedback mechanisms with different numerical resolutions.
Among other uncertainties, the relative importance of FUV radiation compared to other stellar feedback mechanisms and its impact on the global characteristics of the ISM remains to be clarified.

Several previous works have explored the role of FUV radiation in regulating star formation.
For example, \citet{Hu2017} conducted simulations of dwarf galaxies and concluded that FUV-driven PE heating plays a crucial role in shaping the ISM, albeit in conjunction with other feedback processes.
Similarly, \citet{Hopkins2020} and \citet{Smith2021} demonstrated the importance of multichannel feedback, including FUV radiation, in determining the thermal structure of the ISM. While these studies have provided valuable insights, our work advances this understanding by employing a self-consistent, variable FUV radiation model that accounts for local variations in real-time and incorporates additional feedback channels like stellar winds, and additional ISM components like CRs.

In this study, we comprehensively investigate the effects of variable (in space and time) FUV radiation on the SFR and the chemical composition of the multiphase ISM in a stratified galactic patch, utilising magnetohydrodynamic (MHD) simulations conducted within the \textsc{Silcc Project} \citep{Walch2015a, Girichidis2016a, Gatto2017, Peters2017, Girichidis2018a, Rathjen2021, Rathjen2023}.
Our new simulations incorporate a self-consistent modelling of the ISRF, allowing us to explore the local variations in the FUV intensity and its effects on the surrounding ISM.
This paper is organised as follows:
In Sect. \ref{sec:numerics}, we provide an overview of the \textsc{Silcc Project} simulation framework and our methodology to incorporate FUV radiation (Sect. \ref{sec:fuv}) and resulting theoretical predictions (Sect. \ref{sec:theory}).
We present our simulation results in Sect. \ref{sec:results}, focusing on the influence of FUV radiation on the SFR (Sect. \ref{sec:sfr}) and the chemical properties of the ISM (Sect. \ref{sec:gasstructure} - \ref{sec:chemistry}).
We especially focus on the impact of our new self-conistent FUV ISRF treatment on the heating and cooling properties of the ISM in Sect.~\ref{sec:heatcool}.
In Sect. \ref{sec:discussion}, we discuss the implications of our findings and compare them to the existing literature.
The caveats of our models are assessed in Sect. \ref{sec:caveats}.
We conclude the study in Sect. \ref{sec:conclusion}.
To ensure the reading flow, we include additional material in the Appendix.
Appendix \ref{app:tables} presents a tabulated summary of our main findings.
We study the free parameters of our new ISRF model in Appendix \ref{App:distance} and detail how the ISRF affects our chemical network in Appendix \ref{app:chemistry}.
We furthermore complement the analysis presented in the main body of the text in Appendices \ref{app:PECR}, \ref{app:long} and \ref{app:WIMpeak}.

\section{Numerical methods}\label{sec:numerics}

This work expands upon \citet{Rathjen2023} and employs the same parameters and methods unless otherwise stated.
The new method developed to treat the variable FUV field is described in detail.
All other numerical methods are briefly summarised in the following.

We simulate the evolution of the multiphase ISM with the adaptive mesh refinement (AMR) code \textsc{FLASH} v4.6.2 \citep{Fryxell2000}.
We solve the MHD equations using a modified three-wave solver based on \citet{Bouchut2007} and \citet{Waagan2011}.
Multiple modules are included to model a variety of physical processes such as:


(\textit{i}) a non-equilibrium chemical network based on \citet{Nelson1997, Glover2007} to follow the abundances of seven species (atomic (H), molecular (H$_2$), and ionised (H$^+$) hydrogen, carbon monoxide (CO) and ionised carbon (C$^+$), atomic oxygen (O), and free electrons (e$^-$)) and to treat the gas heating, cooling, and molecule formation.
The model incorporates standard hydrogen ionisation processes, including photoionisation, collisional ionisation, cosmic ray and X-ray ionisation.
H$_2$ formation is modelled primarily as occurring on dust grain surfaces, following \citet{Hollenbach1989}.
H$_2$ destruction occurs through several mechanisms: cosmic ray ionisation, collisional dissociation in hot gas, and most significantly, photodissociation by the ISRF.
The model also accounts for the self-shielding of CO and H$_2$, as well as shielding by dust.
A fixed dust-to-gas ratio of 0.01 is assumed, with dust temperature calculated self-consistently under the thermal equilibrium assumption.
For an overview of the most prominent heating and cooling mechanisms in our framework as well as a detailed discussion on the ISRF's role in determining dust temperature, H$_2$ and CO dissociation, see Appendix \ref{app:chemistry}.
We refer to \citet{Walch2015a} for further details on the chemical network implementation;

(\textit{ii}) gravity due to an external stellar potential and self-gravity evaluated with an Octtree-based method \citep{Wunsch2018};

(\textit{iii}) CRs, which are being injected in the shocks of SN remnants with an efficiency of 10~per~cent (i.e. $E_\mathrm{CR} = 0.1 \times E_\mathrm{SN} = 10^{50}$~erg).
They are modelled as an additional relativistic fluid with a non-isotropic advection-diffusion scheme \citep{Girichidis2018}.
This fluid adds CR pressure, $P_\mathrm{CR}$, and CR energy density, $u_\mathrm{CR}$, terms to the MHD equations.
It is characterised by its own adiabatic index ($\gamma_\mathrm{CR} = 4 / 3$) which contributes to a total effective adiabatic index of $\left(\gamma_\mathrm{eff} = (\gamma_\mathrm{th}~P_\mathrm{th} + \gamma_\mathrm{cr}~P_\mathrm{CR}) / (P_\mathrm{th} + P_\mathrm{CR})\right)$.
We account for adiabatic cooling and hadronic cooling losses \citep[see][for details]{Girichidis2018, Rathjen2021, Rathjen2023};

(\textit{iv}) star formation of individual massive stars.
The stars are tracked with a subgrid model for sink particles that represent star clusters \citep{Gatto2017} with a sink particle accretion radius of $r_\mathrm{accr} = 3 \Delta x \approx 12$~pc and an accretion threshold density of $\rho_\mathrm{thr} = 2.1 \times 10^{-21}$~$\mathrm{g\,cm^{-3}}$.
The N-body dynamics of the sink particles is integrated with a fourth-order Hermite integrator \citep{Dinnbier2020};

(\textit{v}) stellar feedback from massive stars in the form of supernovae \citep{Gatto2015}, stellar winds \citep{Gatto2017}, and ionising (EUV) radiation \citep{Peters2017, Haid2018, Rathjen2021} that is treated with the backward ray-tracing scheme \textsc{TreeRay} \citep{Wunsch2021}.

In all previous \textsc{Silcc Project} publications, the ISRF was set to a constant value that got attenuated using the \textsc{TreeRay/OpticalDepth} algorithm \citep{Wunsch2018}, which follows the same principle idea as the \textsc{TreeCol} algorithm developed by \citet{Clark2012}.
For each grid cell in the computational domain, the \textsc{TreeRay/OpticalDepth} module computes the column densities of total gas, H$_2$, and CO using a \textsc{HEALPix} tessellation of the unit sphere with $n_\mathrm{rays} = 48$ directions.
It then computes 3D averages and stores these quantities.
For example, the local $A_\mathrm{V, 3D}$ is:
\begin{align}\label{eq:AV3D}
 A_\mathrm{V, 3D} = -\frac{1}{2.5} \ln \left[\frac{1}{n_\mathrm{rays}}\sum^{n_\mathrm{rays}}_{\mathrm{ray}=1}\exp\left(-2.5\frac{N_\mathrm{H, ray}}{1.87\times10^{21}\mathrm{cm}^{-2}}\right)\right],
\end{align}
where $N_\mathrm{H, ray}$ is the total gas column density along each direction.
The $A_\mathrm{V, 3D}$ is used to attenuate the ISRF.
The columns of H$_2$ and CO are subsequently used to calculate the amount of (self-)shielding of these species, and the total gas column is used to compute the dust attenuation within the chemistry network \citep[see][]{Walch2015a}.

The FUV ISRF is typically measured in Habing units:
\begin{align}\label{eq:G0unit}
 G_0 = \frac{u_\mathrm{FUV}}{5.29\times10^{-14}\,\mathrm{erg\,cm}^{-3}},
\end{align} with the energy density, $u_\mathrm{FUV}$, in the FUV photon energy range between 5.6 and 13.6 eV.
In these units, the standard ISRF in the solar neighbourhood has been estimated as $G_0 = 1.7$ \citep{Draine1978}.
For studies of different galactic environments presented in \citet{Rathjen2023}, the ISRF has been scaled with the gas surface density in a way that it increases linearly with the typical SFR predicted by \citet{KennicuttJr.1998}.

In this study, we assume a lower background ISRF of $G_\mathrm{bg} = 0.0948$.
This value is derived for a cosmic UV background taken from \citet{Haardt2012} plus an assumed static, preexisting, low-mass stellar population with a stellar surface density of $\Sigma_\star = 30\,\mathrm{M_\odot\,pc^{-2}}$.
We include a model for an old stellar population to account for the gravitational potential of those stars.
The old stellar population is assumed to be at solar metallicity and does not evolve in time.
Those stars do not exert any feedback in our model and are not tracked.

This paper extends the stellar feedback model to consider the additional FUV radiation from all formed star cluster sink particles. Sect.~\ref{sec:fuv} describes how the star cluster FUV luminosity is calculated.
In any case, the FUV field from a star or star cluster, like any other radiation field, is diluted as a function of the radial distance $R$ from the source according to the standard inverse square law,
\begin{align}
 G_\mathrm{clus}(R) \propto R^{-2}.
\end{align}
Together with the background, the local unattenuated strength of the ISRF is
\begin{align}\label{eq:G0}
 G_0 = \left(\sum_{_{N_\mathrm{cluster}}}G_\mathrm{clus}\right) + G_\mathrm{bg}.
\end{align}
Additionally, dust effectively absorbs the FUV light.
Hence, the spatially varying FUV radiation field of all star clusters as well as $G_\mathrm{bg}$ must be attenuated.
The local dust attenuation is computed in every grid cell using the \textsc{TreeRay/OpticalDepth} module \citep{Wunsch2018}.
From each cell in the computational domain, the distance over which the local attenuation, i.e. the visual extinction $A_\mathrm{V, 3D}$, is computed is limited to $d = 50\,\mathrm{pc}$, which is the typical distance between massive stars in the solar neighbourhood.
We justify the choice of $d$ in Sect. \ref{sec:theory} and Appendix \ref{App:distance}.
For the effective strength of the local dust-attenuated \citep{Dishoeck1988} ISRF, we use
\begin{align}
 G_\mathrm{eff} = G_0 \times \exp(-2.5\,A_\mathrm{V}),
\end{align}
with the 3D-averaged $A_\mathrm{V, 3D}$ provided by \textsc{TreeRay/OpticalDepth}. This is an approximation of the direct line-of-sight visual extinction between a gas cell and the FUV emitting cluster, which we make due to computational memory considerations.
The resulting effective local FUV field, $G_\mathrm{eff}$, is variable in space and time.

To be consistent with the calculation of the local dust attenuation, we also limit the maximum distance a star cluster can contribute a non-zero FUV field to any surrounding cell to $R = d = 50\,\mathrm{pc}$.
Hence, only cells with cell centres within radius $R = d$ of any given cluster are illuminated with the sum $G_0$ (see Eq.~\ref{eq:G0}).
The distance $d$ is a free model parameter and its implications are discussed in Appendix \ref{App:distance}.

In our chemical network, $G_\mathrm{eff}$ is included in several chemical reactions (e.g. to determine the C$^+$ abundance) and, most importantly, in several heating rates such as PE heating.
All relevant rates and reactions that depend on the strength of the FUV field can be found in Appendix \ref{app:chemistry}.

\subsection{Far-ultravioloet luminosity of star clusters}\label{sec:fuv}

This subsection describes how the FUV radiation field from star clusters with different masses is computed using \textsc{Starburst99} \citep{Leitherer1999}.
We show several tests that validate our treatment of the FUV radiation.
Previously, we only explicitly tracked massive stars with $M \geq 9\,\mathrm{M_\odot}$ in our star cluster sink particles.
We follow the evolution of those stars \citep{Ekstrom2012} and account for their individual stellar feedback \citep{Gatto2017, Haid2018, Rathjen2021}.
Lower-mass stars have been aggregated within the star cluster sink particle and were only considered for their mass. Since they do not exhibit strong stellar winds, do not explode as supernovae, and only have a negligible contribution to the EUV radiation field, their impact has so far been negligible in our modelling.

\begin{figure}
 \centering
 \includegraphics[width=.95\linewidth]{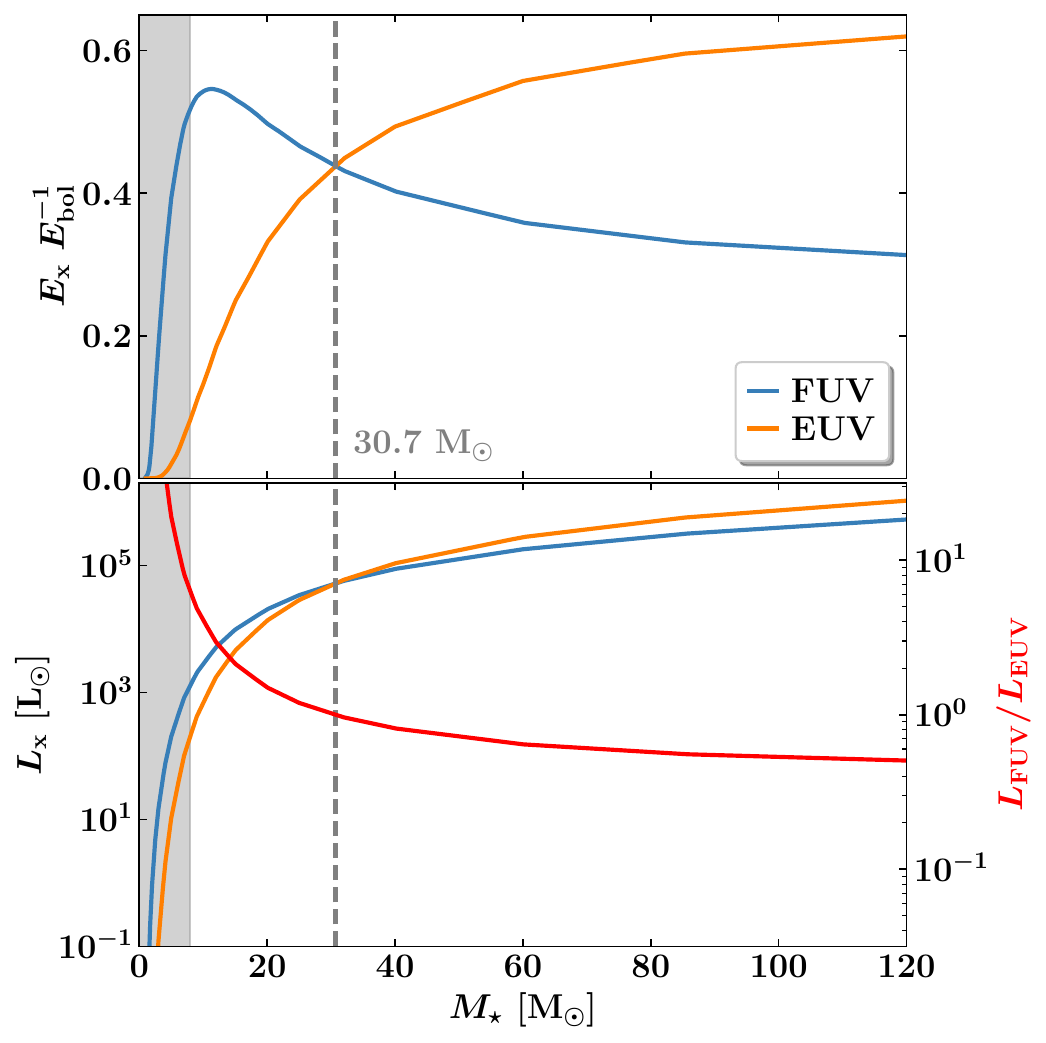}
 \caption{Assessment of the energetic importance of FUV radiation compared to EUV radiation from stars.
 \textit{Top:} Total energy in the FUV plus EUV bands, $E_\mathrm{x}$, normalised to the total bolometric energy of the respective star, $E_\mathrm{bol}$, for different stars with stellar mass $M_\star$.
 \textit{Bottom:} Total luminosity in the two respective energy bands as a function of stellar mass and the ratio of those luminosities (red line, right-hand $y$-axis).
 The FUV and EUV energy outputs are computed by assuming the stars as black bodies with effective temperatures taken from the Geneva stellar evolution tracks for stars that just entered the zero-age main sequence \citep{Ekstrom2012}.
 The grey-shaded area indicates the range of lower-mass stars ($< 9\,\mathrm{M_\odot}$) which have not been individually tracked in our stellar evolution models.
 The vertical dashed line indicates the mass for which the EUV radiation band overtakes the FUV radiation band in terms of energy output.
 Even though FUV radiation cannot ionise atomic hydrogen, it is evident that the FUV radiation band is energetically important and cannot be neglected, especially for lower-mass stars.}
 \label{fig:FUV-EUV_Mstar}
\end{figure}

In Fig.~\ref{fig:FUV-EUV_Mstar} (top panel), we show the fraction of FUV and EUV energy relative to the bolometric radiative energy as a function of the stellar mass $M_{\star}$ of the source.
In the bottom panel, we plot the corresponding FUV and EUV luminosities as a function of $M_{\star}$, and the right-hand $y$-axis shows the ratio of the two.
The grey band indicates the regime of intermediate- and low-mass stars $M_{\star} < 9\;{\rm M}_{\odot}$.
The dashed vertical line shows $M_{\star} = 30.7\;{\rm M}_{\odot}$, where we find that FUV and EUV radiation are energetically equally important.
It is obvious that stars with masses below $9\;{\rm M}_{\odot}$ cannot be neglected for the total FUV luminosity of a star cluster as it is usually done for EUV radiation.
For a {$10^5$ M$_\odot$} star cluster with a Salpter IMF sampled between $1\--120$~M$_\odot$ at a cluster age of $\tau_\mathrm{cluster}=0$~Myr, the low-mass stars contribute only $\sim3$~per~cent of the total FUV luminosity.
However, this changes rapidly with time as the initially dominating massive stars die off and the low-mass star FUV luminosity accounts already for $\sim16$~per~cent after 10 Myr and for $\sim50$~per~cent of the total FUV output of that cluster after $\sim23$~Myr.
Therefore, we need to modify our star cluster subgrid model to account for low- and intermediate-mass stellar population.

Whenever we accrete $120~\mathrm{M_\odot}$ of gas onto a star cluster sink particle, we form one massive star with $9\,\mathrm{M_\odot} \leq M_\star < 120\,\mathrm{M_\odot}$, randomly sampled from an IMF with a Salpeter-IMF slope in the high-mass regime \citep{Gatto2017}.
The left-over mass, $M_\mathrm{lo} = (120 \,\mathrm{M_\odot} - M_\star)$, is assumed to form low- and intermediate-mass stars with $<9\,\mathrm{M_\odot}$ that we do not track individually.
We use \textsc{Starburst99} to calculate the FUV luminosities of the low-mass stellar population of any given mass $M_\mathrm{lo}$.
For the \textsc{Starburst99} model, we assume a total star cluster mass of $10^6\,\mathrm{M_\odot}$ fully sampled with a Kroupa IMF \citep{Kroupa2001} starting with $0.1\,\mathrm{M_\odot}$ and truncated at $9\,\mathrm{M_\odot}$.
We use the Geneva tracks for solar metallicity and model the evolution to a maximum age of up to 200~Myr (which exceeds the maximum simulated time of $t - t_\mathrm{SF} \approx 180\,\mathrm{Myr}$ for our longest running simulation).
We then integrate the resulting spectra calculated by \textsc{Starburst99} over photon energies between 5.6~eV and 13.6~eV to obtain the total FUV luminosity of that fully sampled $10^6\,\mathrm{M_\odot}$ star cluster. We then scale the resulting luminosity with $M_\mathrm{lo}$ to obtain the amount that any given population of low- and intermediate-mass stars with a given $M_\mathrm{lo}$ would contribute.

In addition, we also have to account for the FUV luminosity of our explicitly tracked massive stars (prior, we only tracked their EUV contribution for photon energies larger than 13.6 eV).
We obtain the FUV contribution of each massive star with mass $M_\star$ by integrating a black body spectrum with an effective temperature of that star (taken from the Geneva tracks) over the FUV energy range and dividing by the fully integrated black body spectrum (see Fig. \ref{fig:FUV-EUV_Mstar}).
We then multiply this ratio with the total bolometric luminosity of that star (taken from the Geneva tracks) which yields the star's FUV luminosity.
Adding the two FUV components together, i.e. the total FUV luminosity of a massive star ($M_\star$) with an accompanying low- to intermediate-mass stellar population with mass $\mathrm{M}_\mathrm{lo}$, we obtain the time-dependent total FUV luminosity per unit 120 $\mathrm{M_\odot}$ of formed stars.

\subsection{Theoretical considerations for the far-ultraviolet radiation field}\label{sec:theory}

\begin{figure}
 \centering
 \includegraphics[width=.95\linewidth]{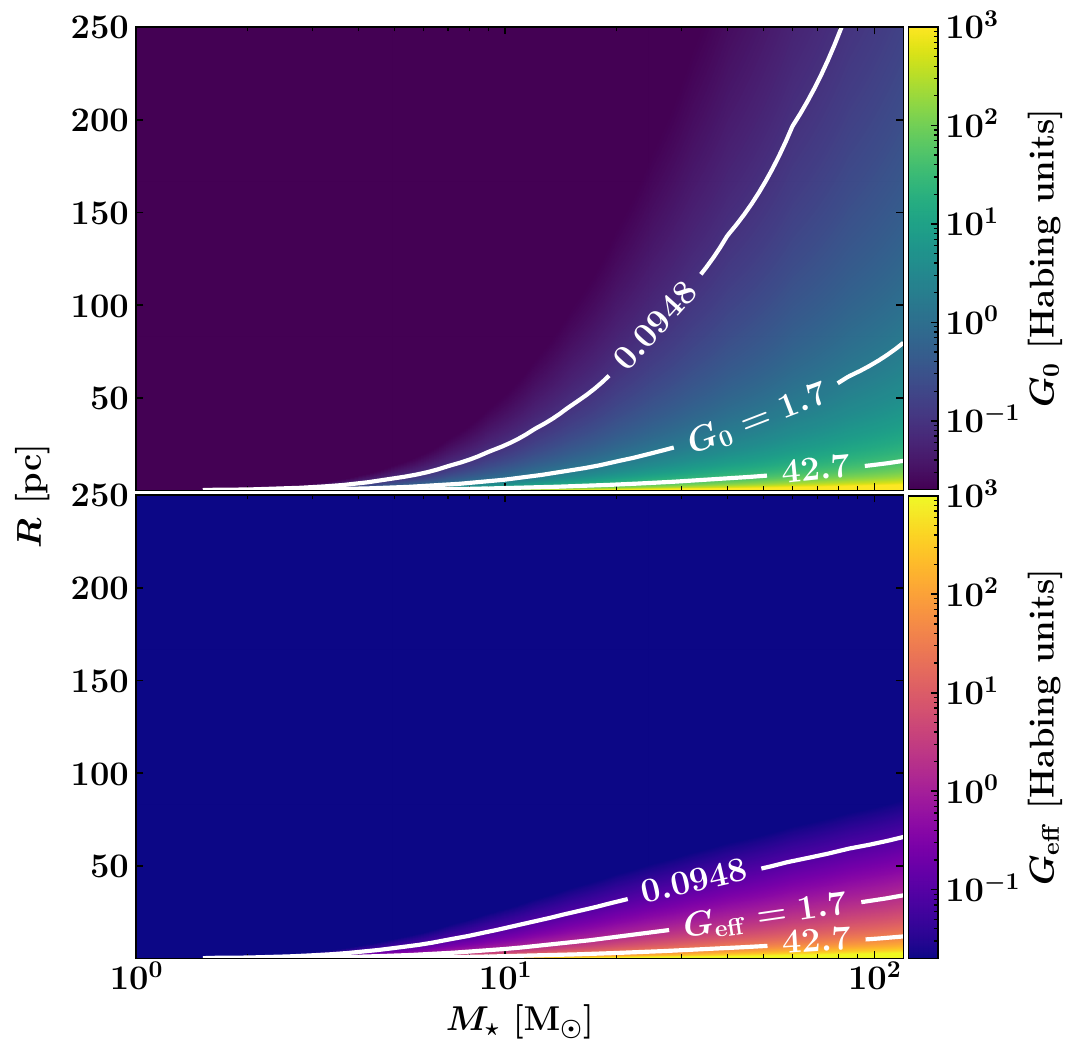}
 \caption{Strength of the FUV radiation field for the unattenuated ($G_0$, top panel) and attenuated ($G_\mathrm{eff}$, bottom panel) cases as a function of distance $R$ to a given single star and its respective mass, $M_\star$.
 For the attenuation, we assume a visual extinction of $A_V = 1$ at a distance of 50 pc (i.e. the attenuation gets stronger further out and weaker closer in).
 This corresponds to a uniform hydrogen gas with a number density of $n_\mathrm{H}\approx 12\,\mathrm{cm^{-3}}$.
 The white contours indicate the assumed background FUV ISRF of $G_\mathrm{bg} = 0.0948$, and, for comparison, the canonical solar neighbourhood value of $G_0 = 1.7$ as well as the value of the constant ISRF previously used in simulations with high gas surface density ($\Sigma100$ with $G_0 = 42.7$, see Table \ref{tab:sims}). All values are in Habing units (see Sect.~\ref{sec:numerics}).}
 \label{fig:R_Mstar_G0Geff}
\end{figure}

\begin{figure}
 \centering
 \includegraphics[width=.95\linewidth]{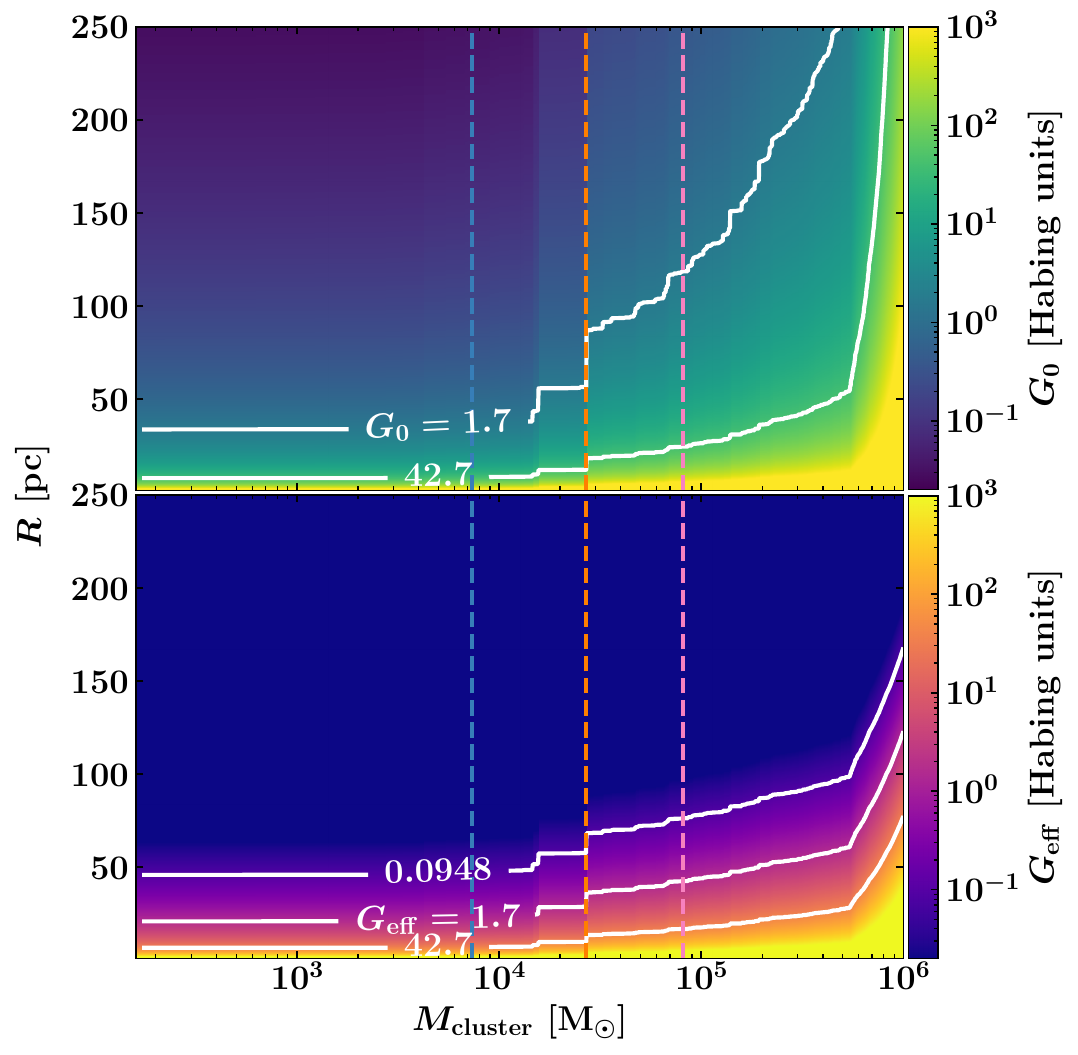}
 \caption{Same as in Fig. \ref{fig:R_Mstar_G0Geff} but for star clusters with masses up to $M_{\rm cluster} = 10^6\,\mathrm{M_\odot}$.
 We sample an IMF with a Salpeter-like slope for stars more massive than $1\,\mathrm{M_\odot}$ to populate the star cluster.
 All stars are assumed to be on the zero-age main sequence.
 As a reference, we indicate the average mass of the formed star clusters in our simulations with varying initial conditions (see text and Table \ref{tab:sims} for details).
 The unattenuated radiation field ($G_0$, top panel) never drops below our assumed background value $G_\mathrm{bg}$ for star clusters more massive than $200\,\mathrm{M_\odot}$.
 However, the dust attenuation (bottom panel) substantially reduces the local FUV field as a function of $R$.}
 \label{fig:R_Mcluster_G0Geff}
\end{figure}

We give predicted model outputs for the variable FUV radiation field and how it would heat the ISM.
In Fig. \ref{fig:R_Mstar_G0Geff}, we show the unattenuated, $G_0$, (top panel) and dust-attenuated, $G_\mathrm{eff}$, (bottom panel) strength of the FUV radiation field generated by a single massive star of mass $M_\star$ as a function of distance $R$ to that star.
Throughout this section, we always assume a visual extinction of $A_\mathrm{V} = 1$ at a distance of $R = d = 50\,\mathrm{pc}$ for the dust-attenuated FUV field $G_\mathrm{eff}$.
This would translate to an uniform environemantal density of $n_\mathrm{H} \approx 12\,\mathrm{cm^{-3}}$.
We indicate with white contour lines the galactic background $G_\mathrm{bg} = 0.0948$, the canonical solar neighbourhood value of $G_0 = 1.7$, and $G_0 = 42.7$, which is the highest value for the ISRF parameter in our static $G_0$ models (see Table \ref{tab:sims}).

We extend this consideration in Fig. \ref{fig:R_Mcluster_G0Geff}, where we do the same analysis as in Fig. \ref{fig:R_Mstar_G0Geff} but this time for fully sampled star clusters with masses ranging between $M_\mathrm{cluster} = 200 - 10^{6}\,\mathrm{M_\odot}$.
The single stars, as well as the sampled clusters, are assumed to have just entered the main sequence ($t_\mathrm{zams} = 0\,\mathrm{Myr}$).
Both, distance to the emitting source and attenuation by dust play a critical role in setting the strength of the FUV radiation field.
For fairly massive star clusters with $M_\mathrm{cluster} \approx 10^4\,\mathrm{M_\odot}$, $G_\mathrm{eff}$ drops to roughly $G_\mathrm{bg}$ at a distance of 50 pc.
Without attenuation by dust, the strength of the ISRF would not fall below $G_\mathrm{bg}$ at distances over $R > 250\,\mathrm{pc}$ (which is equal to half the box size of our computational domain with $L_x = L_y = 500\,\mathrm{pc}$).
We also indicate the maximum average star cluster masses in our simulations as vertical dashed lines, colour-coded by the initial conditions of the respective model (blue, orange, and pink dashed lines for runs with gas surface densities $\Sigma_{\mathrm{gas}}=10,\,30,\,\mathrm{and}\,100\,\mathrm{M}_{\odot}\,\mathrm{pc}^{-2}$, see Table \ref{tab:sims} for details).

\begin{figure}
 \centering
 \includegraphics[width=.95\linewidth]{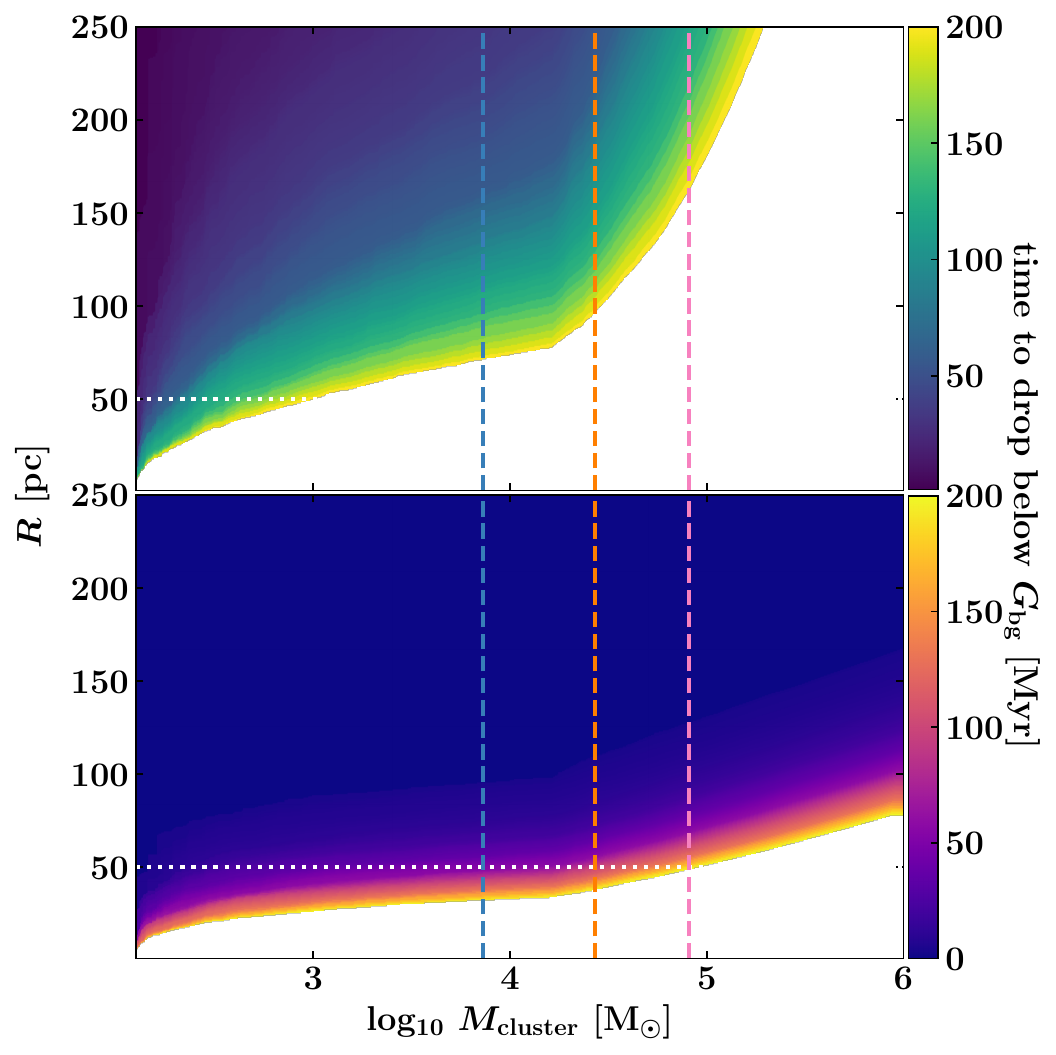}
 \caption{Time evolution of the strength of the FUV radiation field.
 We show the time after which the radiation field in the unattenuated case (top panel) and attenuated case (bottom panel) drop below $G_\mathrm{bg}$ as a function of cluster mass, $M_\mathrm{cluster}$, and distance $R$.
 White areas mean that the radiation field strength never drops below $G_\mathrm{bg}$.
 The vertical dashed lines show the maximum average mass of the formed star clusters in runs with different initial gas surface densities (see text and Table \ref{tab:sims}).
 The horizontal white dotted line indicates an FUV propagation distance of $d = 50$~pc.
 We again assume an $A_\mathrm{V} = 1$ at $d = 50$~pc, which translates to density of $n_\mathrm{H} \approx 12$~$\mathrm{cm^{-3}}$.}
 \label{fig:R_Mcluster_time}
\end{figure}

The stellar population within a star cluster is not static in time.
Less massive stars live longer but have a lower luminosity output compared to the high-mass stars which have an expected lifetime between $\tau_\mathrm{life} \sim 3 - 30\,\mathrm{Myr}$ \citep[see e.g.][on how feedback evolves with cluster age]{Agertz2013}.
This shapes the radiation spectrum of the stellar cluster and has therefore also an impact on the overall energy density in the FUV band.
In Fig. \ref{fig:R_Mcluster_time}, we study after which time the FUV field strength of a stellar cluster with initial mass $M_\mathrm{cluster}$ at distance $R$ drops below $G_\mathrm{bg}$.
The top panel shows the unattenuated case while the bottom panel shows the case accounting for extinction by dust.
The region coloured in white indicates that for that configuration of $M_\mathrm{cluster}$ and $R$ the FUV field strength would not drop below $G_\mathrm{bg}$ within 200 Myr of stellar evolution within the cluster.
The long-lived lower-mass stars provide a significant amount of the total FUV energy density.
Only very massive star clusters ($M_\mathrm{cluster} \gtrsim 3\times10^5\,\mathrm{M_\odot}$) would have a non-negligible contribution to the FUV radiation field at distances $R > 50\,\mathrm{pc}$ when they enter the zero-age main sequence.
Those super-massive star clusters are rare.
The overall maximum cluster mass in our different models ranges between $\mathrm{max}(M_\mathrm{cluster}) \approx 3 \times 10^4 \mathrm{-\,} 3 \times 10^5\,\mathrm{M_\odot}$.
The colour-coded vertical dashed lines in Fig. \ref{fig:R_Mcluster_G0Geff} and Fig. \ref{fig:R_Mcluster_time} indicate the average star cluster mass in our simulations with varying initial conditions (see Table \ref{tab:sims} for details).
Typical star clusters in our simulation, even at the most extreme conditions, emit a FUV radiation field which drops below $G_\mathrm{bg} = 0.0948$ at a distance of $d = 50$~pc under the assumption of a moderate visual extinction of $A_\mathrm{V} = 1$ at that distance.
This is especially true when also considering the time evolution of the star cluster.
These figures demonstrate that limiting the maximum propagation range of the clusters' FUV radiation field to $d = 50$~pc is a reasonable choice.
This limit is compatible with dust attenuation calculations and would not significantly underpredict the far-away FUV ISRF.
We discuss and test this assumption and model parameter further in Appendix \ref{App:distance}.

\begin{figure}
 \centering
 \includegraphics[width=.95\linewidth]{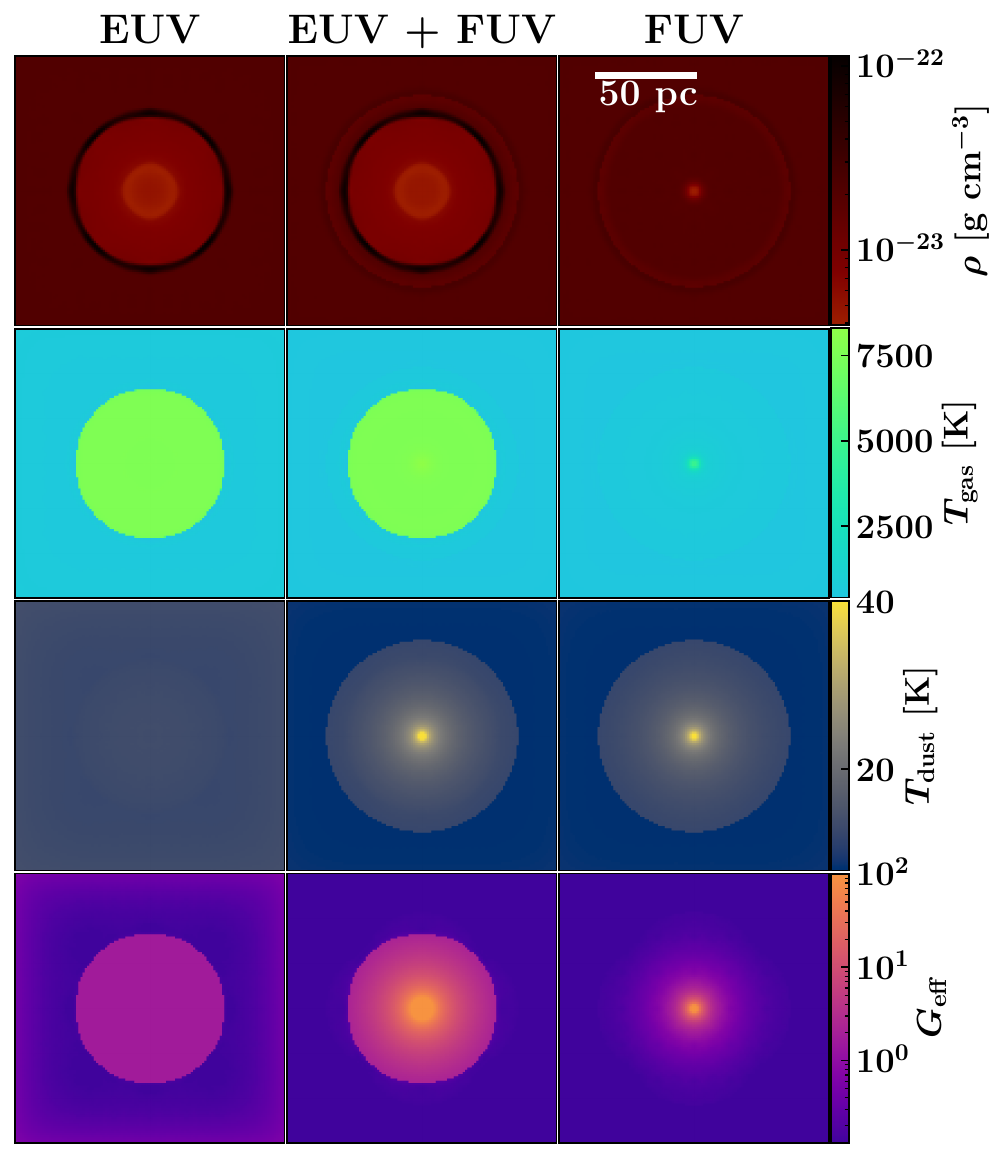}
 \caption{Isolated test of a single 46 M$_\odot$ star in an uniform and purely atomic hydrogen medium with $n_\mathrm{H} = 12\,\mathrm{cm^{-3}}$ shown after 2 Myr of evolution.
 From top to bottom, we plot the total gas density, $\rho$, the kinetic gas temperature, $T_\mathrm{gas}$, the dust temperature, $T_\mathrm{dust}$, and $G_\mathrm{eff}$.
 For this particular test, we compare models with only ionising radiation (\textit{EUV}, first column), ionising plus FUV radiation (\textit{EUV + FUV}, second column), and only self-consistent FUV radiation (\textit{FUV}, third column).
 For the calculation of $G_\mathrm{eff}$ we use a value of $G_0 = 1.7$ in the \textit{EUV}-only case (first column).}
 \label{fig:HII_FUV}
\end{figure}

We study the impact of a self-consistent treatment of the FUV radiation field on the ISM with an idealised test of an isolated single $46\,\mathrm{M_\odot}$ star in a uniform medium in Fig. \ref{fig:HII_FUV}.
The simulation box encompasses $(150\,\mathrm{pc})^3$ on a 128$^3$ grid ($dx \approx 1.17$ pc).
This spatial resolution is a factor 3.3 higher than in the ISM-scale simulations.
The gas is assumed to consist of purely atomic hydrogen with a hydrogen number density of $n_\mathrm{H} \approx 12\,\mathrm{cm^{-3}}$ with an initial gas temperature of $T_\mathrm{gas} \approx 160\,\mathrm{K}$.
The initial dust temperature is $T_\mathrm{dust} = 10\,\mathrm{K}$.
This density is chosen so that the visual extinction at a distance of 50 pc is $A_\mathrm{V}(50\,\mathrm{pc}) = 1$.
All stellar feedback is turned off, except for ionising radiation (EUV, first column), FUV plus EUV radiation (middle column), and only FUV radiation (right column).
We let the system evolve for 2 Myr and show (from top to bottom) slices of the gas density, $\rho$, the gas temperature, $T_\mathrm{gas}$, the dust temperature $T_\mathrm{dust}$ and the strength of the FUV radiation field, $G_\mathrm{eff}$.
All quantities are presented as slices through the domain centre.
We use a value of $G_0 = 1.7$ when calculating $G_\mathrm{eff}$ in the \textit{EUV}-only case.
The FUV radiation field strongly impacts the dust temperature, which is very sensitive to the strength of $G_\mathrm{eff}$.
However, in the FUV-only case, the gas remains largely unaffected.
The median temperature around the star only increases to $T_\mathrm{gas}\approx650$ K in the FUV-only test as compared to $T_\mathrm{gas}\approx7600$ K for EUV + FUV.
The temperature right at the centre reaches $T_\mathrm{gas}^\mathrm{c} = 4800$ K (FUV) and $T_\mathrm{gas}^\mathrm{c} = 8300$ K (EUV + FUV).
The PE heating is negligible compared to the effects of photoionisation, which heats the gas to the expected $T \sim 8000\,\mathrm{K}$.

\begin{figure}
 \centering
 \includegraphics[width=.95\linewidth]{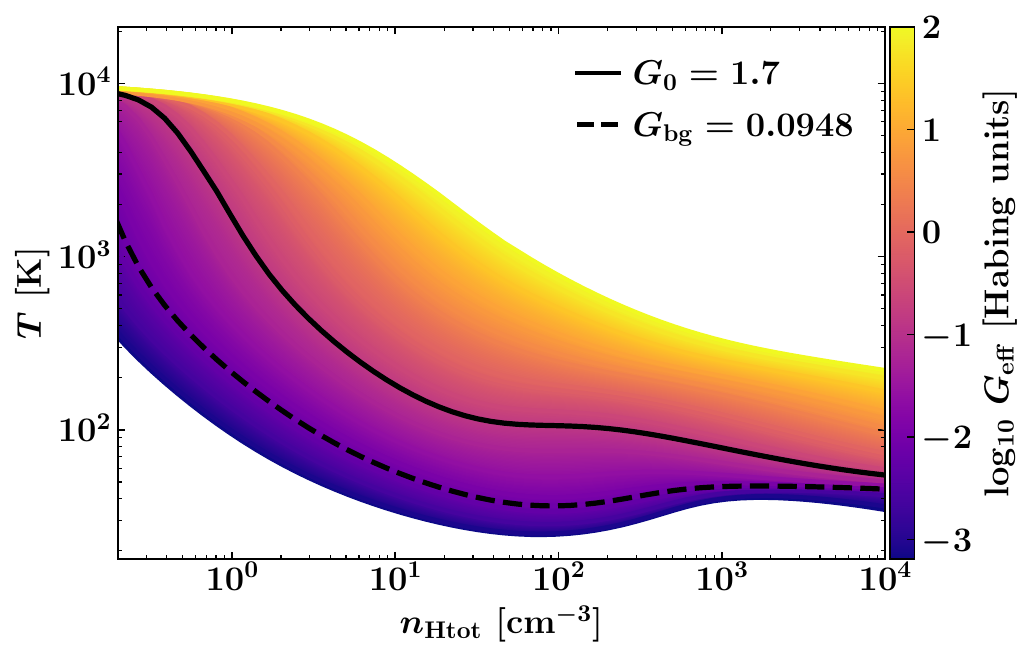}
 \caption{Equilibrium cooling curves computed with our chemical network.
 For all calculations, we assume a constant shielding by an external atomic hydrogen column density of $N_\mathrm{H} = 1.87 \times 10^{21}\,\mathrm{cm^{-2}}$ which translates to a visual extinction of $A_\mathrm{V} = 1$.
 We show the cooling curves over a range of $G_0 \approx 10^{-2} \-- 10^{3}$ ($G_\mathrm{eff} \approx 10^{-3} \-- 10^{2}$) in Habing units.
 Additionally, the gas is heated through the ionisation of hydrogen by CNMs with a CNM ionisation rate of $\zeta_\mathrm{CR} = 3\times10^{17}\,\mathrm{s}^{-1}$.
 The solid black line shows the cooling curve for the canonical value $G_0 = 1.7$ while the value corresponding to the model with $G_0 = G_{\mathrm{bg}} = 0.0948$ is depicted with a dashed black line.
 At higher number densities, necessary for star formation ($n_\mathrm{H} > 10^3\,\mathrm{cm^{-3}}$), the equilibrium temperatures differ only by a factor of $4.3 \pm 0.4$, even though $G_0$ varies over four orders of magnitude.}
 \label{fig:equibCurve}
\end{figure}

To understand the rather marginal impact of PE heating (as compared to heating by photoionisation), we explore equilibrium cooling curves derived for a range of $G_0 = 10^{-2} - 10^3$ using our chemical network at solar-neighbourhood metallicity in Fig. \ref{fig:equibCurve}.
This calculation assumes constant shielding by an external atomic hydrogen column density of $N_\mathrm{H} = 1.87\times10^{21}\,\mathrm{cm^{-2}}$ ($A_\mathrm{V} = 1$) and additional heating by CRs using a $\zeta_\mathrm{CR} = 3\times10^{17}\,\mathrm{s}^{-1}$.
We highlight the cooling curve for $G_0 = 1.7$ and $G_0 = G_\mathrm{bg} = 0.0948$ with a solid and dashed black line
Due to the exponential attenuation, the equilibrium temperature is a weak function of $G_0$, especially for higher densities.
At densities larger than our required threshold density for star formation, $n_\mathrm{thr} \approx 10^3\,\mathrm{cm^{-3}}$, a change in $G_0$ over four orders of magnitude results in a change of the equilibrium temperature by just a factor of $(4.3 \pm 0.4)$.
In strong FUV radiation fields ($G_0 \sim 50-100$), within exposed (unshielded) environments and at moderately diffuse gas densities (total hydrogen nuclei number density $n_\mathrm{Htot} \sim 1\,\mathrm{cm^{-3}}$), PE heating can raise gas temperatures to levels comparable to those produced by photoionisation ($T \sim 10^4\,\mathrm{K}$).
We want to note that all simulations presented in this work are carried out under the assumption of solar metallicity as well as a fixed dust-to-gas ratio of 1~per~cent.
Changes in dust properties and metallicity will have impact on the cooling and heating rates.
At lower metallicity, fewer metals will be present which reduces the radiative cooling efficiency.
At the same time, the amount of dust is expected to be lower, too, which decreases the PE heating efficiency.
In \citet{Brugaletta2025}, we explore this complex interplay and limit us to fixed metallicity and dust properties in the present work.

\begin{figure}
 \centering
 \includegraphics[width=.95\linewidth]{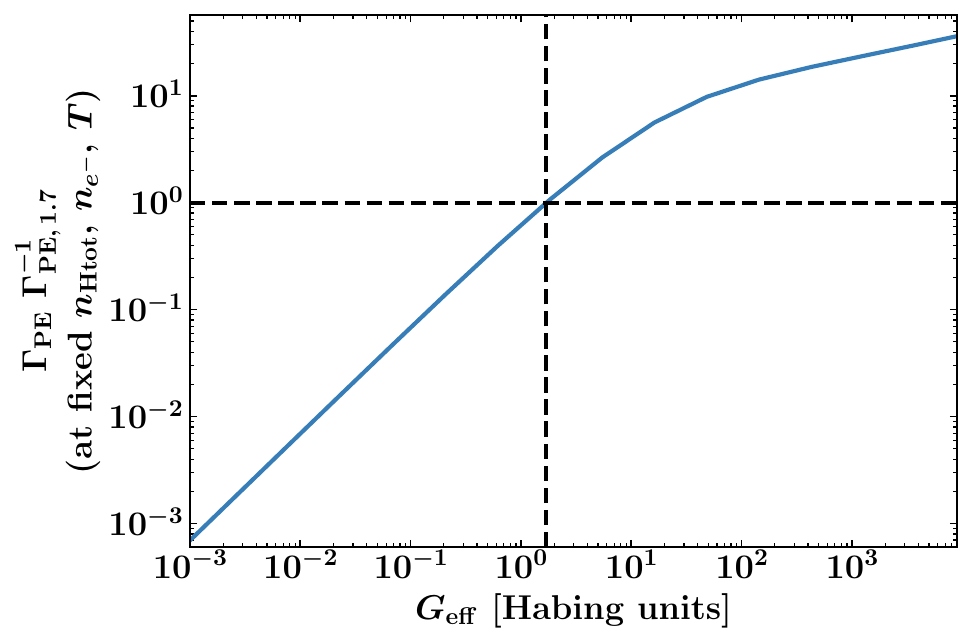}
 \caption{PE heating rate, $\Gamma_\mathrm{PE}$, as a function of $G_\mathrm{eff}$, normalised to $\Gamma_\mathrm{PE,\,1.7}$ for $G_\mathrm{eff} = 1.7$ at fixed hydrogen number density, $n_\mathrm{Htot}$, electron number density, $n_\mathrm{e^-}$, and temperature, $T$.
 We indicate the values for $G_\mathrm{eff} = 1.7$ for reference with dashed lines.
 The PE heating rate first scales linearly with $G_\mathrm{eff}$, but then increases more slowly for $G_\mathrm{eff} \gtrsim 10$.}
 \label{fig:heatingRates}
\end{figure}

We further investigate the PE heating rate, $\Gamma_\mathrm{PE}$, normalised to the $\Gamma_\mathrm{PE}$ at constant $G_0 = 1.7$, $\Gamma_\mathrm{PE,\,1.7}$, as a function of $G_\mathrm{eff}$.
For $\Gamma_\mathrm{PE}$, we follow the prescription of \citet{Bakes1994} and \citet{Bergin2004}:

\begin{align}
 \Gamma_\mathrm{PE} &= 1.3 \times 10^{-24} \cdot \epsilon \cdot G_\mathrm{eff} \cdot n_\mathrm{H, \text{tot}} \label{eq:heatingRate}\\
 \epsilon &= 0.049 \cdot \left(1 + \left(\Psi / 963\right)^{0.73}\right)^{-1} \notag\\
 &+ \left(0.037 \cdot \left(T / 10^{4}\right)^{0.7}\right) \cdot \left(1 + 4 \times 10^{-4} \cdot \Psi\right) \\
 \Psi &= G_\mathrm{eff} \cdot \sqrt{T} \cdot n_\mathrm{e^-}^{-1},
\end{align}

with the PE heating efficiency, $\epsilon$ \citep{Wolfire2003}, electron number density, $n_\mathrm{e^-}$, and total hydrogen number density, $n_\mathrm{H, tot}$.
In Fig. \ref{fig:heatingRates}, we show the normalised $\Gamma_\mathrm{PE} \Gamma_\mathrm{PE,\,1.7}^{-1}$ as a function of $G_\mathrm{eff}$ at fixed hydrogen and electron number densities and gas temperature.
Dashed lines indicate the normalised $\Gamma_\mathrm{PE}$ where $G_\mathrm{eff} = 1.7$.
For low FUV radiation field strengths, $\Gamma_\mathrm{PE}$ scales linearly with $G_\mathrm{eff}$.
For $G_\mathrm{eff} \gtrsim 10$, this behaviour changes significantly and $\Gamma_\mathrm{PE}$ flattens out.
This explains the marginal heating capabilities of the FUV radiation field beyond temperatures of $\sim 10^3\,\mathrm{K}$.

\subsection{Initial conditions and parameters of the simulations}\label{sec:params}

\begin{figure*}
 \centering
 \includegraphics[width=.95\linewidth]{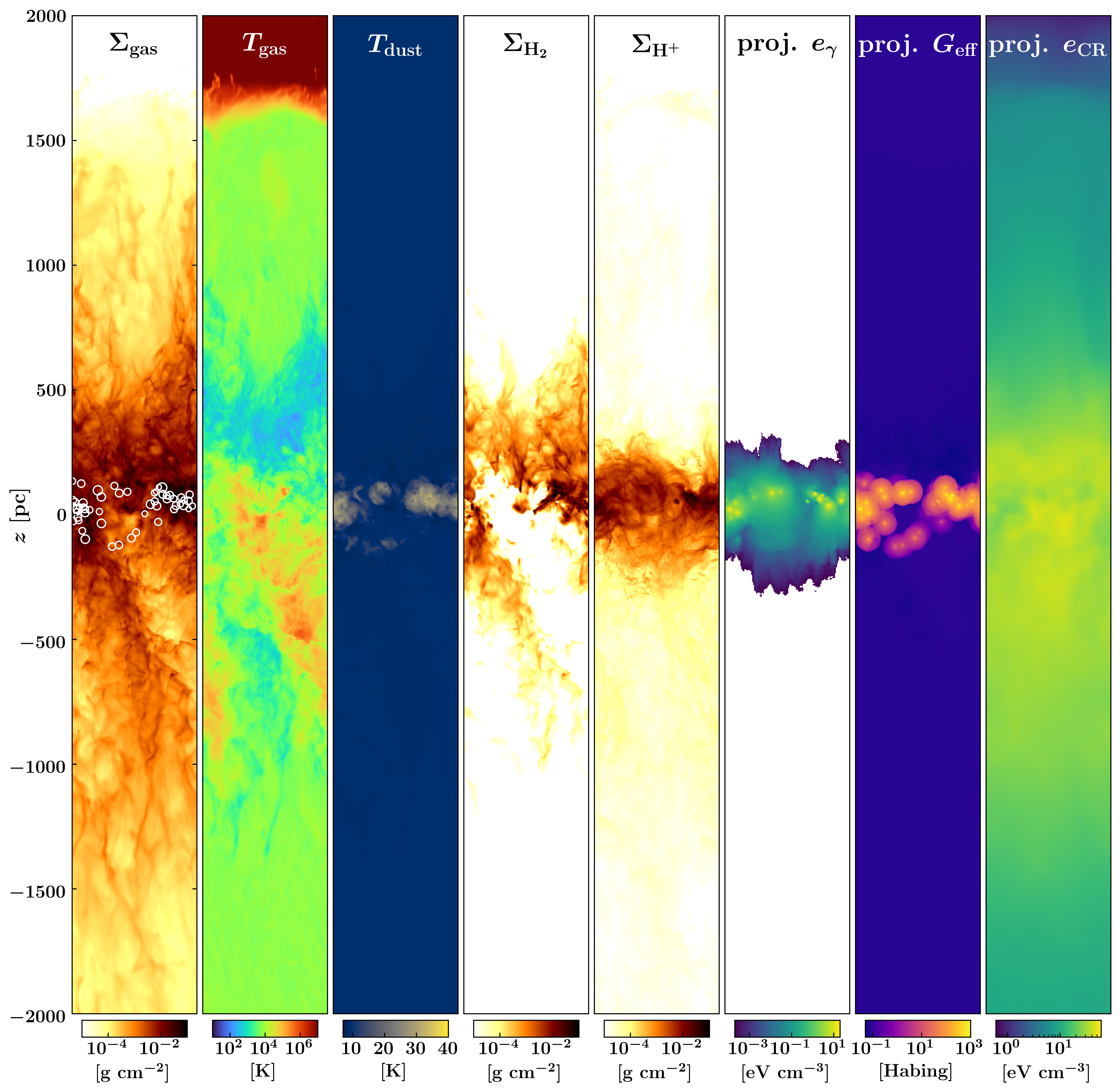}
 \caption{Overview of $\mathrm{\Sigma100vFUV}$ at $t-t_\mathrm{SF} = 30$ Myr.
 Shown are the edge-on views of the total gas ($\Sigma_\mathrm{gas}, $\nth{1} panel), molecular hydrogen ($\Sigma_\mathrm{H_2}$, \nth{4} panel) and ionised hydrogen ($\Sigma_\mathrm{H^+}$, \nth{5} panel) surface densities, as well as mass-weighted gas ($T_\mathrm{gas}$, \nth{2} panel) and dust ($T_\mathrm{dust}$, \nth{3} panel) temperatures, ionising photon energy density ($e_\gamma$, \nth{6} panel), the effective $G_0$ field ($G_\mathrm{eff}$, \nth{7} panel), and CNM energy density ($e_\mathrm{CR}$, \nth{8} panel) in projection.
 The star-forming galactic ISM is concentrated around the midplane.
 White circles in the \nth{1} panel indicate active star clusters.}
 \label{fig:overview}
\end{figure*}

We discuss a series of nine high-resolution MHD simulations of the multiphase ISM in a stratified galactic patch.
The simulations are part of the \textsc{Silcc Project} simulation framework and closely follow the setup of \citet{Rathjen2023}.
The key addition to \citet{Rathjen2023} is the new module \textsc{AdaptiveG0} to evaluate the ISRF FUV radiation field produced by star clusters.
The set of simulations is designed to explore the complex and non-linear effects of incorporating a self-consistent FUV radiation field generated by stellar clusters -- the novel \textsc{AdaptiveG0} module -- on the SFR and ISM chemistry.

The computational domain of the simulated galactic patch spans ($500\times500\times\pm4000)\,\mathrm{pc^3}$ with a base AMR grid resolution of $dx \approx 3.9\,\mathrm{pc}$ in the midplane.
We implement periodic boundary conditions along the $x$- and $y$-directions and outflow boundaries along the elongated $z$-direction.
With this setup, we aim to mimic a patch cut out of a galactic disc.
Initially, the gas is in hydrostatic equilibrium and is set up with a Gaussian density distribution around the midplane.
Together with the initial gas surface density, $\Sigma_\mathrm{gas}$, we vary the thickness of the Gaussian gas disc, $\sigma_\mathrm{gas}$, the initial magnetic field strength, $|\mathbf{B}|$, the strength of the initial turbulent driving, $v_\mathrm{rms}$, and the CR ionisation rate, $\zeta_\mathrm{CR}$, to model different galactic environments.
The initial turbulent driving in each simulation prevents the gas disc from collapsing into a single dense sheet and triggers star formation.
The artificial turbulent driving is turned off once star formation begins and the ISM self-regulates via stellar feedback.

We present three simulations with $\Sigma_{\mathrm{gas}}=10,\,30,\,\mathrm{and}\,100\,\mathrm{M}_{\odot}\,\mathrm{pc}^{-2}$ to model different galactic environments, and which include the \textsc{AdaptiveG0} module.
These are compared to three simulations with the same initial conditions but using the static ISRF\footnote{We note that we account for extinction by dust in both cases.
In the case of a static ISRF, the specified strength, e.g. $G_0 = 1.7$ for model $\mathrm{\Sigma010}$, represents the baseline FUV ISRF, which remains constant in space and time but undergoes dust attenuation in each computational cell.
Locally, the ISRF cannot exceed this baseline value, though it may be significantly reduced in regions with high optical depth. For higher initial $\Sigma_\mathrm{gas}$, the baseline value is also increased.}.

Both sets of simulations include all major stellar feedback channels, stellar winds, EUV radiation, and SNe.
CR acceleration in SN remnants with 10 per cent efficiency ($E_\mathrm{CR} = 0.1\,E_\mathrm{SN} = 10^{50}\,\mathrm{erg}$) and their anisotropic diffusion along the magnetic field is considered in all models.

Lastly, we present three more simulations, two with \textsc{AdaptiveG0} and one with a static ISRF. In these models, we turn off the stellar wind and EUV radiation feedback.
In \citet{Rathjen2021} and \citet{Rathjen2023}, we show that these early stellar feedback mechanisms are likely the primary self-regulation processes for star formation.
By omitting them, we can further investigate how capable FUV radiation from young massive star clusters is in reducing further gas accretion onto the cluster and therefore reducing the SFR.
In two of the three models, $\mathrm{\Sigma010vFUV\dagger}$ and $\mathrm{\Sigma010\dagger}$, SN feedback is still turned on.
In the model $\mathrm{\Sigma010vFUVnoSN\dagger}$, we also switch off SN feedback and only consider the FUV radiation feedback channel.
All simulations\footnote{We note that the models $\mathrm{\Sigma010}$, $\mathrm{\Sigma030}$, $\mathrm{\Sigma100}$ have previously been published in \citet{Rathjen2023} under the same name and model $\mathrm{\Sigma010\dagger}$ in \citet{Rathjen2021} as \textit{S}.
We include them here for comparison between the models.}
and their varying initial parameters are summarised in Table \ref{tab:sims}.
All simulations employ a fixed dust-to-gas ratio of 1~per~cent and assume solar metallicity.

\setlength{\tabcolsep}{1pt}
\begin{table}
 \centering
 \caption{(\textit{Top}) List of simulations with their varying initial parameters.
 From left to right, we give the name of each simulation, the initial gas surface density, $\Sigma_\mathrm{gas}$, whether we use the novel \textsc{AdaptiveG0} module or a constant FUV ISRF with $G_0$, the initial strength of the magnetic field, $|\mathbf{B}|$, the thickness of the initial Gaussian gas density profile, $\sigma_\mathrm{gas}$, the target root mean square velocity of the initial turbulent driving, $v_\mathrm{rms}$, and the value of the constant CR ionisation rate, $\zeta_\mathrm{CR}$.
 (\textit{Bottom}) Stellar feedback processes included in each simulation.
 Except for the last three models, all simulations utilise all stellar feedback processes (winds, ionising radiation, SNe), and CRs.
 Models without ionising radiation, stellar winds nor CRs are labelled with the $\dagger$-suffix ($\mathrm{\Sigma010vFUV\dagger}$ and $\mathrm{\Sigma010\dagger}$).
 The model $\mathrm{\Sigma010vFUV\dagger}$ includes SNe and \textsc{AdaptiveG0}.
 The model $\mathrm{\Sigma010\dagger}$ includes only SNe.
 Finally, the model $\mathrm{\Sigma010vFUVnoSN\dagger}$ only includes \textsc{AdaptiveG0} without any other feedback processes.}
 \begin{tabular}{lcccccc}
	\hline
	Name & $\Sigma_\mathrm{gas}$ & $G_0$ & |\textbf{B}| & $\sigma_\mathrm{gas}$ & $v_\mathrm{rms}$ & $\zeta_\mathrm{CR}$\\
	 & [M$_\odot$ pc$^{-2}$] & [Habing] & [$\mu\mathrm{G}$] & [pc] & [kms$^{-1}$] & [s$^{-1}$]\\
	\hline\hline
	$\mathrm{\Sigma010vFUV}$ & 10 & \textsc{AdaptiveG0} & 6 & 30 & 10 & $3 \times 10^{-17}$ \\
	$\mathrm{\Sigma030vFUV}$ & 30 & \textsc{AdaptiveG0} & 10 & 37 & 15 & $9 \times 10^{-17}$ \\
	$\mathrm{\Sigma100vFUV}$ & 100 & \textsc{AdaptiveG0} & 19 & 60 & 30 & $3 \times 10^{-16}$ \\
	\hline
	$\mathrm{\Sigma010}$ & 10 & 1.7 & 6 & 30 & 10 & $3 \times 10^{-17}$ \\
	$\mathrm{\Sigma030}$ & 30 & 7.9 & 10 & 37 & 15 & $9 \times 10^{-17}$ \\
	$\mathrm{\Sigma100}$ & 100 & 42.7 & 19 & 60 & 30 & $3 \times 10^{-16}$ \\
	\hline
	$\mathrm{\Sigma010vFUV\dagger}$ & 10 & \textsc{AdaptiveG0} & 6 & 30 & 10 & $3 \times 10^{-17}$ \\
	$\mathrm{\Sigma010\dagger}$ & 10 & 1.7 & 6 & 30 & 10 & $3 \times 10^{-17}$ \\
	$\mathrm{\Sigma010vFUVnoSN\dagger}$ & 10 & \textsc{AdaptiveG0} & 6 & 30 & 10 & $3 \times 10^{-17}$ \\
 \hline
 \end{tabular}
\setlength{\tabcolsep}{3.25pt}
  \begin{tabular}{lccccc}
	\hline
	Name & SNe & Stellar & Ionising & CR & \textsc{AdaptiveG0}\\
 	     &    & winds & radiation & acceleration & \\
	\hline\hline
	$\mathrm{\Sigma010vFUV}$ & \checkmark & \checkmark & \checkmark & \checkmark & \checkmark \\
	$\mathrm{\Sigma030vFUV}$ & \checkmark & \checkmark & \checkmark & \checkmark & \checkmark \\
	$\mathrm{\Sigma100vFUV}$ & \checkmark & \checkmark & \checkmark & \checkmark & \checkmark \\
	\hline
	$\mathrm{\Sigma010}$ & \checkmark & \checkmark & \checkmark & \checkmark & $\times$ \\
	$\mathrm{\Sigma030}$ & \checkmark & \checkmark & \checkmark & \checkmark & $\times$ \\
	$\mathrm{\Sigma100}$ & \checkmark & \checkmark & \checkmark & \checkmark & $\times$ \\
	\hline
	$\mathrm{\Sigma010vFUV\dagger}$     & \checkmark & $\times$ & $\times$ & $\times$ & \checkmark \\
	$\mathrm{\Sigma010\dagger}$         & \checkmark & $\times$ & $\times$ & $\times$ & $\times$ \\
	$\mathrm{\Sigma010vFUVnoSN\dagger}$ & $\times$   & $\times$ & $\times$ & $\times$ & \checkmark \\
 \hline
 \end{tabular}
 \label{tab:sims}
\end{table}

We give a general overview of our setup in Fig. \ref{fig:overview}.
We define the onset of star formation, $t_\mathrm{SF}$, as the moment when the ISM starts to regulate itself and show the run $\mathrm{\Sigma100vFUV}$, 30 Myr after the onset of star formation, at $t - t_\mathrm{SF} = 30$~Myr.
Due to formatting reasons, we chose only to show the simulation domain for $|z| \leq 2\,\mathrm{kpc}$ (instead of the full box, $z = \pm 4\,\mathrm{kpc}$).
Horizontally, we show the full extent ($x = 500\,\mathrm{pc}$).
We present, from left to right: the total gas column density, $\Sigma_\mathrm{gas}$, the projected mass-weighted gas temperature, $T_\mathrm{gas}$, and projected mass-weighted dust temperature, $T_\mathrm{dust}$, the molecular hydrogen column density, $\Sigma_\mathrm{H_2}$, the ionised hydrogen column density, $\Sigma_\mathrm{H^+}$, a volume-weighted projection of the ionising radiation energy density, $e_\gamma$, a volume-weighted projection of the ISRF strength, $G_\mathrm{eff}$, and finally a volume-weighted projection of the CR energy density, $e_\mathrm{CR}$.
Star cluster sink particles are represented as white circles in the first panel.
The depicted size scales with the star clusters' mass and does not represent the star clusters' physical extent.
The ISM is multiphase with molecular hydrogen (\nth{4} panel) forming in cold regions ($T < 300\,\mathrm{K}$, blueish regions in the \nth{2} panel).
Most of the volume is filled by the WNM ($300 < T \leq 3\times10^5\,\mathrm{K}$, $\chi_\mathrm{ion} < 0.5$) and the warm ionised medium (WIM, $300 < T \leq 3\times10^5\,\mathrm{K}$, $\chi_\mathrm{ion} \geq 0.5$) at temperatures around $T \approx 10^4\,\mathrm{K}$.
Overlapping SN remnants generate pockets of hot gas with temperatures exceeding $T \gtrsim 3 \times10^5\,\mathrm{K}$.
Strong CR-supported outflows, which can lift the gas to heights up to 2~kpc, are present.

\section{Results}\label{sec:results}

\subsection{Star formation}\label{sec:sfr}

\begin{figure}
 \centering
 \includegraphics[width=.95\linewidth]{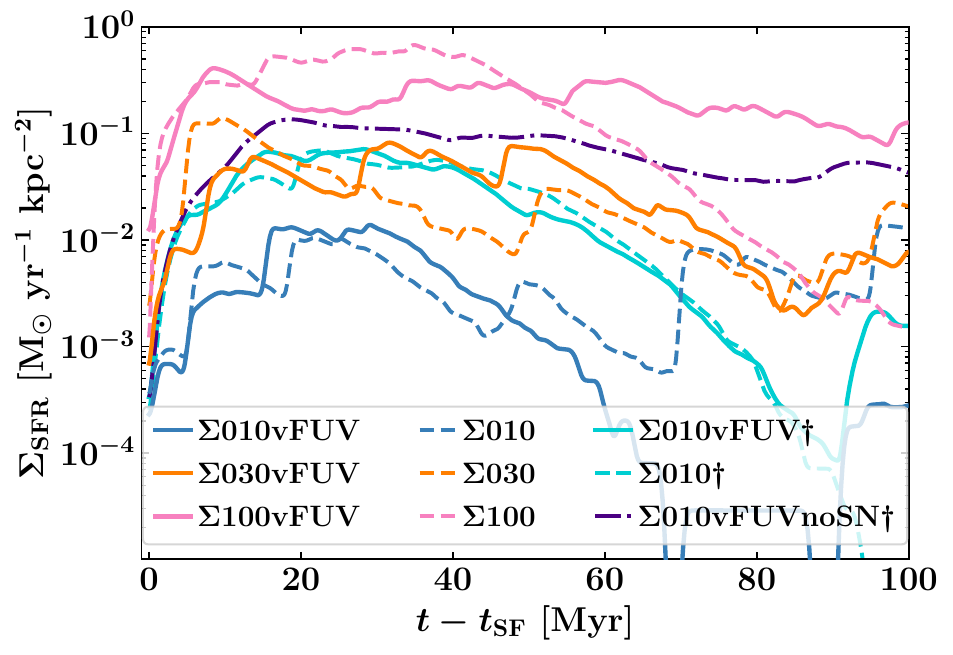}
 \caption{Star formation rate surface density, $\Sigma_\mathrm{SFR}$, as a function of simulated time after the onset of star formation, $t - t_\mathrm{SF}$.
 Runs with the \textsc{AdaptiveG0} model are shown in solid lines while their static $G_0$ counterparts are depicted with dashed lines.
 We also include two simulations with only SN feedback (cyan dashed) and SN+FUV feedback (cyan solid).
 In SN-only runs, an adaptive $G_0$ makes virtually no difference to the SFR (see similarity of the two cyan lines).
 The purple dash-dotted line shows a model at solar neighbourhood conditions without any feedback but the variable FUV radiation via \textsc{AdaptiveG0}.
 Self-consistent FUV feedback does not seem to efficiently regulate star formation and produce $\Sigma_\mathrm{SFR}$ in line with predictions motivated by observations.}
 \label{fig:sfr}
\end{figure}

In Fig. \ref{fig:sfr}, we show the time evolution of the star formation rate surface density, $\Sigma_\mathrm{SFR}$.
We compute $\Sigma_\mathrm{SFR}$ by summing over all active massive stars, $i$, at a given time, $t$, account for the \mbox{120 M$_\odot$} of accreted gas to form each individual massive star (see Sect. \ref{sec:fuv}), and divide by the lifetime of that massive star $t_{\mathrm{OB,} i}$.
This gives
\begin{equation}
    \Sigma_\mathrm{SFR}(t) = A^{-1}~\sum\limits_{i=1}^{N_\star} 120~\mathrm{M}_\odot~t_{\mathrm{OB,} i}^{-1},
\end{equation}
 with the surface area of the midplane ISM, $A = (500~\mathrm{pc})^2$.
This approach is motivated by the aim of directly comparing $\Sigma_\mathrm{SFR}$ to observational measurements, such as those derived from H$\alpha$ emission, which is primarily dependent on the presence of massive OB stars.
It is favoured over the calculation of an instantaneous $\Sigma_\mathrm{SFR}$, which is highly sensitive to the choice of a time bin $\Delta t$.
However, both methods for determining $\Sigma_\mathrm{SFR}$ yield consistent results and align well when comparing integrated values \citep[see e.g.][]{Gatto2017}.
The data is presented up to $t - t_\mathrm{SF}=100$~Myr.
Note that $t_\mathrm{SF}$ is different for each model (between $\sim5\--20$~Myr).
Solid lines indicate models that include the \textsc{AdaptiveG0} module, while dashed lines represent the respective models with static $G_0$.
The two models shown with cyan colour are the runs without early stellar feedback in the form of stellar winds and ionising radiation.
We do not observe any distinct trends that would indicate the impact of the locally varying FUV radiation on $\Sigma_\mathrm{SFR}$ across different galactic environments.
Models with and without \textsc{AdaptiveG0} closely follow each other.
For each pair of models with equal $\Sigma_\mathrm{gas}$, there are periods in time where $\Sigma_\mathrm{SFR}$ of one model is higher/lower than the other, but those differences are most likely due to the highly non-linear stochastic nature of the star formation process.
We want to especially highlight the nearly identical behaviour of the models $\mathrm{\Sigma010vFUV\dagger}$ and $\mathrm{\Sigma010\dagger}$ with respect to $\Sigma_\mathrm{SFR}$.
Massive star formation in those models is purely regulated by SNe and their $\Sigma_\mathrm{SFR}$ is about one order of magnitude higher than their counterparts including stellar winds and ionising radiation ($\mathrm{\Sigma010}$ and $\mathrm{\Sigma010vFUV}$).
This is a clear indication of the importance of HII regions and massive-star winds in the process of regulating star formation. At the same time, the impact of the variable FUV field seems to be minimal.
The model without any stellar feedback besides variable FUV radiation, $\mathrm{\Sigma010vFUVnoSN\dagger}$, stabilises at $\Sigma_\mathrm{SFR} \approx 10^{-1}\,\mathrm{M_\odot\,yr^{-1}\,kpc^{-2}}$, which is approximately twice as high as that of the $\mathrm{\Sigma010vFUV\dagger}$ and $\mathrm{\Sigma010\dagger}$ models.

\begin{figure}
 \centering
 \includegraphics[width=.95\linewidth]{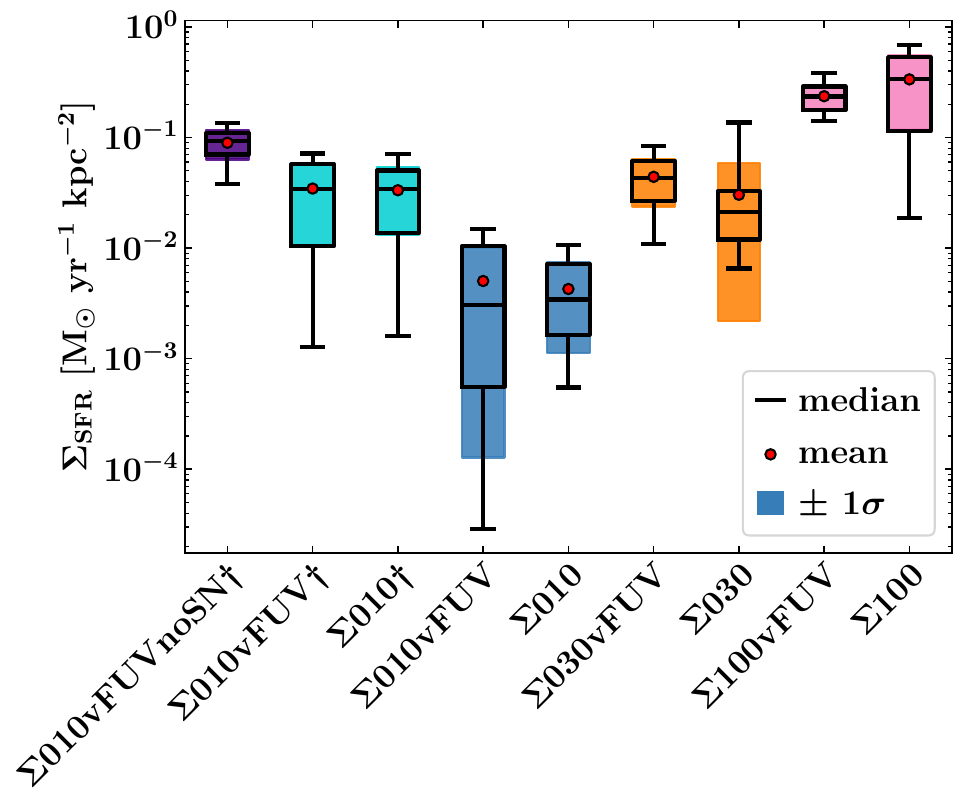}
 \caption{Star formation rate surface densities summarised as box plots.
 We calculate the values for a time evolution starting 10 Myr after the onset of star formation to minimise the possible impact of the numerical initial conditions on the first star formation episode.
 Red dots indicate the averaged $\Sigma_\mathrm{SFR}$ values.
 A standard deviation of $\pm\,1\sigma$ is shown as coloured shaded areas.
 The boxes span from the \nth{25} to the \nth{75} percentile and the whiskers indicate the minimum and maximum value of the time evolution.
 No clear trends for the impact of the variable FUV radiation are visible between the different simulation models.}
 \label{fig:sfr_sum}
\end{figure}

To better quantify the results of Fig. \ref{fig:sfr}, we show the distribution of $\Sigma_\mathrm{SFR}$ for each model in Fig. \ref{fig:sfr_sum} over a time frame of $t-t_\mathrm{SF} = 10\-100$~Myr.
We start only 10~Myr after the onset of star formation to reduce the possible impact of the initial conditions.
Each box plot in Fig. \ref{fig:sfr_sum} encompasses the $\Sigma_\mathrm{SFR}$ data within the \nth{25}- and \nth{75}-percentiles.
Thick black lines inside each box plot indicate the median value, whereas the time-averaged mean is shown as red dots.
Coloured shaded areas show the standard deviation of the mean ($\pm 1\sigma$).
The whiskers show the minimum and maximum values of the time evolution.
The integrated values of $\Sigma_\mathrm{SFR}$ do not differ with statistical significance.
Between $\mathrm{\Sigma010vFUV}$ and $\mathrm{\Sigma010vFUV\dagger}$, $\Sigma_\mathrm{SFR}$ differs by one order of magnitude.
Therefore, these results suggest that $\Sigma_\mathrm{SFR}$ is regulated by ionising radiation and hot wind bubbles, rather than by non-ionising radiation.
Observations of local galactic star-forming regions obtained by \citet{Leroy2008} estimate $\Sigma_\mathrm{SFR} \approx 7.89\times10^{-3}$~M$_\odot$~yr$^{-1}$~kpc$^{-2}$ for $\Sigma_\mathrm{gas} \approx 10$~M$_\odot$~pc$^{-2}$.
These observations, however, have a large scatter, especially at conditions similar to the solar neighbourhood.
Models without stellar wind and ionising radiation feedback (purple and cyan) overestimate $\Sigma_\mathrm{SFR}$ relative to observational predictions.

The variable FUV radiation field causes a highly non-uniform PE heating rate within the computational domain. However, the impact of the FUV radiation feedback is subdominant to photoionisation for supporting the gas against further collapse, thus regulating the accretion of gas onto the cluster and, as a consequence, the SFR.
The distribution of $\Sigma_\mathrm{SFR}$ over time in models without photoionisation and stellar winds ($\mathrm{\Sigma010vFUV\dagger}$ and $\mathrm{\Sigma010\dagger}$) is nearly identical.
There are no consistent trends observed when comparing the models with the \textsc{AdatpiveG0} module to those without it across different galactic environments.
The \textsc{AdaptiveG0} models can have either higher or lower $\Sigma_\mathrm{SFR}$ compared to the static $G_0$ models.
This variable relationship holds true for all $\Sigma_\mathrm{gas}$ realisations.

We analyse the small-scale fluctuations of star formation and define a burstiness parameter, $\beta_\mathrm{SF}$, as the ratio of the interquartile range (IQR) and the median of $\Sigma_\mathrm{SFR}$,
\begin{align}
    \beta_\mathrm{SF} \equiv \mathrm{IQR(\Sigma_\mathrm{SFR})}~\mathrm{Median^{-1}(\Sigma_\mathrm{SFR})}.
\end{align}
When averaged over all initial $\Sigma_\mathrm{gas}$, the burstiness of star formation increases with the addition of \textsc{AdaptiveG0} from $\beta_\mathrm{SF} = 1.3\pm0.2$ to $\beta_\mathrm{SF} = 1.5\pm1.2$.
However, as can be seen at the large uncertainty estimate, this result is strongly skewed by the sharp decrease of $\Sigma_\mathrm{SFR}$ in $\mathrm{\Sigma010vFUV}$ starting at $t - t_\mathrm{SF} \sim 50\,\mathrm{Myr}$.
When excluding the solar neighbourhood models, we measure a decrease in $\beta_\mathrm{SF}$ by nearly a factor of 2 when accounting for a self-consistent FUV treatment from $\beta_\mathrm{SF}^\prime=1.2\pm0.2$ in the static $G_0$ models to $\beta_\mathrm{SF}^\prime=0.7\pm0.2$ in \textsc{AdaptiveG0} models.
Another possibility is to measure $\beta_\mathrm{SF}$ only for the first 50 Myr of star formation to exclude the likely unrelated $\Sigma_\mathrm{SFR}$ dip in $\mathrm{\Sigma010vFUV}$.
Doing so results in $\beta_\mathrm{SF}^\prime=1.1\pm0.5$ for the static $G_0$ models and $\beta_\mathrm{SF}^\prime=0.7\pm0.3$ for \textsc{AdaptiveG0} models.
A potential explanation for this phenomenon could be a homogenisation of the star-forming gas through the additional PE heating.
We argue that the decrease of $\Sigma_\mathrm{SFR}$ in $\mathrm{\Sigma010vFUV}$ does not stem from potentially FUV-regulated local star formation but from large-scale gas dynamical effects.
We confirm this hypothesis in Appendix \ref{app:long}, showing how the solar neighbourhood runs evolve over a longer time frame with $\beta_\mathrm{SF}$ decreasing between $\mathrm{\Sigma010}$ and $\mathrm{\Sigma010vFUV}$.

We give a summary of the $\Sigma_\mathrm{SFR}$ data in Table \ref{tab:SFR} in Appendix \ref{app:tables}.

\subsection{Gas structure}\label{sec:gasstructure}

\begin{figure}
 \centering
 \includegraphics[width=.95\linewidth]{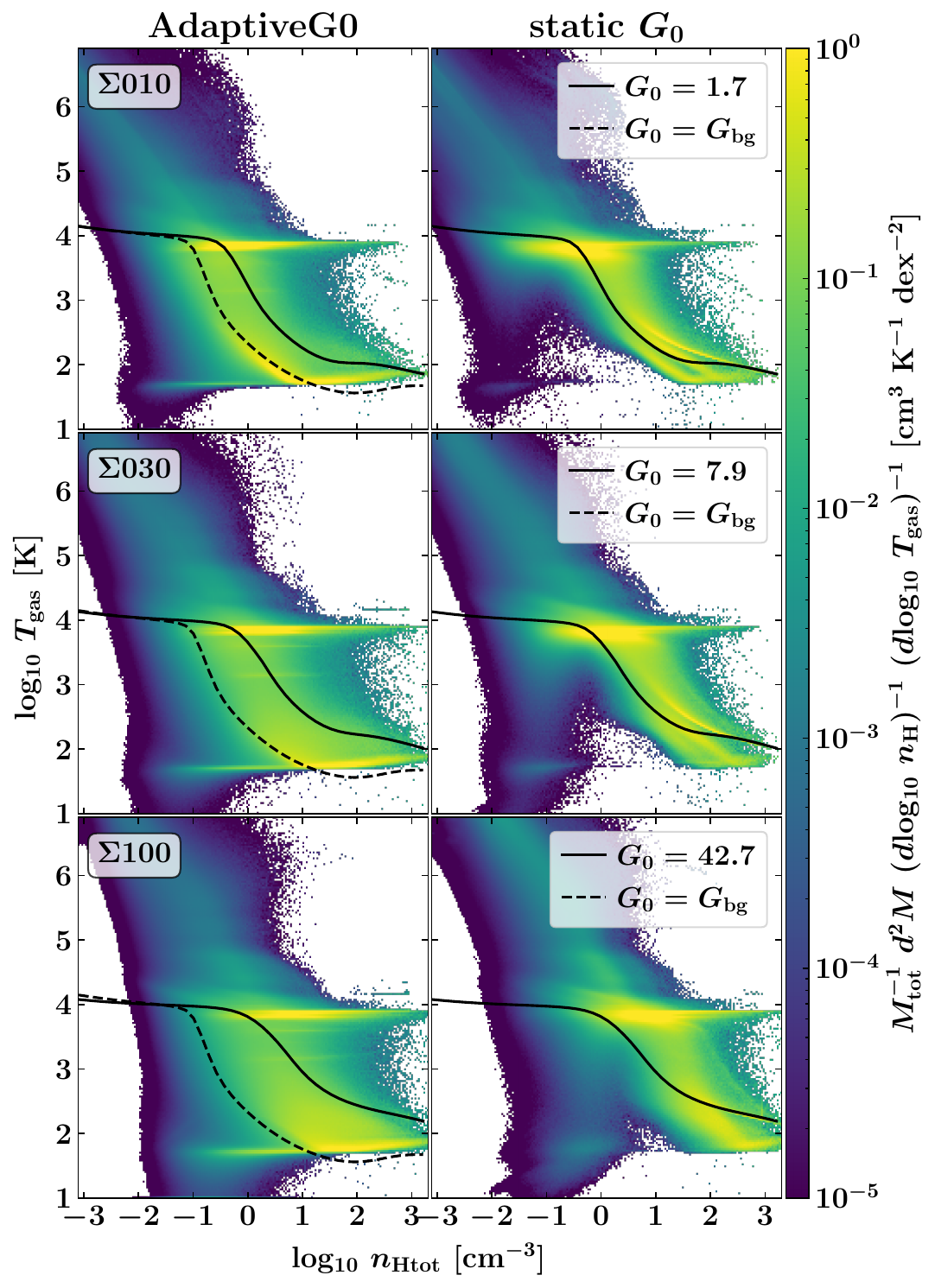}
 \caption{Joint probability density functions showing the gas mass distribution in the plane of the gas temperature, $T_\mathrm{gas}$ and the total hydrogen nuclei number density, $ n_\mathrm{Htot}$.
 We only show data near the disc midplane (defined as $|z| \leq 500$ pc). We cover the time range from $t - t_\mathrm{SF} = 10\,\mathrm{Myr}$ till the end of the simulated time at $t - t_\mathrm{SF} = 100\,\mathrm{Myr}$, to exclude initial condition effects.
 We show the static $G_0$ models in the right column and indicate the equilibrium cooling curve for the given static $G_0$ value as a solid black line.
 For the models with the \textsc{AdaptiveG0} module (left column), we also indicate the equilibrium cooling curve for the corresponding static $G_0$ (solid line), as well as the equilibrium cooling curve for the background $G_\mathrm{bg} = 0.0948$ (dashed line).
 For both cooling curves we assume an external shielding with $A_\mathrm{V} = 1$, similar to Fig. \ref{fig:equibCurve}.
 The wider range of potential cooling curves (also see Fig. \ref{fig:equibCurve}) leads to a smeared-out temperature-density PDF for temperatures between a couple tens to 10$^4$ K.}
 \label{fig:dens_temp_pdf}
\end{figure}

\begin{figure}
 \centering
 \includegraphics[width=.95\linewidth]{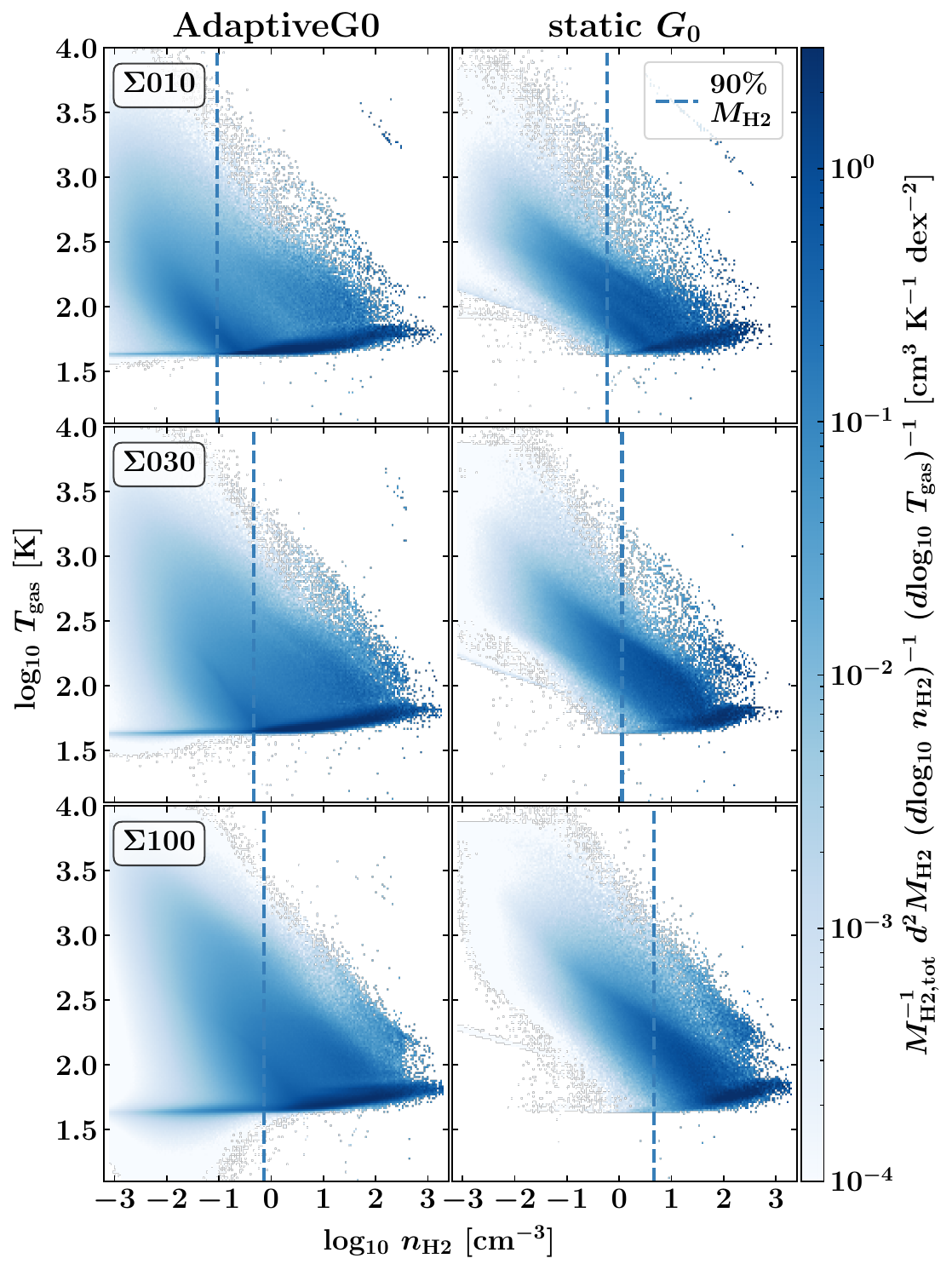}
 \caption{Same as Fig. \ref{fig:dens_temp_pdf} but for $T_\mathrm{gas}$ and molecular hydrogen number density, $n_\mathrm{H2}$.
 We indicate the H$_2$ number density above which $90\,\mathrm{per\,cent}$ for the total molecular hydrogen gas mass, $M_\mathrm{H2}$, resides with vertical dashed lines.
 This threshold density is approximately one order of magnitude lower in the \textsc{AdaptiveG0} cases compared to the static $G_0$ counterparts.
 The variable FUV radiation field promotes the formation of a diffuse molecular hydrogen phase with temperatures below 100 K.}
 \label{fig:H2_temp_pdf}
\end{figure}

We investigate the effects of a time- and space-varying ISRF on the gas phase dynamics of the ISM.
In Fig. \ref{fig:dens_temp_pdf}, we display mass-weighted joint probability density functions (PDFs) of the gas temperature, $T_\mathrm{gas}$, and the total number density of hydrogen nuclei, $ n_\mathrm{Htot}$.
The left column shows models with \textsc{AdaptiveG0}, while the right column shows corresponding models with a static $G_0$.
As a reference, equilibrium cooling curves are displayed for as solid black lines for $G_0 = [1.7, 7.9, 42.7]$, corresponding to $G_\mathrm{eff} = [0.14, 0.65, 3.51]$ (compare with Fig. \ref{fig:equibCurve}) and for $G_0 = G_\mathrm{bg} = 0.0948$.
The gas phase distribution in models with a static $G_0$ closely matches the equilibrium configuration, except for the HII region branch at $T\sim10^4\,\mathrm{K}$.
Hot, low-density gas (upper left part of the phase diagrams) is shock-heated by SNe.
The marginal amount of cold, low-density gas results from CRs, accelerated in SN remnants, and their diffusion.
Additional CR pressure near embedded star clusters leads to gas expansion via adiabatic processes.
However, after multiple SNe explode in an area, overlapping remnants thermalise the gas, rapidly heating it to temperatures above $T \gtrsim 3 \times 10^5,\mathrm{K}$.
This scenario changes with a variable FUV radiation field.
In static $G_0$ models, most of the gas mass ($67.1\pm3.2$~per cent) resides above their respective solid black $G_0$ equilibrium cooling curve in the $n_\mathrm{Htot}$-$T_\mathrm{gas}$ plane\footnote{We want to note that these cooling curves are all a function of the external visual extinction, $A_\mathrm{V}$.
The curves depicted in Fig. \ref{fig:dens_temp_pdf} have been calculated for $A_\mathrm{V} = 1$.
Gas in more shielded regions would end up at a lower equilibrium temperature while more diffuse gas would be able to be heated more.}.
The majority of this gas is photoionised within HII regions.
For \textsc{AdaptiveG0} models, this is significantly different, with a total gas mass fraction (MF) of $31.2\pm3.9$~per cent above the corresponding equilibrium phase.
Due to the broad spatial variations of the FUV ISRF, the gas phase distribution broadens and smears out between indicated equilibrium states (solid line and dashed line).
A distinct branch of cold ($T < 300\,\mathrm{K}$) and diffuse ($ n_\mathrm{Htot} \lesssim 1\,\mathrm{cm^{-3}}$) gas containing a significant MF emerges.
As in the static $G_0$ case, CR pressure partially contributes to this.
However, this diffuse, cold (and potentially molecular) gas mostly resides far from young massive star clusters, where it is not exposed to strong FUV radiation and now it is no longer heated by a static radiation field with $G_0$.

We further analyse this cold diffuse gas branch by examining another density-temperature phase diagram in Fig. \ref{fig:H2_temp_pdf}, focusing this time on molecular hydrogen gas density ($n_\mathrm{H2}$).
As a reference, the threshold molecular hydrogen density above which $90\,\mathrm{per\,cent}$ of all molecular hydrogen resides are indicated with vertical dashed lines in Fig. \ref{fig:H2_temp_pdf}.
In both scenarios, with and without \textsc{AdaptiveG0}, most molecular hydrogen gas is in a cold, dense gas phase\footnote{We note that we do not explicitly follow the gas phases of individual chemical species with tracer particles.
For each cell in the computational domain, we know the total gas density, temperature, and MF of each chemical species included in the network (see Sect.~\ref{sec:numerics}).
Gas mixing on scales smaller than $\sim (4\,\mathrm{pc})^3$ (corresponding to the resolution of the AMR grid in the midplane ISM) is not resolved.}.
Similar to Fig. \ref{fig:dens_temp_pdf}, a self-consistent FUV radiation field broadens the temperature-density distribution in the ISM.
Based on the $90\,\mathrm{per\,cent}$ H$_2$ density threshold in the static $G_0$ models (dashed blue vertical line in Fig. \ref{fig:H2_temp_pdf}), we define diffuse H$_2$ as $n_\mathrm{H2} < 2$~cm$^{-3}$ and dense/compact H$_2$ as $n_\mathrm{H2} \geq 2$~cm$^{-3}$.
With this definition, the average MF of diffuse H$_2$ gas weighted by the total H$_2$ mass in the \textsc{AdaptiveG0} models is $22.5\pm7.1$~per cent and $8.9\pm6.2$~per cent in the static $G_0$ models.
We identify a clear anti-correlation of the diffuse H$_2$ MF with $\Sigma_\mathrm{gas}$, as it ranges from $40.1\pm12.2$~per cent in $\Sigma010\mathrm{vFUV}$ down to $20.4\pm2.7$~per cent in $\Sigma100\mathrm{vFUV}$.
A similar anti-correlation exists for the static $G_0$ models.
The diffuse, cold (and molecular) gas is present in the \textsc{AdaptiveG0} models but is mostly absent in the static $G_0$ runs.
Nonetheless, most of the H$_2$ gas still exists in high-density environments, especially at higher $\Sigma_\mathrm{gas}$.

\begin{figure}
 \centering
 \includegraphics[width=.95\linewidth]{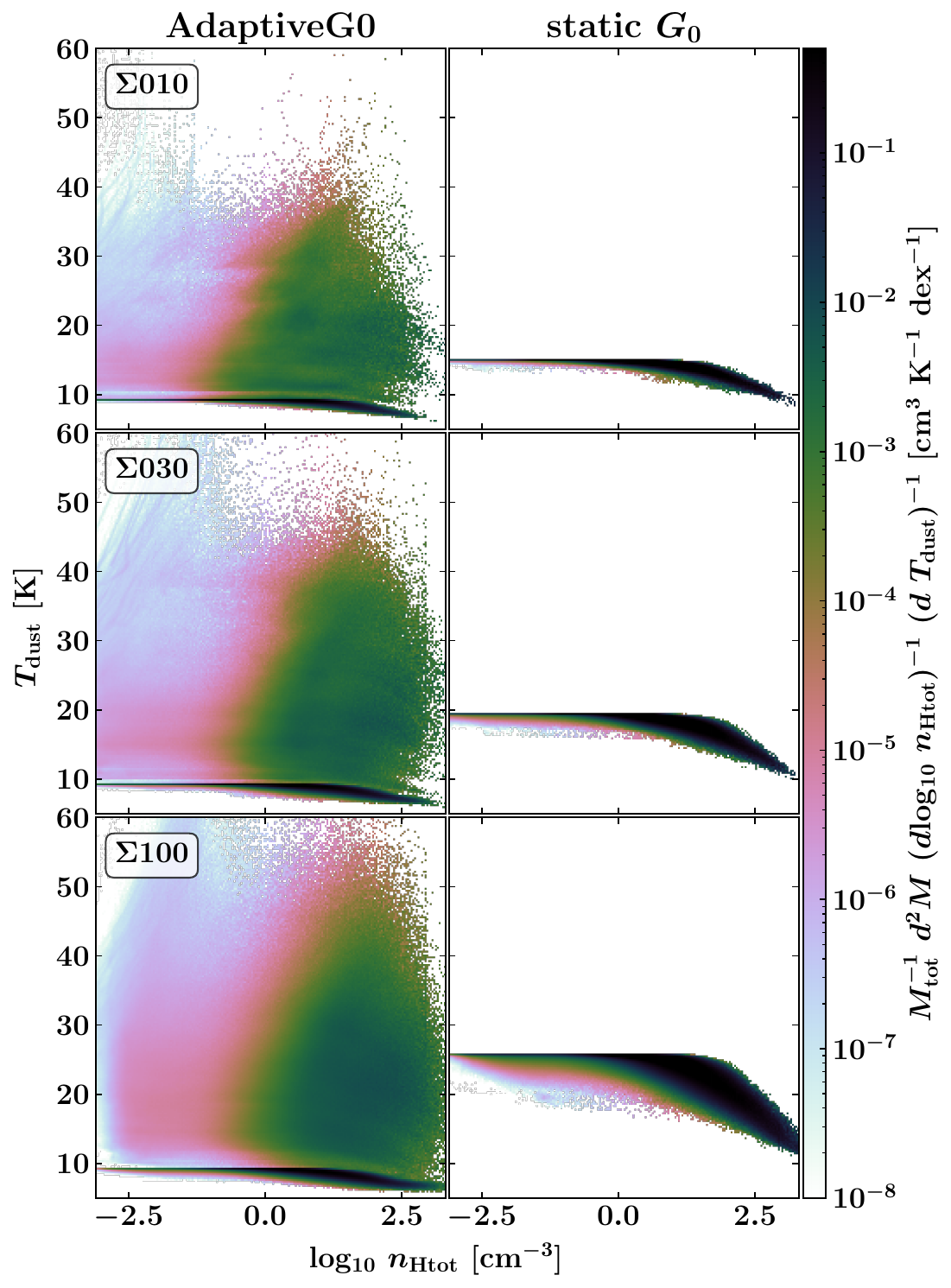}
 \caption{Joint probability density functions showing the mass distribution in the dust temperature, $T_\mathrm{dust}$, and $ n_\mathrm{Htot}$ plane.
 Since we employ a constant gas-to-dust mass ratio of one per cent, $ n_\mathrm{Htot}$ is a tracer for $n_\mathrm{dust}$.
 A large portion of the dust in all \textsc{AdaptiveG0} models sits at lower temperatures compared to the static models due to lesser heating in regimes of $G_\mathrm{eff} = G_\mathrm{bg}$.
 However, dust closer to star-forming regions does not show a flat distribution as in the static models and can be heated up to 60 K via the PE effect.
 A wider spread in dust temperature can substantially impact synthetic observables deduced from our models.}
 \label{fig:dust-temp}
\end{figure}

We focus on the dust density and temperature distribution in Fig. \ref{fig:dust-temp}, similar to Fig. \ref{fig:dens_temp_pdf}.
We can show $T_\mathrm{dust}$ as a function of $ n_\mathrm{Htot}$ because we adopt a constant dust-to-gas MF of $1\,\mathrm{per\,cent}$ in our chemical network.
For a static radiation field, the dust temperature has a sharp upper limit between $T_\mathrm{dust} \approx 15 - 25\,\mathrm{K}$, depending on the strength of the static $G_0$ field, which scales with the initial $\Sigma_\mathrm{gas}$ of each model.
The temperature distributions are almost flat, with a slight trend to colder temperatures at higher densities because of an increasing extinction.
With \textsc{AdaptiveG0}, the majority of the dust in the computational domain (by mass) has a temperature of $T_\mathrm{dust} \approx 10\,\mathrm{K}$, which reflects the temperature reached through heating with the background FUV radiation field $G_\mathrm{bg} = 0.0948$.
However, a significant amount of dust is heated to temperatures around $T_\mathrm{dust} \approx 60\,\mathrm{K}$.
There is no clear correlation between the gas density (and therefore also dust density) and dust temperature.
Still, most of the dust (by mass) resides in gas reservoirs with densities above $ n_\mathrm{Htot} \gtrsim 1\,\mathrm{cm^{-3}}$.
The variations in $T_\mathrm{dust}$ are negligible for the gas dynamics in our models.
Nonetheless, observables are strongly dependent on $T_\mathrm{dust}$ \citep[see e.g.][]{Bisbas2022}.
The dust temperature we recover for the bulk of the gas in the $\textsc{AdaptiveG0}$ models is systemically slightly below those seen in Milky Way observations of $T_\mathrm{dust} \approx 15 - 20\,\mathrm{K}$ \citep{Marsh2017}.
The reason behind this is that in our chemical network so far, we only account for dust heating by EUV and FUV but not by far-infrared radiation (FIR).
We discuss this discrepancy further in Sect. \ref{sec:caveats}.

\subsection{Heating and cooling}\label{sec:heatcool}

The thermal balance in the warm neutral ISM is governed by a complex interplay of various heating and cooling processes.
In the absence of photoionisation, PE heating is the primary contributor to the local heating rate, $\Gamma$.
The cooling rate, $\Lambda$, in the WNM is determined by multiple processes, with the most significant being the fine-structure cooling of atomic oxygen (O) and singly-ionised carbon (C$^+$), through OI and CII transitions, respectively.
Lyman-alpha cooling, resulting from electronic de-excitation, also plays a substantial role, while in cooler environments, H$_2$ rovibrational transitions become an important cooling mechanism.
An overview of all relevant heating and cooling processes in our chemical network, as a function of ISM density, is provided in Appendix Fig.~\ref{fig:HeatCool}.

\subsubsection{Cooling rates}\label{sec:cool}

\begin{figure}
 \centering
 \includegraphics[width=.95\linewidth]{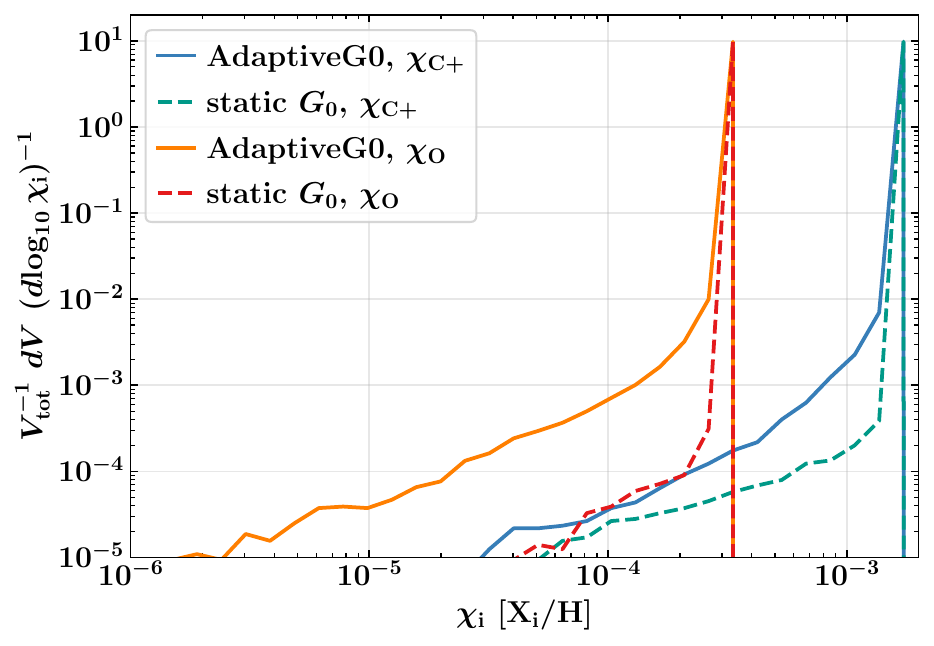}
 \caption{Volume-weighted PDFs of the abundances of atomic oxygen (O) and singly-ionised carbon (C$^+$), derived from simulations using \textsc{AdaptiveG0} (solid lines) and static $G_0$ (dashed lines).
 Each abundance bin represents the average value across simulations with varying $\Sigma_\mathrm{gas}$.
 While the peak abundance values remain similar between the two models, the \textsc{AdaptiveG0} simulations exhibit a broader abundance distribution for both O and C$^+$, indicating greater variability.}
 \label{fig:abundoc}
\end{figure}

In Fig.~\ref{fig:abundoc}, we present the volume-weighted PDFs of the abundances of the main coolants in the WNM, specifically $\chi_\mathrm{C+}$ and $\chi_\mathrm{O}$.
The data are compiled from simulations with varying initial $\Sigma_\mathrm{gas}$, and the PDFs are shown for both \textsc{AdaptiveG0} and static $G_0$ models.
The PDFs exhibit similar peak abundances for simulations with and without \textsc{AdaptiveG0}, however, the distributions in the \textsc{AdaptiveG0} runs are broader, indicating a wider spread of abundances.
This broader distribution reflects the more realistic treatment of the FUV radiation field in \textsc{AdaptiveG0}, which varies spatially and temporally according to the local distribution of stellar sources.
As a result, the ISM in \textsc{AdaptiveG0} simulations encompasses both strongly irradiated and well-shielded regions, leading to greater chemical diversity.
In contrast, simulations with static $G_0$ apply uniform irradiation, resulting in a more homogenised abundance distribution.
Overall, the abundance of $\chi_\mathrm{C+}$ increases slightly by a factor of $1.3\pm0.5$, while $\chi_\mathrm{O}$ increases by a factor of $1.2\pm0.3$ when accounting for \textsc{AdaptiveG0}.
This increase scales with inital $\Sigma_\mathrm{gas}$, as the $\Sigma_\mathrm{gas} = 100$\,M$_\odot$ pc$^{-2}$ models show the highest increase by a factor of 1.7 and 1.6 for $\chi_\mathrm{C+}$ and $\chi_\mathrm{O}$, respectively.
Furthermore, the peak values of the abundance distributions remain nearly identical, with $\chi_\mathrm{C+} \approx 2\times10^{-3}$ and $\chi_\mathrm{O} \approx 3\times10^{-4}$.
\textsc{AdaptiveG0} models exhibit a pronounced tail towards lower abundances in the abundance distribution, which leads to the slight increase of the integrated abundances.

\begin{figure}
 \centering
 \includegraphics[width=.95\linewidth]{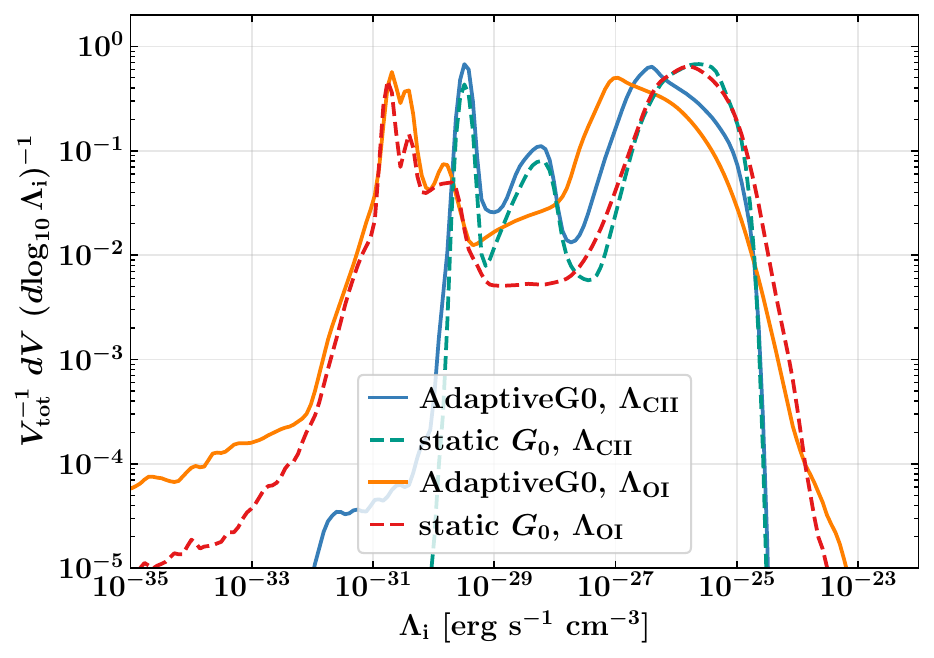}
 \caption{Volume-weighted PDFs of the cooling rates ($\Lambda_i$) for C$^+$ and O, obtained from simulations using \textsc{AdaptiveG0} (solid lines) and static $G_0$ (dashed lines).
 The PDFs illustrate the distribution of cooling rates for the two main cooling agents in the WNM, highlighting the differences in cooling efficiency between the \textsc{AdaptiveG0} and static $G_0$ models.}
 \label{fig:LambdaCIIOI}
\end{figure}

In Fig.~\ref{fig:LambdaCIIOI}, we present the volume-weighted PDFs of the cooling rates ($\Lambda_i$) for C$^+$ and O, comparing simulations using \textsc{AdaptiveG0} (solid lines) and static $G_0$ (dashed lines).
The distributions exhibit a roughly bimodal structure for both $\Lambda_\mathrm{CII}$ and $\Lambda_\mathrm{OI}$, with the two peaks separated by up to four to six orders of magnitude.
Models using static $G_0$ display a higher maximum cooling rate, exceeding those of \textsc{AdaptiveG0} by approximately one order of magnitude.
When integrated over the volume of the midplane ISM and averaged across simulations with varying initial $\Sigma_\mathrm{gas}$, the CII and OI cooling rates decrease in the \textsc{AdaptiveG0} models, reaching $38^{51}_{35}$\,per\,cent and $17^{28}_{16}$\,per\,cent (median$^\mathrm{\nth{75}percentile}_\mathrm{\nth{25}percentile}$), respectively.
This reduction may seem counterintuitive, given the slight increase in the average abundances of O and C$^+$ in \textsc{AdaptiveG0} (see Fig.~\ref{fig:abundoc}).
However, the cooling rate in optically thin gas generally follows the relation:
\begin{align}
\Lambda_i \propto \chi_i n_\mathrm{H} n_\mathrm{e} \gamma(T),
\end{align}
where $\gamma(T)$ is the temperature-dependent collisional excitation rate coefficient.
Since the abundance enters linearly, the cooling rate can decrease even with a slight increase in abundance if the ISM conditions evolve differently, as is the case between the \textsc{AdaptiveG0} and static $G_0$ models.
Furthermore, the increase in abundance happens mostly in the low-abundance tail of the distribution (see Fig.~\ref{fig:abundoc}.
The absolute number density of C$+$ and especially of O remains small and has therefore only a minimal positive effect on the cooling rate.

\subsubsection{Heating rates}\label{sec:heating}

As already evident from the negligible difference in SFR between the $\textsc{AdaptiveG0}$ and static $G_0$ models (see Sect.~\ref{sec:sfr}), the contribution of PE heating to the suppression of star formation is minor compared to that of local photoionisation (PI) heating.
To quantify this, we compute the ratio of PI to PE heating, $\chi_\mathrm{PI/PE}$, for all cells within a distance of $d_\star = 100$\,pc from the nearest star cluster sink particle.
The data is compiled across all simulations with varying $\Sigma_\mathrm{gas}$.
For the \textsc{AdaptiveG0} models, we find $\chi_\mathrm{PI/PE} = 9.1^{17.4}_{7.6}$ (median$^\mathrm{\nth{75}percentile}_\mathrm{\nth{25}percentile}$), while for the static $G_0$ models, the ratio increases to $\chi_\mathrm{PI/PE} = 46.9^{452.3}_{15.5}$.
The self-consistent treatment of the FUV radiation field enhances $\Gamma_\mathrm{PE}$ in the vicinity of stellar clusters, yet $\Gamma_\mathrm{PI}$ remains approximately an order of magnitude stronger.

\begin{figure}
 \centering
 \includegraphics[width=.95\linewidth]{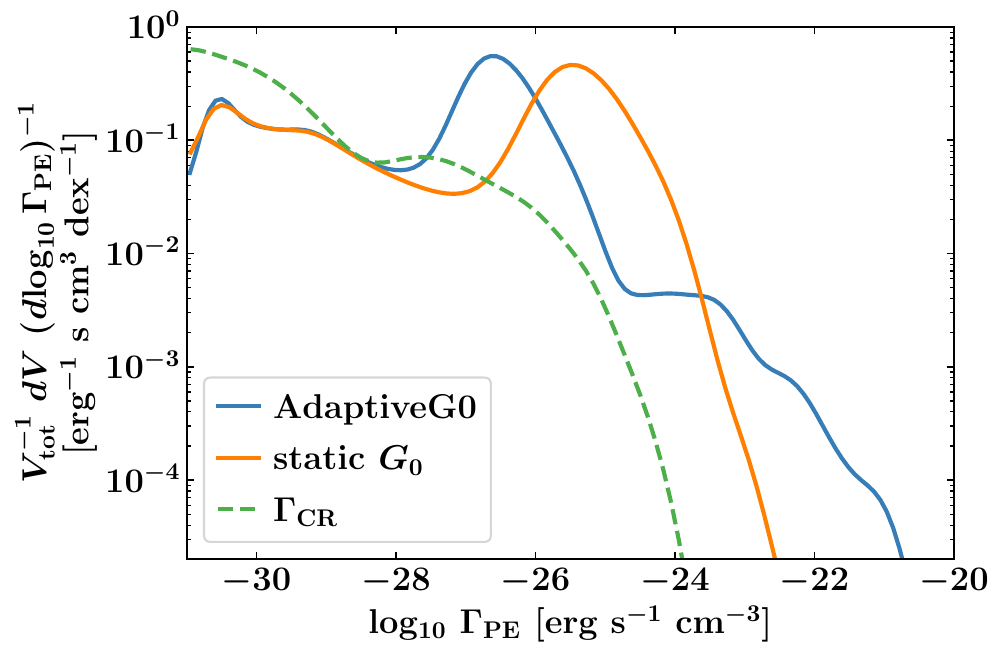}
 \caption{Probability density functions of $\Gamma_\mathrm{PE}$ for the cases with \textsc{AdaptiveG0} and static $G_0$.
 The data for this plot were taken from representative snapshots of the $\mathrm{\Sigma010}$ model around the peak of star formation.
 We also show the heating caused by CR ionisation, $\Gamma_\mathrm{CR}$, with a dashed green line.
 The model with a static $G_0$ field reaches significantly lower maximum $\Gamma_\mathrm{PE}$ values but has the peak of the distribution at higher heating rates than the model with \textsc{AdaptiveG0}.
 CR heating is subordinate to the peak heating rates due to the PE effect on dust grains.}
 \label{fig:heatingDist}
\end{figure}

We want to focus on the impact of the \textsc{AdaptiveG0} implementation on $\Gamma_\mathrm{PE}$ and take a representative snapshot from the static $G_0$ simulation $\mathrm{\Sigma010}$ and re-run the simulation with the \textsc{AdaptiveG0} module for 0.1 Myr starting from this snapshot.
The time of the representative snapshot has been chosen so that its instantaneous SFR is close to the globally averaged SFR at $t_\mathrm{SF} = 19.8\,\mathrm{Myr}$.
By doing so, we obtain the closest possible comparison between the two realisations.
In Fig. \ref{fig:heatingDist}, we show a volume-weighted PDF of $\Gamma_\mathrm{PE}$ measured in the snapshots with static $G_0$ (orange line) and for the same gas configuration with \textsc{AdaptiveG0} (blue line).
The heating due to CR ionisation, $\Gamma_\mathrm{CR}$, is shown as a dashed green line for comparison.
In Appendix \ref{app:PECR}, we show the distribution of $\Gamma_\mathrm{CR}$ in relation to $\Gamma_\mathrm{PE}$.
The gas density structures in the presented snapshots are close to identical to each other and the differences in the heating rates are due to the differences in the ISRF due to the different treatment of the FUV radiation between \textsc{AdaptiveG0} and static $G_0$.

In general, $\Gamma_\mathrm{PE}$ strongly depends on the strength of the FUV radiation field (see Eq. \ref{eq:heatingRate}).
The $\Gamma_\mathrm{PE}$ distribution in Fig. \ref{fig:heatingDist} extends nearly 10 orders of magnitude.
At the low-energy end of the distribution, the $\Gamma_\mathrm{PE}$ values are virtually identical between both simulations.
These $\Gamma_\mathrm{PE}$ correspond to regions with the most external and self-shielding, e.g. dense cores which have not yet begun star formation.
In the \textsc{AdaptiveG0} models, the gas is predominantly exposed to a $\Gamma_\mathrm{PE}$ that is 1.5 to 2 orders of magnitude lower compared to a static $G_0$ field (see the shift between the peaks of the distributions in Fig. \ref{fig:heatingDist}).
However, for a static $G_0$, the amount of gas subjected to high heating rates steeply declines. At the same time, the \textsc{AdaptiveG0} models reach maximum heating rates of about two orders of magnitude higher than their static counterparts.

\begin{figure}
 \centering
 \includegraphics[width=.95\linewidth]{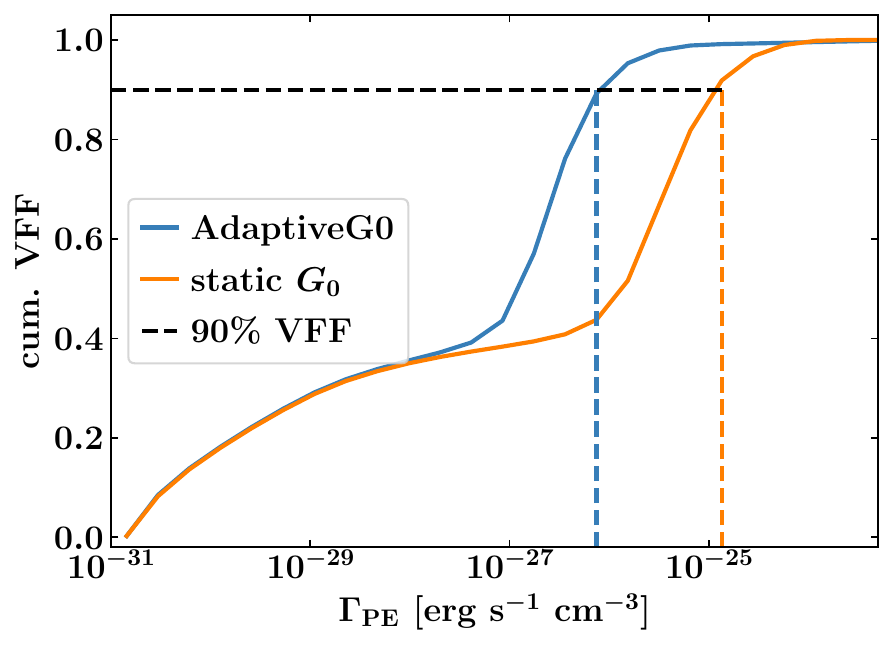}
 \caption{Cumulative volume-filling factors, VFF, of $\Gamma_\mathrm{PE}$, for the case with \textsc{AdaptiveG0} (blue) and static $G_0$ (orange).
 We indicate the maximum heating rate to which $90\,\mathrm{per\,cent}$ of the midplane ISM volume is exposed (dashed vertical lines).
 Even though the integrated amount of PE heating for a \textsc{AdaptiveG0} model far exceeds the equivalent energy in a static $G_0$ model, most of the volume is exposed to lesser PE heating.}
 \label{fig:vffHeat}
\end{figure}

To understand how the increased PE heating affects the ISM in star-forming regions, we examine the cumulative volume-filling factor (VFF) of $\Gamma_\mathrm{PE}$ as shown in Fig. \ref{fig:vffHeat}.
We compare the static $G_0$ model (orange) with the \textsc{AdaptiveG0} model (blue), similar to the comparison in Fig. \ref{fig:heatingDist}, and plot the cumulative VFF of $\Gamma_\mathrm{PE}$ within the midplane ISM ($|z| \leq 500\,\mathrm{pc}$).
Dashed lines indicate the maximum heating rate that affects 90 per cent of the midplane ISM.
Despite a significantly higher total energy input in the \textsc{AdaptiveG0} case, most of the gas (90 per cent) experiences a PE heating rate approximately 1.5 dex lower than in the static $G_0$ model.
The excess energy input is concentrated in small volumes near the star cluster sink particles, which are also exposed to hydrogen-ionising EUV radiation and stellar winds.
These mechanisms can heat the gas to temperatures exceeding $T \gtrsim 8 \times 10^3\,\mathrm{K}$, beyond which even the strongest FUV radiation fields cannot further increase the temperature (see Fig. \ref{fig:equibCurve}).

\begin{figure}
 \centering
 \includegraphics[width=.95\linewidth]{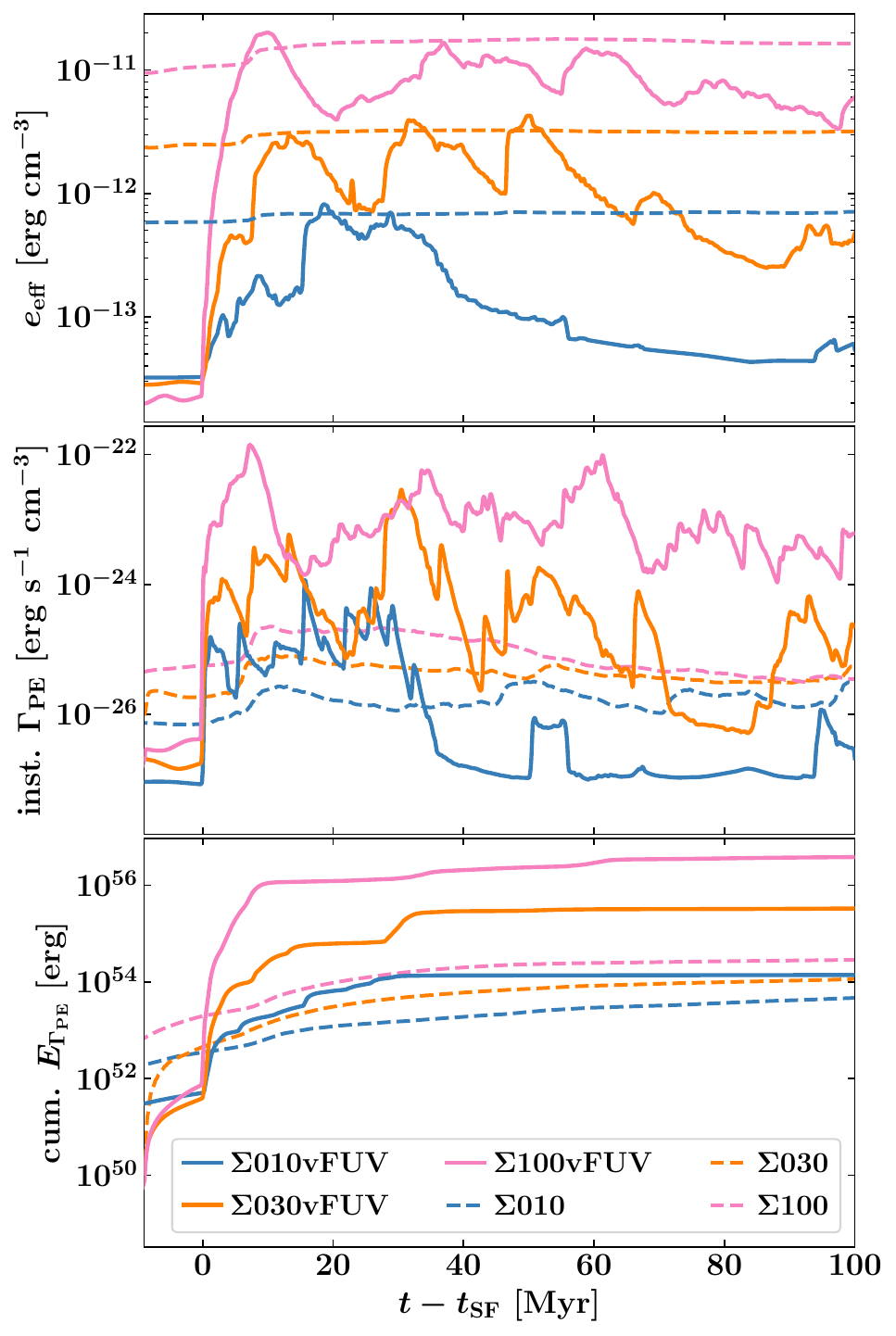}
 \caption{\textit{Top:} Effective FUV radiation field energy density, $e_\mathrm{eff} = G_\mathrm{eff} \times V \times u_\mathrm{Habing}$, with gas volume, $V$, and $u_\mathrm{Habing} = 5.29\times10^{-14}\,\mathrm{erg\,cm^{-3}}$, as a function of time for our different models.
 We include data from before the onset of star formation since PE heating already takes place during these periods, due to the background ISRF with a strength of either $G_\mathrm{bg}$ or the respective values of the static models (see Table \ref{tab:sims}).
 Variations in $e_\mathrm{eff}$ for the static G$_0$ models arise from varying amounts of shielding experience by the otherwise constant FUV radiation field employed in those models.
 \textit{Middle:} Instantaneous PE heating rate, $\Gamma_\mathrm{PE}$.
 PE heating is more efficient in \textsc{AdaptiveG0} models.
 \textit{Bottom:} Cumulative energy injected by PE heating, $E_{\Gamma_\mathrm{PE}}$, over time.
 The total energy injected into the medium via PE heating for \textsc{AdaptiveG0} models vastly exceeds the corresponding values in the static $G_0$ models.
 The amount of PE heating correlates with gas surface density (and therefore also with SFR).}
 \label{fig:cumHeat}
\end{figure}

We present the time evolution of the effective FUV radiation field energy density available for PE heating, $e_\mathrm{eff} = G_\mathrm{eff} \times V \times u_\mathrm{Habing}$, with gas volume, $V$, and $u_\mathrm{Habing} = 5.29\times10^{-14}\,\mathrm{erg\,cm^{-3}}$, for our models with varying initial $\Sigma_\mathrm{gas}$ in Fig. \ref{fig:cumHeat} (top panel), along with the instantaneous volume-averaged midplane $\Gamma_\mathrm{PE}$ (middle panel) and the cumulative thermal energy injected through PE heating (bottom panel).
Unlike previous analyses, we include data from before the onset of star formation, $t_\mathrm{SF}$.
This inclusion is motivated by the PE heating originating from the static $G_0$ field independent of star formation in one case, and the heating from the background $G_\mathrm{bg}$ in the other case.
The midplane $e_\mathrm{eff}$ is consistently higher in static $G_0$ models compared to \textsc{AdaptiveG0}.
Averaged over the full time-evolution and all three $\Sigma_\mathrm{gas}$ models, $e_\mathrm{eff}$ in \textsc{AdaptiveG0} is $\sim 41\pm11\,\mathrm{per\,cent}$ of that in static $G_0$.
As expected, $\Gamma_\mathrm{PE}$ evolves flat in the static $G_0$ case with only minor variations due to local changes in the column densities.
For the \textsc{AdaptiveG0} models, $\Gamma_\mathrm{PE}$ ramps up drastically with the onset of star formation and strongly varies over time.
At higher surface densities ($\mathrm{\Sigma030vFUV}$ and $\mathrm{\Sigma100vFUV}$), more energy is constantly injected through PE heating than in their corresponding static $G_0$ counterparts.
This demonstrates that PE heating is more efficient in \textsc{AdaptiveG0} models with spatially and temporally varying FUV radiation fields, achieving higher overall energy input despite lower available FUV radiation energy density.
For the solar neighbourhood model ($\mathrm{\Sigma010vFUV}$), PE heating injects more energy into the medium during episodes of strong star formation.
However, it imparts less thermal energy than the static $G_0$ model during quiescent periods of star formation.
The cumulative energy injected through PE heating (bottom panel of Fig. \ref{fig:cumHeat}) shows that after the onset of star formation, all models with a self-consistent treatment of the FUV radiation field generated by stellar clusters significantly surpass the energy injection of PE heating in a static model.
The difference scales with the system's initial gas surface density, and therefore with the SFR, and can reach up to three orders of magnitude.
However, despite this substantial increase in energy injection into the surroundings of massive star clusters, the overall SFR appears to be unaffected (see Fig. \ref{fig:sfr} and Fig. \ref{fig:sfr_sum}).
FUV photons are absorbed by dust grains in HII regions, but the resulting PE heating is minor compared to the dominant heating from EUV photons ionising hydrogen.
The gas in HII regions is mainly heated by the thermalisation of high-energy electrons produced by photoionisation of hydrogen atoms. In contrast, FUV-driven PE heating becomes more important in the surrounding PDRs where the gas is neutral or only partially ionised.

\subsubsection{Thermal balance}

\begin{figure}
 \centering
 \includegraphics[width=.95\linewidth]{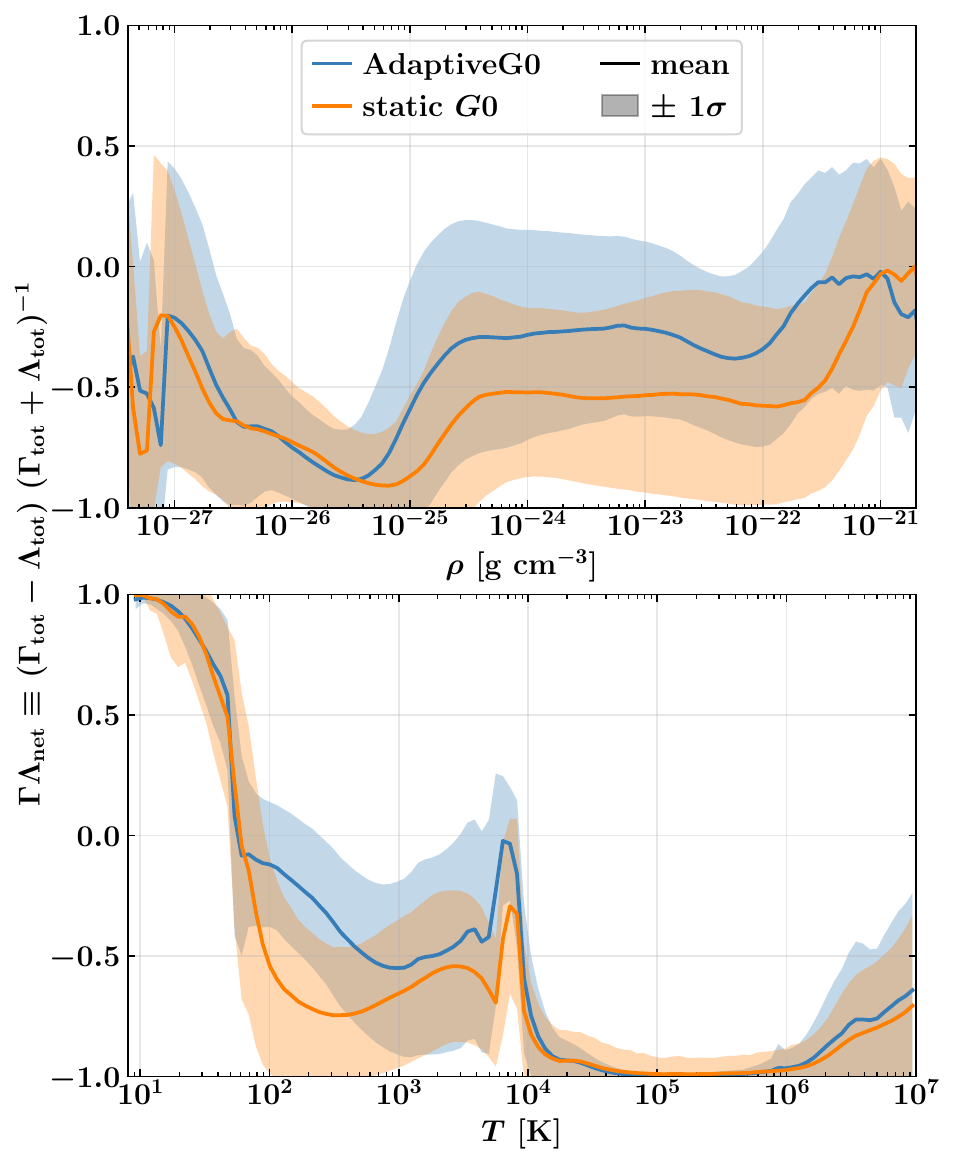}
 \caption{Volume-weighted and normalised net heating/cooling rate, $\Gamma\Lambda_\mathrm{net}$, shown as a function of density ($\rho$, top panel) and temperature ($T$, bottom panel) for simulations using \textsc{AdaptiveG0} (blue) and static $G_0$ (orange).
 The solid lines represent the mean values averaged across all initial $\Sigma_\mathrm{gas}$ simulations, while the shaded regions indicate the $\pm1\sigma$ range, capturing the variability among the simulations.
 Most of the ISM is out of heating and cooling equilibrium except for gas at $T\approx10^4$~K and towards high densities.
 \label{fig:ThermalBalance}}
\end{figure}

In the previous two subsections, we analysed the contributions of the main heating and cooling agents in the WNM and their differences between \textsc{AdaptiveG0} and static $G_0$ models.
We now introduce a volume-weighted and normalised net heating/cooling rate, $\Gamma\Lambda_\mathrm{net} \equiv (\Gamma_\mathrm{tot} - \Lambda_\mathrm{tot})\,(\Gamma_\mathrm{tot} + \Lambda_\mathrm{tot})^{-1}$, which we present as a function of density and temperature in Fig.~\ref{fig:ThermalBalance} for simulations using both \textsc{AdaptiveG0} and static $G_0$.
This metric accounts for all heating and cooling processes included in our chemical network, providing a comprehensive view of the thermal balance.
We refer the reader to Fig.~\ref{fig:HeatCool} for a compilation of all heating and cooling processes (minus PI heating) in our chemical network.
A value of $\Gamma\Lambda_\mathrm{net} = -1$ indicates a regime dominated entirely by cooling, while $\Gamma\Lambda_\mathrm{net} = +1$ signifies dominance by heating processes.
A value of $\Gamma\Lambda_\mathrm{net} = 0$ denotes an exact balance between heating and cooling, which is achieved in \textsc{AdaptiveG0} models only at $T\approx10^4$~K, corresponding to HII regions, and towards high densities ($\rho \gtrsim 3\times10^{-22}$~g~cm$^{-3}$).
At temperatures below $T \approx 100$~K, efficient cooling becomes increasingly difficult, and the thermal structure is dominated by heating, primarily from CR ionisation and PE heating.
It is important to note that $\Gamma\Lambda_\mathrm{net}$ provides information about the relative balance between heating and cooling but not about the absolute strength of either.
This relative imbalance across most temperature and density regimes underscores the importance of non-equilibrium chemistry for accurately capturing the thermal dynamics of the ISM.
In \textsc{AdaptiveG0}, $\Gamma\Lambda_\mathrm{net}$ is systematically higher across most temperatures and densities, remaining close to or below zero except for the very low-temperature range between $T \approx 10 - 100$~K.
This outcome is consistent with the increased energy input from PE heating in the self-consistent FUV treatment of \textsc{AdaptiveG0}, as illustrated in Fig.~\ref{fig:cumHeat}.

\subsection{Gas phases}\label{sec:chemistry}

\begin{figure*}
 \centering
 \includegraphics[width=.92\linewidth]{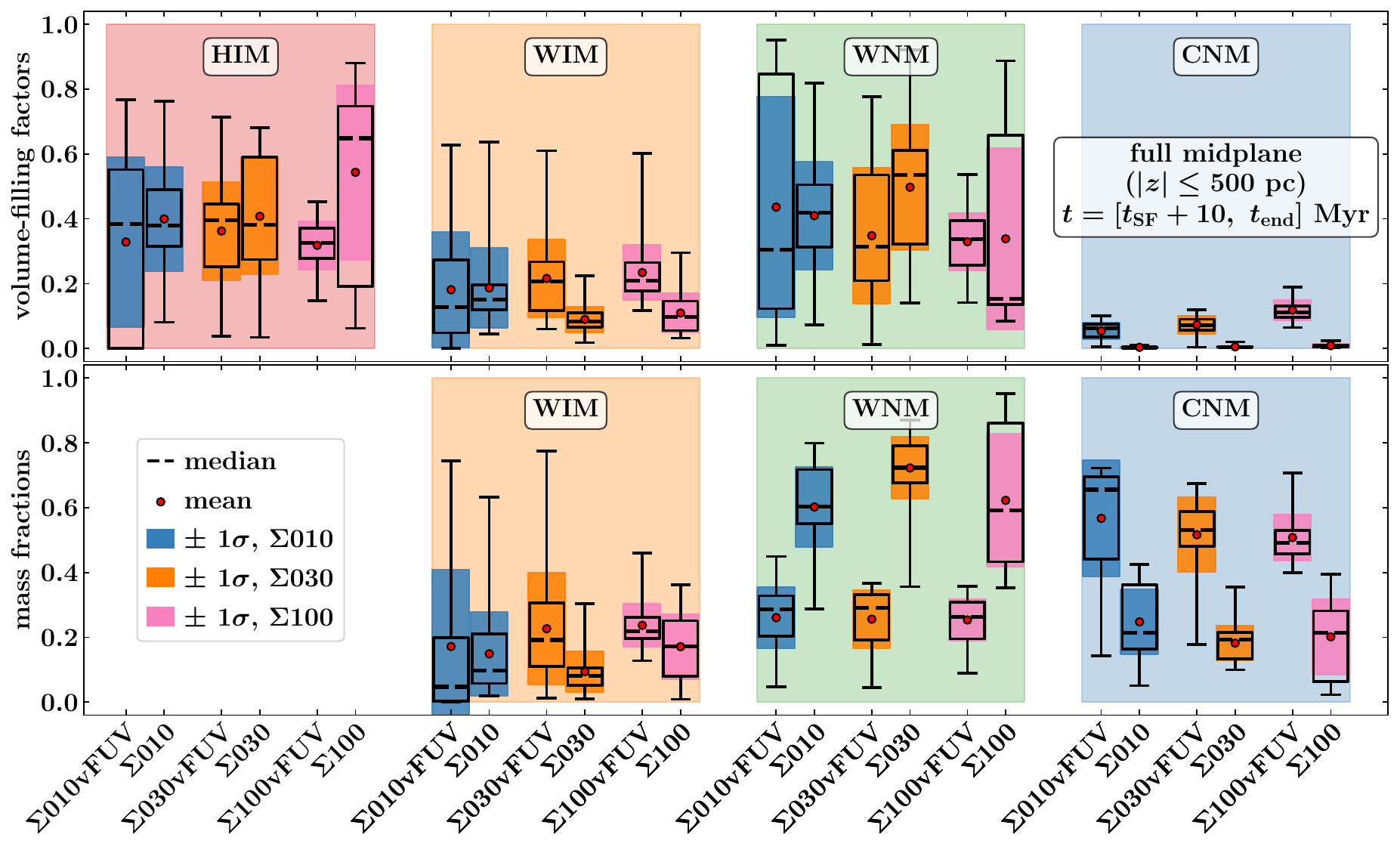}
 \includegraphics[width=.92\linewidth]{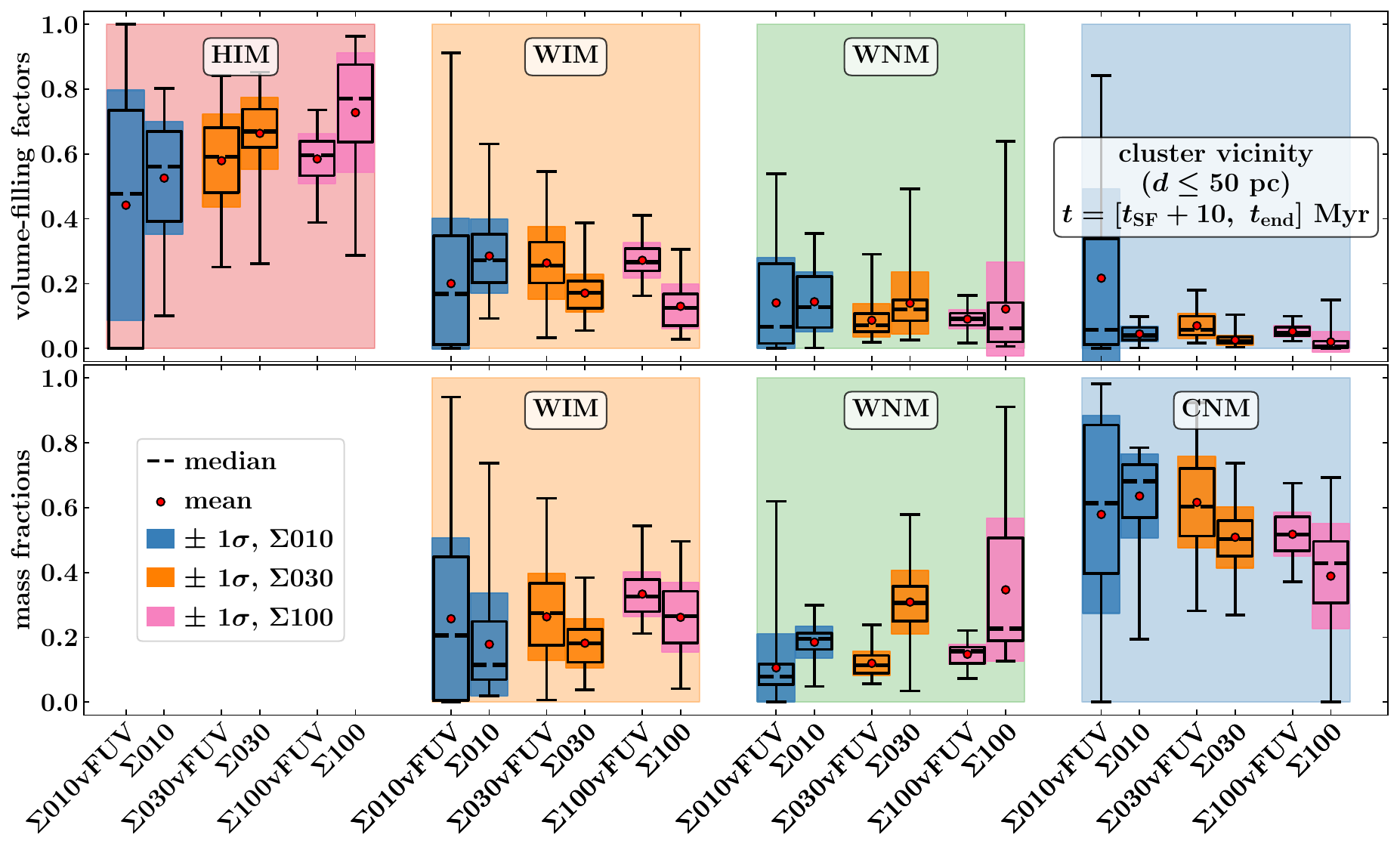}
 \caption{\textit{Top figure:} MFs (\textit{top panel}) and VFFs (\textit{bottom panel}) of the hot (HIM, $T > 3\times 10^5\,\mathrm{K}$), warm ionised (WIM, $300 < T \leq 3\times10^5\,\mathrm{K}$, $\chi_\mathrm{ion} \geq 0.5$), warm neutral (WNM, $300 < T \leq 3\times10^5\,\mathrm{K}$, $\chi_\mathrm{ion} < 0.5$), and cold neutral (CNM, $T \leq 300\,\mathrm{K}$) gas phases within the midplane ($|z| \leq 500\,\mathrm{pc}$).
 The data is processed in the same way as in Fig. \ref{fig:sfr_sum} with the median, the \nth{25} percentile and \nth{75} percentile shown as a box plot, the whiskers indicating the minimum and maximum values and the red dots and shaded area indicating the time-average with $\pm\,1\sigma$ standard deviation.
 A self-consistent FUV treatment boosts the MFs of the warm neutral and cold gas phases.
 \textit{Bottom figure:} Same overview of the MF and VFF of the different gas phases as above but this time only considering the gas in the vicinity of the stellar clusters ($d \leq 50\,\mathrm{pc}$, same as the used FUV injection radius).
 Similar trends develop for the WNM in the cluster vicinity compared to the full midplane.
 \label{fig:chemistry}}
\end{figure*}

We detail the gas phase distribution in Fig. \ref{fig:chemistry}, comparing the full midplane ISM (upper part) and the gas near stellar clusters ($d < 50\,\mathrm{pc}$, lower part).
The gas is classified into shock-heated hot ionised medium (HIM, $T > 3\times10^5\,\mathrm{K}$), WIM ($300 < T \leq 3\times10^5\,\mathrm{K}$, ionisation degree $\chi\equiv n(\mathrm{H}^+)~n(\mathrm{H}_\mathrm{tot})^{-1} \geq 0.5$), WNM ($300 < T \leq 3\times10^5\,\mathrm{K}$, $\chi < 0.5$), and CNM ($T \leq 300\,\mathrm{K}$).
The data is presented using box plots similar to Fig. \ref{fig:sfr_sum}, where boxes represent the \nth{25} and \nth{75} percentiles.
The thick horizontal line shows the median and the red dot indicates the mean value over time.
Whiskers mark the maximum and minimum values, while shaded areas denote the standard deviation of the mean.
The colour-coding represents the initial $\Sigma_\mathrm{gas}$ of the model.
In each pair of models with the same $\Sigma_\mathrm{gas}$, the \textsc{AdaptiveG0} model is on the left and the static $G_0$ model is on the right.
The negligible MFs of the HIM gas phase are not shown.
By analysing both the full midplane ISM and the vicinity of star clusters, we can better understand how the ISRF influences star formation and the overall chemistry and dynamics of the ISM.

The most striking difference in the midplane ISM is the significant increase in the CNM MF at the expense of the WNM\footnote{We do not employ the use of tracer particles which prohibits us from giving definitive answers about how the gas transitions between the different phases in detail.
We can only explain the overall evolution given the integrated quantities.}.
This increase in the potential molecular CNM aligns with the findings in Sect. \ref{sec:gasstructure}.
We measure a significant CNM VFF ranging from $\mathrm{VFF} = 5 - 12\,\mathrm{per\,cent}$, scaling with $\Sigma_\mathrm{gas}$.
The absence of artificial PE heating in static $G_0$ models, particularly in regions distant from massive star clusters, has two significant effects.
Firstly, it facilitates increased molecular gas formation.
Secondly, it prevents the destruction of diffuse molecular gas that has formed in denser regions and subsequently dispersed, by avoiding excessive heating of this gas.
Additionally, dissociation by Lyman-Werner photons (11.2 eV - 13.6 eV), which is included in our one-band FUV energy range (5.6 eV - 13.6 eV), is decreased in those regions.
The CRs, modelled as a distinct relativistic fluid using an advection-diffusion approximation, contribute an extra pressure component to the system.
This CR pressure, which permeates the ISM, provides further support to the diffuse molecular gas phase, helping it resist compression from external forces and maintain pressure equilibrium.

\begin{figure}
 \centering
 \includegraphics[width=.92\linewidth]{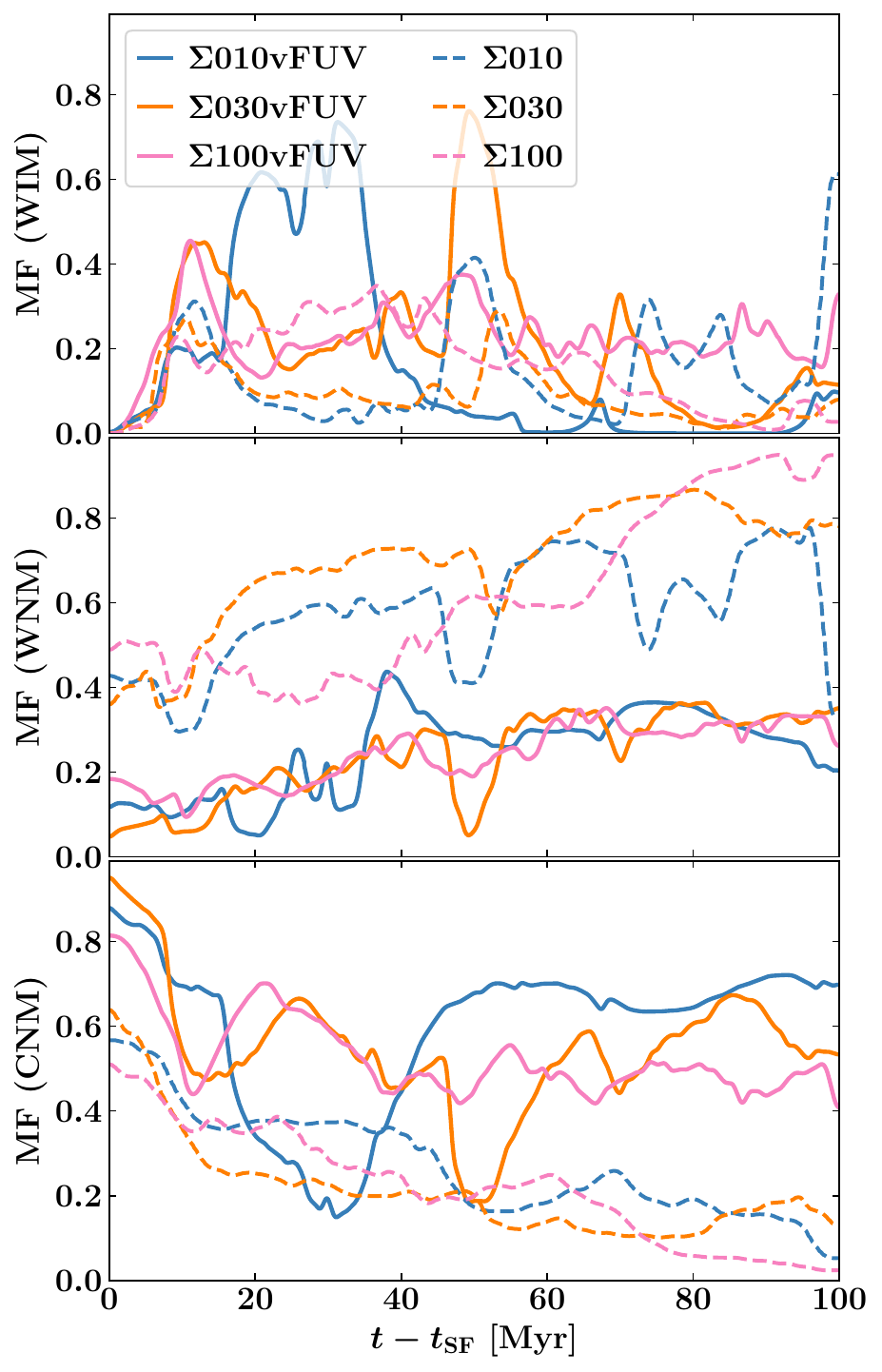}
 \caption{Time evolution of the midplane ($|z| \leq 500\,\mathrm{pc}$) MFs of the WIM (top), WNM (middle), and CNM (bottom).
 In \textsc{AdaptiveG0} models, the amount of gas in the thermally unstable phase is reduced and added to the CNM.
 This adjustment leads to a noticeable redistribution of gas phases over time, with the \textsc{AdaptiveG0} models showing more pronounced fluctuations in the CNM MF compared to the static $G_0$ models.
 The two outlier peaks in the $\mathrm{\Sigma010vFUV}$ and $\mathrm{\Sigma030vFUV}$ WIM MF are discussed in Appendix \ref{app:WIMpeak}.}
 \label{fig:mf_time}
\end{figure}

In Fig. \ref{fig:mf_time}, we show the time evolution of the midplane MFs for the different thermal phases.
We summarise and quantify the average and median MFs and VFFs in Table \ref{tab:mf} and Table \ref{tab:vff} in Appendix \ref{app:tables}.
The \textsc{AdaptiveG0} simulations have a systematically lower amount of gas in the thermally unstable regime and an increased amount of gas in the CNM throughout the simulated time.
However, we also see two deviations of this behaviour with temporary peaks in the WIM MF of the models $\mathrm{\Sigma010vFUV}$ and $\mathrm{\Sigma030vFUV}$.
The self-consistent FUV radiation field treatment does not cause these outliers, they result from the systems' underlying gas dynamics.
While the total stellar mass in each system is an important factor, the individual mass of those stars (determined stochastically by our star formation model) is crucial since more massive stars significantly contribute to photoionisation, which likely sustains the WIM.
Additionally, the spatial distribution of the stars strongly impacts the efficiency of stellar feedback.
To explain the WIM peaks, we further analyse these aspects in Appendix \ref{app:WIMpeak}.

\section{Discussion}\label{sec:discussion}

\subsection{The impact of far-ultraviolet radiation on star formation in the literature}\label{sec:litdisc}

In the literature, the relative importance of the FUV ISRF for self-regulating star formation is ambiguous. We discuss our findings in the context of the numerical work of other groups.

The closest comparison to our simulation framework can be made with the TIGRESS simulation suite.
In the "classic" model \citep{Kim2017}, the multiphase ISM is simulated in a tall-box setup, similar to ours.
A major difference to the \textsc{Silcc Project} framework is the inclusion of large-scale galactic rotation realised with a shearing box.
This enables them to simulate larger systems over multiple galactic orbits since they can model the impact of the large-scale galactic shearing motions.
These large-scale motions, which can efficiently redistribute gravitational energy, have been argued to be another regulatory mechanism for star formation and a driver for ISM turbulence \citep[see e.g.][]{Krumholz2018, FortuneBashee2024}, as well as for large-scale magnetic dynamos \citep[see e.g.][]{Gent2024}.
The TIGRESS simulation incorporates stochastically sampled star formation with a sink particle approach.
The stellar feedback is realised through SNe and PE heating by a time-dependent ambient FUV field.
Their model assumes that $\Gamma_\mathrm{PE}$ scales with the FUV luminosity surface density, $\Gamma_\mathrm{PE} \propto \Sigma_\mathrm{FUV} \equiv L_\mathrm{FUV} / (L_xL_y)$.
The total FUV luminosity, $L_\mathrm{FUV}$, is integrated over all present star particles and therefore spatially constant.
This will inevitably lead to increased artificial PE heating regions without star formation.
They report a $\Sigma_\mathrm{SFR} \approx 5 \times 10^{-3}\,\mathrm{M_\odot\,yr^{-1}\,kpc^{-2}}$ for their solar neighbourhood model with an initial $\Sigma_\mathrm{gas} = 13\,\mathrm{M_\odot\,pc^{-2}}$ at $dx \approx 4\,\mathrm{pc}$ spatial resolution.

In \citet{Kim2023a}, the model was updated to TIGRESS-NCR (non-equilibrium cooling and radiation), including the propagation of UV radiation through radiative transfer.
This is coupled to a chemical network primarily involving hydrogen and carbon photo-chemistry, accounting for CR ionisation \citep{Kim2023}.
TIGRESS-NCR achieves a complexity in photo-chemistry similar to our work presented in \citet{Rathjen2021} but does not account for other stellar feedback mechanisms like stellar winds and the transport of CRs.
CR transport, with a focus on its impact on galactic outflows, has been added to the TIGRESS framework in \citet{Armillotta2024}.
For radiation propagation, \citet{Kim2023a} implemented a direct ray-tracing method with three energy bands (PE photons, Lyman-Werner photons, and Lyman continuum), which accounts for the attenuation by dust along the rays.
In \citet{Kim2023a}, two configurations are presented with $\Sigma_\mathrm{gas} = 12\,\mathrm{M_\odot\,pc^{-2}}$ and $\Sigma_\mathrm{gas} = 50\,\mathrm{M_\odot\,pc^{-2}}$, respectively.
The sources of the UV are the star cluster particles with a mass-weighted age $t_\mathrm{age} < 20\,\mathrm{Myr}$.
This is a good approximation for the EUV but might drastically underpredict the FUV radiation emitted by older low-mass stars.
To save computational costs, \citet{Kim2023a} terminate the ray transfer of the FUV above a height of $|z| = 300\,\mathrm{pc}$ and switch to an analytical plane-parallel assumption from there.
They test different limits in the horizontal maximum propagation distance, $d_\mathrm{xy}$ (between 512 pc and 4096 pc), as well as a termination of the FUV radiation transfer when the luminosity of an individually transported photon package becomes lower than the total FUV luminosity in the system times a prefactor, $\epsilon_\mathrm{pp} = 10^{-9} \- 10^{-7}$.
This FUV efficiency cut-off can lead to underestimating the contribution of low-luminosity sources.
While these tests of varying propagation distance and FUV cut-off did not achieve convergence in total FUV radiation energy density at larger heights, the computationally motivated parameter choices had minimal impact on the thermal and dynamical balance of the ISM, particularly near the midplane.
This result aligns with our assumptions in this study regarding the limited propagation distance of the FUV under consideration of the star-forming capabilities of the midplane ISM.

Their models run for up to 700 Myr, facilitated by the shearing box setup.
Without this, the galactic context could not be considered, and simulating a galactic patch model over multiple orbit periods would cease to make physical sense.
Their analysis is carried out over a 200 Myr time stretch starting approximately 150 Myr after the onset of star formation.
Regarding star formation, they observe a median $\Sigma_\mathrm{SFR}$ ($\pm$\nth{16}, \nth{84} percentile) of $\Sigma_\mathrm{SFR} = [2.8^{+1.5}_{-1.0}, 29.0^{+20.3}_{-8.5}]\,\mathrm{M_\odot\,yr^{-1}\,kpc^{-2}}$.
The gas surface densities during this time frame are $\Sigma_\mathrm{gas} = [10.6^{+0.3}_{-0.2}, 36.2^{+1.4}_{-0.6}]\,\mathrm{M_\odot\,pc^{-2}}$.
These results are similar to our findings (see Table \ref{tab:SFR}), which was expected due to their inclusion of non-equilibrium chemistry and ionising radiative transfer.
Compared to the results from \citet{Kim2017}, which used a simplistic stellar feedback model with only SNe and a planar FUV ISRF scaled by the SFR, $\Sigma_\mathrm{SFR}$ is reduced by roughly a factor of two in the solar neighbourhood model (labelled \textit{R8} in the TIGRESS framework).
This is another indicator of the importance of self-consistent early stellar feedback.
\citet{Kim2023a} also provide a study of the VFF and MF of the thermal gas phases with a more refined distinction of a total of nine (partly overlapping) phases compared to our four (CNM, WNM, WIM, HIM).
For comparison, we group their cold molecular and cold neutral medium as CNM, unstable neutral and warm neutral medium as WNM, warm photoionised, warm collisional ionised and warm-hot ionised medium as WIM, and finally hot ionised as HIM \citep[see Table 3 in][]{Kim2023a}.
Derived from their values in Table 3, we acquire the averaged MF and VFF ($\pm1\sigma$) of $\mathrm{MF^{KO23}_{CNM}} = 26(\pm4)\,\mathrm{per\,cent}$, $\mathrm{MF^{KO23}_{WNM}} = 65(\pm9)\,\mathrm{per\,cent}$, $\mathrm{MF^{KO23}_{WIM}} = 8(\pm3)\,\mathrm{per\,cent}$, $\mathrm{VFF^{KO23}_{CNM}} = 1.5(\pm0.4)\,\mathrm{per\,cent}$, $\mathrm{VFF^{KO23}_{WNM}} = 65(\pm7)\,\mathrm{per\,cent}$, $\mathrm{VFF^{KO23}_{WIM}} = 16(\pm4)\,\mathrm{per\,cent}$, and $\mathrm{VFF^{KO23}_{HIM}} = 18(\pm7)\,\mathrm{per\,cent}$.
The inclusion of self-shielding leads to a higher $\mathrm{VFF^{KO23}_{CNM}}$ in TIGRESS-NCR compared to TIGRESS-classic.
Additionally, they see an increase in the $\mathrm{VFF^{KO23}_{WIM}}$ caused by photoionisation.
Compared to observational estimates provided in \citet{Tielens2005}, the TIGRESS-NCR framework underpredicts the MF of the CNM and the WIM ($\mathrm{MF^{obs}_{CNM}} \approx 48\,\mathrm{per\,cent}$, $\mathrm{MF^{obs}_{WIM}} \approx 14\,\mathrm{per\,cent}$).
Our chemical network and stellar feedback model yield predictions much closer to the observed estimates ($\mathrm{MF_{CNM}} = 52(\pm19)\,\mathrm{per\,cent}$, see Table \ref{tab:mf}).
The spatially and temporally variable FUV ISRF ensures that the thermal balance in the volume-filling warm and cold ISM is self-consistent.
Another feature that differs between the TIGRESS and \textsc{Silcc} models is the VFF of the hot ionised gas phase.
TIGRESS-NCR predicts $\mathrm{VFF^{KO23}_{HIM}} = 18(\pm7)\,\mathrm{per\,cent}$, while we find $\mathrm{VFF^{}_{HIM}} = 45(\pm19)\,\mathrm{per\,cent}$, in closer agreement with observed estimates of $\mathrm{VFF^{obs}_{HIM}} \approx 50\,\mathrm{per\,cent}$ \citep{Tielens2005}.

The galactic patch simulations by \citet{Butler2017} use a model with $\Sigma_\mathrm{gas} = 17\,\mathrm{M_\odot}\,\text{pc}^{-2}$.
This model includes subgrid star formation, supernova feedback, and a moment-method radiative transfer scheme that treats the radiation field as a fluid rather than using explicit ray tracing.
They consider four energy bands: one for $\mathrm{H}_2$-dissociating Lyman-Werner photons and three for EUV photons, distinguishing the ionisation of helium (He).
They systematically toggle different stellar feedback mechanisms on and off and find that in starburst regions, $\mathrm{H}_2$-dissociating radiation has the strongest impact on reducing the SFR compared to a model without any feedback.
These effects are less pronounced when considering the full kpc-sized domain.
However, they do not examine the impact of the EUV radiation band alone, which we argue has the strongest effect on self-regulating star formation.
They report that the reduction in $\Sigma_\mathrm{SFR}$ by supernovae is comparable to that by FUV radiation.
Nevertheless, the chemical state of the ISM differs significantly between the SN-only and FUV-only runs.
They state that models including both the FUV and EUV radiation bands and supernovae match observations best, similar to our claims.
For their starburst region with all feedback channels considered, they show an average $\Sigma_\mathrm{gas} \approx 20\,\mathrm{M_\odot\,pc^{-2}}$ and $\Sigma_\mathrm{SFR} \approx 3 \times 10^{-2}\,\mathrm{M_\odot\,yr^{-1}\,kpc^{-2}}$, which is aligned with our $\mathrm{\Sigma030vFUV}$ and $\mathrm{\Sigma030}$ models (average $\Sigma_\mathrm{SFR} = [44\pm20, 30\pm28]\,\mathrm{M_\odot\,yr^{-1}\,kpc^{-2}}$, respectively).
Only supernovae or only FUV radiation feedback leads to highly elevated SFRs of the order $10^{-1}\,\mathrm{M_\odot\,yr^{-1}\,kpc^{-2}}$ in that region.\newline

Dwarf galaxy setups are the focus of extensive numerical studies, as they enable high enough resolution (spatial or mass, depending on the code approach) due to their lower total mass \citep[e.g.][]{Hu2017, Emerick2018, Lahen2019, Smith2021, Deng2024, Sugimura2024}.
\citet{Hu2017} studied the effects of a variable ISRF in smoothed particle hydrodynamics (SPH) dwarf galaxy simulations and compared the efficacy of PE heating and supernovae (SNe).
Their treatment of the FUV ISRF is similar to ours, integrating the FUV luminosity over all present star particles and accounting for geometric attenuation ($G_\mathrm{eff} \propto R^{-2}$) and dust attenuation.
Photoionisation is modelled using a Str\"omgren-sphere approach rather than direct radiative transfer.
They arrive at similar conclusions as \citet{Butler2017}, finding that PE heating achieves a similar reduction in the SFR as SNe but with a drastically different ISM composition.
However, they also emphasise the strong non-linear coupling of different stellar feedback processes, making it impractical or unphysical to rule out the importance of a single feedback channel.
We further emphasise that the effects of PE heating, which depend on the hydrogen number density and non-linearly on the electron number density, might be vastly overestimated in models that do not account for stellar winds or the proper propagation of EUV radiation and the development of HII regions around young massive stars.
Photoionisation and stellar winds disperse the star-forming environment, leading to a drastic reduction in the ambient density.
PE heating appears to be subdominant.
Models that do not account for EUV radiation and stellar winds might see a stronger influence of PE heating while also lacking the formation mechanisms for the WIM.
It is thus important to account for all early stellar feedback channels in addition to SNe to achieve a realistic model of the multiphase ISM with locally self-regulated star formation.

The parsec-scale low-mass dwarf galaxy simulations by \citet{Emerick2018} follow detailed stellar feedback on a star-by-star basis and utilise an adaptive ray-tracing radiative transfer method for the ionising radiation, as well as SNe and wind feedback from both young massive stars and asymptotic giant branch (AGB) stars.
They assume the FUV radiation to be optically thin with local attenuation, similar to our approach.
Similar to \citet{Hu2017} and our findings, \citet{Emerick2018} conclude that multichannel stellar feedback (i.e., early feedback in the form of winds and especially ionising radiation, but also PE heating from FUV radiation plus SNe) is needed to model a realistic evolution of dwarf galaxies.
Their simulations agree with observations regarding star formation, galactic outflows, and the overall composition/metallicity of the ISM.

\citet{Smith2021} simulate isolated dwarf galaxies with parsec-scale simulations in which star formation is modelled by explicitly sampling massive stars from an IMF, enabling the modelling of individual HII regions, PE heating from a spatially varying FUV ISRF, and SN feedback.
Their PE heating approach, similar to \citet[][see above]{Hu2017}, models the FUV luminosity of each star cluster sink particle, accounting for geometric ($G_\mathrm{eff} \propto R^{-2}$) and dust attenuation.
Short-range photoionisation is modelled using an overlapping Str\"omgren-type approach, accounting for anisotropic neutral gas distribution rather than perfectly spherical HII regions.
SNe are implemented through thermal energy injection of order $E_\mathrm{inj} = 10^{51}\,\mathrm{erg}$.
\citet{Smith2021} test various combinations of these feedback channels, finding that photoionisation and SNe independently can regulate star formation to the same level, with photoionisation-regulated star formation being less bursty than SN-regulated star formation.
PE heating alone is insufficient to regulate star formation to that level.
The latter results align well with our findings.

\citet{Deng2024} introduce the RIGEL framework, which utilises high-resolution (1 M$_\odot$) radiation-hydrodynamics simulations to model the multiphase ISM in dwarf galaxies.
Their approach incorporates explicit radiative transfer and metallicity-dependent stellar feedback on a star-by-star basis using \textsc{arepo-rt}.
The framework employs a moment-based radiative transfer solver with M1 closure to explicitly follow radiation distributed across seven spectral energy bins, ranging from infrared to EUV.
Metallicity-dependent photon production rates are derived from zero-age main sequence (ZAMS) stellar models.
Key processes such as photoionisation, molecular hydrogen dissociation, PE heating, and radiation pressure are self-consistently modelled to accurately capture the thermal structure of the ISM.
\citet{Deng2024} find that radiative feedback rapidly disperses molecular clouds within 1 Myr of massive star formation which significantly reduces the star-forming gas reservoir and suppresses subsequent star formation.
Their findings highlight the non-linear coupling between FUV radiation and other stellar feedback mechanisms, including photoionisation and stellar winds, which together regulate star formation and drive galactic outflows.
These results underscore the critical role of resolving FUV radiation to realistically model ISM properties and star formation dynamics in low-mass galaxies.

Motivated by recent high-redshift ($z \gtrsim 8$) observations of the first galaxies by JWST, \citet{Sugimura2024} model metal-poor galaxies with a cosmological zoom-in approach that would evolve into a dwarf galaxy analogue in the Local Group by $z = 0$.
In their study, they focus on the impact of FUV and EUV radiation and distinguish between Pop II and Pop III stars at sub-pc resolution up to a redshift of $z \approx 10$ while including non-equilibrium chemistry.
Their radiative transfer is a moment-based method with M1 closure and allows for four photon energy bins (Lyman-Werner, H-ionising, He-ionising, He$^+$-ionising).
Interestingly, they find a positive effect of FUV radiation from Pop II stars on the SFR.
First, the extremely low-metallicity clouds in the first galaxies are heated to $\sim10^{4}\,\mathrm{K}$ by the diffuse FUV radiation field.
This retards star formation and allows the clouds to accumulate more gas.
As the clouds become very massive, self-shielding begins to occur.
This cools the clouds, which then become gravitationally unstable and trigger rapid star formation because of the large amount of gas available.
This effect can also explain the observed bursty star formation and clumpiness in high-redshift galaxies.
They find that EUV radiation from Pop II stars is not able to photoevaporate the gas within the first galaxies.
However, EUV radiation from Pop III stars can, in turn, enhance the efficiency of the subsequent supernovae.
The results from \citet{Sugimura2024} highlight that the effects of various stellar feedback mechanisms are not static for different systems over cosmological time.
Changes in overall gas mass and metallicity can strongly impact how efficiently feedback regulates star formation and can even cause a qualitative switch from positive to negative feedback.\newline

A similar conclusion about the impact of metallicity is drawn in \citet{Bialy2020}, where an analytical model for the FUV ISRF flux as a function of dust-to-gas ratio, $\Sigma_\mathrm{SFR}$, gas density, scale radius, and position of the observer is developed and applied to a series of galactic disc simulations.
They find that the ISRF flux per $\Sigma_\mathrm{SFR}$ is anti-correlated with the metallicity of the systems, i.e., the $F_\mathrm{ISRF}\,\Sigma^{-1}_\mathrm{SFR}$ is a factor of three to six higher in metal-poor dwarf galaxies than in their Milky Way-like (MW-like) counterparts.
Hence, in metal-poor systems, FUV radiation might be the main regulating mechanism of star formation.
However, potentially stronger effects of EUV radiation or SNe on the SFR are not taken into account.
Nonetheless, these results show that it is important to note that our findings could potentially only apply to systems with solar metallicity and gas surface densities between $\Sigma_\mathrm{gas} = 10 - 100\,\mathrm{M_\odot\,pc^{-2}}$.

The effects of PE heating triggered by FUV radiation on star formation have also been studied in more massive isolated MW-like disc galaxies but at the cost of less sophisticated stellar feedback models.
\citet{Tasker2011} modelled kpc-scale disc galaxies with a galactic-radial dependent PE heating rate with an ISRF strength of $G_0 = 1.7$ and a radial scale length of 8 kpc.
They account for tabulated radiative cooling and star formation with fixed star formation efficiency, $\epsilon_\mathrm{SF}$, but without EUV radiation, stellar winds or SNe.
Compared to their no-feedback model, PE heating suppresses cloud fragmentation and reduces star formation.
However, they show that their measured SFR is still more than an order of magnitude higher than in observed local neighbourhood galaxies.
This is due to the lack of additional stellar feedback processes, especially SNe and ionising radiation.

\citet{Osman2020} use a more refined stellar feedback model in their MW-like isolated disc galaxy simulation and include SNe and PE heating through FUV radiation but still no explicit radiative transfer for hydrogen-ionising radiation.
The strength of their FUV radiation field is estimated with a stellar population synthesis code considering the age and metallicity of the stellar particle.
In addition, they deploy an extensive dust model which considers the formation and destruction of dust.
They can switch the PE heating on and off and vary various model parameters of their dust model.
Depending on the choice of parameters, they report suppression of star formation by factors between zero and five.
Models with higher gas fractions ($f_\mathrm{gas} \sim 0.5$) exhibit a stronger suppression of the SFR.
Additionally, they identify a secondary large-scale effect that regulates the SFR.
In their models, the effectiveness of SNe is enhanced when PE heating is considered, due to the pre-processing of the gas before the SN explosion.
This results in stronger thermally driven ballistic outflows, which deplete the ISM more quickly and subsequently reduce star formation.
However, we argue that the dispersion of gas around single stars solely by FUV radiation is likely to be rather small, as demonstrated in Fig. \ref{fig:HII_FUV}.
Moreover, EUV radiation is capable of evacuating star-forming regions and enabling SNe to become the most efficient with the least amount of radiative cooling losses \citep[see also][]{Rathjen2021}.\newline

The FUV ISRF can also be studied in terms of its effects on the thermal phases and stability of the multiphase ISM.
Our models demonstrate a clearer separation of respective gas phases and a reduction in thermally unstable gas (see Sect. \ref{sec:gasstructure} and \ref{sec:chemistry}).
Accurate modelling of the FUV radiation field is crucial for understanding star formation and the chemical evolution of the ISM.
The time- and space-varying nature of the ISRF is particularly important in this context.
Our findings indicate an increase in cold neutral gas fractions in models with \textsc{AdaptiveG0}, suggesting a higher potential for molecular gas formation in regions shielded from strong FUV radiation.
Variations in heating rates and cumulative energy injection through PE heating significantly impact the ISM's thermodynamic properties.
Previous studies, such as \citet{Hill2018} and \citet{Bialy2020}, have shown that the FUV ISRF's intensity in galaxies determines the thermal and chemical evolution of neutral interstellar gas and is essential for interpreting extragalactic observations and star formation theories.
The isolated galactic disc model with SN feedback and explicit ray-tracing FUV radiation transport (one energy band, PE plus Lyman-Werner photons) by \citet{Benincasa2020} agree with \citet{Hu2017} and our findings that star formation cannot be efficiently regulated via FUV-induced PE heating but that a correct treatment of the FUV ISRF is needed to arrive at realistic ISM phases.
However, since \citet{Benincasa2020} do not model the formation and destruction of molecules, they only account for the thermal balance in the ISM through the direct PE heating by dust grains.
From the observational side, FUV radiation is often used as a tracer for star formation \citep[see e.g.][]{Leroy2008}.
A detailed and self-consistent FUV radiation field in numerical simulations allows us to create synthetic observations and further calibrate and inform observational models.\newline

This study highlights the significant impact of varying ISRF conditions on the ISM's thermodynamic properties. It emphasises the necessity of accurately modelling the FUV radiation field to resolve the ISM's gas phases.
Nevertheless, the impact of the FUV radiation field on massive star formation is negligible for systems at solar metallicity when also accounting for additional stellar feedback mechanisms.
It was demonstrated in \citet{Rathjen2021} that HII regions created by hydrogen-ionising radiation might be the strongest regulator of massive star formation.
Another important component in dispersing massive star-forming regions could be the momentum input from wind-blown bubbles of massive stars \citep[see e.g.][]{Weaver1977}.
However, this has not yet been convincingly demonstrated in ISM scale experiments, most likely due to lacking numerical resolution \citep{Lancaster2024}.
Recent experiments \citep[e.g.][]{Rathjen2021, Lahen2023}, find massive star wind feedback has only minor importance for the regulation of star formation, compared to SNe and ionising radiation.

\subsection{The diffuse molecular gas}\label{sec:diffmol}

The primary objective of this study is to understand the impact of the FUV ISRF on the star formation efficiency in the ISM.
However, a potentially controversial finding from this study is a cold diffuse molecular gas phase in our \textsc{AdaptiveG0} models, not previously observed in parsec-scale numerical experiments of the star-forming multiphase ISM.
This raises the question of whether this cold diffuse gas phase has a physical explanation or is an artefact of our model limitations.

To address this, we analysed the properties of the cold diffuse medium (CDM, $n_\mathrm{H2} < 2$ cm$^{-3}$, $T_\mathrm{gas} < 300$ K) based on the $\mathrm{\Sigma010vFUV}$, $\mathrm{\Sigma030vFUV}$, and $\mathrm{\Sigma100vFUV}$ simulations towards the end at $t - t_\mathrm{SF} \approx 60\,\mathrm{Myr}$.
A comprehensive follow-up study focusing solely on the diffuse molecular gas is being prepared.

For any given time step, we define the star-forming plane as the average $z$-height of the star cluster sink particles weighted by their current mass.
Then we calculate the distance of the diffuse $\mathrm{H}_2$ gas parcels to that plane.
We define $d_\mathrm{CDM}$ as the mean of those distances, weighted by the mass of each diffuse $\mathrm{H}_2$ gas parcel and find $d_\mathrm{CDM} = 162 \pm 62\,\mathrm{pc}$.
This suggests that most of the CDM is located further than $d = 50\,\mathrm{pc}$ from active star clusters and is thus only irradiated by an ISRF with strength $G_\mathrm{bg} = 0.0948$.
However, CDM pockets also exist within the star-forming plane, with overall vertical distances $z_\mathrm{CDM}$ ranging from $\leq 3.9\,\mathrm{pc}$ to $823 \pm 143\,\mathrm{pc}$, averaged over the three models with varying initial $\Sigma_\mathrm{gas}$.

The total mass of the CDM ranges between $M_\mathrm{CDM} \approx 1.0 - 1.4\times10^{4}\,\mathrm{M_\odot}$, which is insignificant compared to the total gas mass in the midplane ISM ($\sim1-10\times10^6\,\mathrm{M_\odot}$).
When we assume a free-fall time $t_\mathrm{ff} = \sqrt{\frac{3\pi}{32G\rho}}$, with gravitational constant $G$ and total gas density, $\rho$, for the diffuse H$_2$ gas with $n_\mathrm{H2}\ll2$~cm$^{-3}$ (see Fig. \ref{fig:H2_temp_pdf}), we can calculate $t_\mathrm{ff}^\mathrm{CDM}\gtrsim25$~Myr.
This suggests that the CDM is highly unlikely to collapse under its self-gravity.

The thermal pressure in the CDM measured in our simulations is $\sim90\,\mathrm{K\,cm^{-3}}$.
Additional support against the external pressure of the WNM comes from the CR component.
The median CR pressure in gas cells below $\sim50\,\mathrm{K}$ and with an $\mathrm{H}_2$ number density between $n_{\mathrm{H}_2} = 10^{-5} - 1\,\mathrm{cm^{-3}}$ is approximately four orders of magnitude higher than the thermal pressure in these regimes.

These CRs originate from supernova remnants embedded in the star-forming ISM and diffuse almost freely through the medium, creating a negative CR pressure gradient along the vertical axis of the simulation domain \citep[see e.g.][]{Girichidis2018}.
Their inefficient cooling provides a long-lasting energy reservoir, leading to additional support, especially in diffuse gas.
This property also enables CRs to support cold gas galactic outflows, as demonstrated in \citet{Rathjen2023}.

\subsection{Caveats and future improvements}\label{sec:caveats}

While this study provides valuable insights into the dynamics of the ISM and its interaction with the ISRF, several caveats and areas for future improvement should be considered.

\subsubsection{Simplified radiative transfer approach}
In the \textsc{AdaptiveG0} module, we employ a simplified approach rather than explicit radiative transfer as it is used for ionising radiation with \textsc{TreeRay/OnTheSpot} \citep{Wunsch2021}.
As a first-order approximation, FUV transport does not need iterative radiative transfer as FUV photons do not excite new FUV photon sources when traversing the ISM.
However, FUV photons scatter off dust grains while traversing the ISM.
Ideally, to treat this scattering properly, each cell should be considered as a potential source of photons.

Our method focuses on calculating the local strength of the ISRF, which is highly dependent on the column density along the line-of-sight and local attenuation.
We use the \textsc{TreeRay/OpticalDepth} module \citep{Wunsch2018}, which utilises the \textsc{HEALPix} algorithm, to calculate this attenuation (see Sect.~\ref{sec:numerics}, Eqs.~\ref{eq:AV3D} and \ref{eq:G0}).
Due to computational limitations, we cannot store the column densities for each cell and each sightline to a source in the current implementation of the simulation code.
Therefore, we calculate and store a 3D-averaged column density for each cell.

Moreover, we limit the FUV photon propagation to $d = 50\,\mathrm{pc}$ from the emitting stellar source, consistent with our attenuation factor calculation (see Appendix \ref{App:distance}, for a detailed analysis of this free model parameter).
This approach is based on our demonstration in Sect. \ref{sec:theory} that the strength of the FUV ISRF powered by a single star or massive star cluster is a strong function of distance to the source.
However, we recognise that this simplified approach has limitations.
Proper treatment of radiation transfer requires knowledge of the attenuation between the source and each cell along each ray.
Our approximate method therefore does not model the exact attenuation of the FUV by dust and also does not capture other relevant physics like dust scattering.
We acknowledge that explicit full radiative transfer would provide a more accurate treatment of the FUV ISRF.\newline

\subsubsection{One FUV energy band}
We currently do not consider multiple energy bands for non-ionising FUV photons.
As described in Sect. \ref{sec:numerics}, we obtain the FUV luminosity of our star cluster sink particles from \textsc{StarBurst99}.
To do so, we integrate over the energy range of $5.6\,\mathrm{eV} \leq E_\gamma < 13.6\,\mathrm{eV}$.
A possible distinction could be two energy bands for the FUV: $5.6\,\mathrm{eV} \leq E_\mathrm{5.6} < 11.2\,\mathrm{eV}$ and $11.2\,\mathrm{eV} \leq E_\mathrm{11.2} < 13.6\,\mathrm{eV}$ \citep[see e.g.][]{Baczynski2015}.
PE heating is expected to be dominated by photons in the $E_\mathrm{5.6}$ band in the dense ISM \citep{Bakes1994}.
Photodissociation of $\mathrm{H}_2$ is triggered by excitation of $\mathrm{H}_2$ with $E_\mathrm{11.2}$ band photons (Lyman-Werner band).

Lower energy photons are only able to dissociate $\mathrm{H}_2$ that is not in the vibrational ground state and are therefore negligible in our modelling \citep{Glover2015}.
The $\mathrm{H}_2$ gas is shielded from the FUV ISRF through the external column densities as well as through self-shielding.
In the current implementation, we assume that the $\mathrm{H}_2$ photodissociation rate scales with the radiation intensity of the whole FUV band.

Adding more energy bands increases the numerical complexity and cost but would also increase the accuracy of the \textsc{AdaptiveG0} model.
To address these limitations and improve the physical accuracy of our simulations, we plan to implement full radiative transfer for the FUV band, among others, by adding additional energy bands explicitly into our backwards-radiative transfer scheme \textsc{TreeRay} in an upcoming iteration of the \textsc{Silcc Project}.

\subsubsection{Chemical complexity and metallicity}
All simulations presented here are carried out under the assumption of solar metallicity.
We do not account for the effects of chemical variations through enrichment by stellar winds, SNe, or galactic inflows of low-metallicity gas.
Furthermore, we do not explicitly follow the evolution of dust but assume a constant dust-to-gas MF of one per cent.
The interaction of the FUV field with the gas and especially dust is strongly affected by the assumed metallicity and dust-to-gas ratio.
As seen in the discussion in Sect. \ref{sec:discussion}, FUV radiation could play a more important role in regulating $\Sigma_\mathrm{SFR}$ in lower metallicity environments.
We explore these effects in \citet{Brugaletta2025}.\newline

\subsubsection{Radiation pressure on dust and dust heating from the far-infrared radiation}
We account for the radiation pressure exerted on the gas by EUV photons with \textsc{TreeRay/OnTheSpot}.
The radiation pressure on dust grains is mostly exerted by infrared (IR) radiation but also overlaps with the FUV radiation band.
It has been demonstrated that radiation pressure on dust by IR radiation from young massive stars can decrease further fragmentation around the star and add to the self-regulation of star formation \citep{Klepitko2023}.
Furthermore, our chemical network only considers dust heating by EUV and FUV radiation.
We do not account for dust heating by the FIR and lower energy photons.
This could lead to a slight underprediction of $T_\mathrm{dust}$ for our models.
The dust temperature in the bulk of the gas of the \textsc{AdaptiveG0} models is $T_\mathrm{dust} \approx 10\,\mathrm{K}$, whereas Milky Way observations suggest temperatures of $T_\mathrm{dust} \approx 15-20\,\mathrm{K}$ \citep{Marsh2017}.
The impact of this discrepancy is negligible for the objectives of this study since the slightly higher dust temperature would not strongly alter the dynamics of the ISM or its efficacy in forming stars.

\section{Conclusion}\label{sec:conclusion}

We presented a series of nine magnetohydrodynamic simulations\footnote{As mentioned in Sect. \ref{sec:params}, three out of the total nine simulations presented here have been previously published in \citet{Rathjen2023} and one it \citet{Rathjen2021}.} of the multiphase interstellar medium (ISM) in varying galactic environments.
The presented simulations are carried out within the \textsc{Silcc Project} simulation framework.
Five of the simulations utilise the novel \textsc{AdaptiveG0} module for a self-consistent treatment of the spatially and temporally varying far-ultraviolet interstellar radiation field (FUV ISRF) that is powered by the stellar component.
Our primary focus was on understanding how a self-consistent, variable FUV radiation field influences the formation of massive stars, chemical composition, and dust properties in a stratified galactic ISM patch.
The main findings of our study are as follows:

\begin{enumerate}
 \item \textbf{Impact of FUV Radiation on Star Formation:} The simulations show that while FUV radiation can reach local intensities up to $G_0 \approx 10^4$ (in Habing units), its overall impact on regulating star formation is minor compared to other stellar feedback mechanisms, such as ionising UV radiation, stellar winds, and supernovae.
 Both static and adaptive FUV radiation fields result in similar SFRs, indicating that the presence of FUV radiation alone does not significantly alter the star formation rate (SFR).
 At solar neighbourhood conditions, the star formation rate surface densities, $\Sigma_\mathrm{SFR}$, only vary marginally between the model with \textsc{AdaptiveG0}, $\Sigma_\mathrm{SFR} = (5.0\pm4.9)\times10^{-3}\,\mathrm{M_\odot\,yr^{-1}\,kpc^{-2}}$ (mean$\pm1\sigma$), and with a simple static FUV radiation field, $\Sigma_\mathrm{SFR} = (4.3\pm3.1)\times10^{-3}\,\mathrm{M_\odot\,yr^{-1}\,kpc^{-2}}$ (mean$\pm1\sigma$).
 The models with only SN feedback or with SNe and the FUV treatment but without stellar winds and ionising radiation ($\mathrm{\Sigma010vFUV\dagger}$ and $\mathrm{\Sigma010\dagger}$), exhibit an order of magnitude higher SFR but without a significant difference between them.
 Their $\Sigma_\mathrm{SFR}$ is $(35\pm2.4)\times10^{-3}\,\mathrm{M_\odot\,yr^{-1}\,kpc^{-2}}$ and $(33\pm2.0)\times10^{-3}\,\mathrm{M_\odot\,yr^{-1}\,kpc^{-2}}$, respectively.
 A model at solar neighbourhood conditions without any feedback but with \textsc{AdaptiveG0} ($\mathrm{\Sigma010vFUVnoSN\dagger}$) shows an average $\Sigma_\mathrm{SFR}$ approximately three times higher, around $\Sigma_\mathrm{SFR} \approx 10^{-2}\,\mathrm{M_\odot\,yr^{-1}\,\mathrm{kpc}^{-2}}$.
 In this model, star formation is only weakly regulated.
 A measurable difference between the two models is the slight decrease of the star formation burstiness, $\beta_\mathrm{SF}$.
 For \textsc{AdaptiveG0} models, $\beta_\mathrm{SF}$ averaged over all $\Sigma_\mathrm{gas}$ realisations decreases to $59\pm23$~per cent (without $\Sigma010$ simulations), or to $85\pm35$~per cent (measured for the first 50 Myr of star formation) as compared to the static $G_0$ models.\newline

 \item \textbf{Heating Rates:} Photoelectric (PE) heating rates in the \textsc{AdaptiveG0} models show a wider distribution compared to static $G_0$ models, featuring higher peak rates but lower heating rates across most of the ISM volume.
 The cumulative energy injected by PE heating in the \textsc{AdaptiveG0} models significantly exceeds that in static models, particularly in regions close to star clusters hosting young massive stars.
 PE heating is more efficient in \textsc{AdaptiveG0} models as it can reach higher overall heating rates with less overall available FUV radiation energy density for PE heating.
 In the direct comparison between the two cases for nearly identical gas density structures, we see that the increase in PE heating energy results from the locally significantly higher FUV radiation field (see Sect. \ref{sec:heating}).
 Less gas further away from active star-forming regions is artificially exposed to extensive PE heating as it would be in the static case.
 This has implications for the chemical composition of the gas in the midplane ISM and especially on the gas' ability to form and sustain a cold neutral gas medium (CNM).\newline

 \item \textbf{Cooling Rates and Thermal Balance:} In \textsc{AdaptiveG0} models, cooling rates for singly-ionised carbon (C$^+$) and atomic oxygen (O) exhibit a broader distribution with a distinct bimodal structure, while their maximum cooling rates are generally lower compared to static $G_0$ models, despite slightly higher average abundances.
 This reduction in cooling efficiency is driven by the spatially and temporally variable FUV radiation field, which alters the local thermal and ionisation conditions of the ISM.
 The net heating/cooling rate, $\Gamma\Lambda_\mathrm{net}$, primarily indicates a dominance of cooling ($\Gamma\Lambda_\mathrm{net} < 0$) across most density and temperature ranges in both the \textsc{AdaptiveG0} and static $G_0$ models.
 However, \textsc{AdaptiveG0} models consistently exhibit higher $\Gamma\Lambda_\mathrm{net}$ values, approaching but remaining below zero, indicating increased heating capabilities compared to static $G_0$.
 This behaviour is a direct consequence of the spatially and temporally variable FUV radiation field in \textsc{AdaptiveG0}, which enhances PE heating in regions near star-forming clusters while reducing heating in shielded areas.\newline

 \item \textbf{Chemical Properties:} The space- and time-varying FUV radiation field promotes the formation of a diffuse molecular hydrogen gas phase and increases the mass fraction (MF) of the CNM outside the vicinity of stellar clusters.
 A detailed analysis of the temperature-density phase diagrams reveals that the variable FUV field broadens the equilibrium states of the gas phases, particularly promoting the presence of cold, diffuse gas.
 This diffuse molecular gas phase is supported against compression from external pressure by the additional cosmic ray pressure component.
 The MF of the diffuse molecular gas (compared to the total H$_2$ gas mass) in the \textsc{AdaptiveG0} models is $22.5\pm7.1$~per cent.
 We measure a systematic reduction of the MF in the warm neutral medium (WNM) by $36\pm6$ percentage points, independent of the initial gas surface density, $\Sigma_\mathrm{gas}$.
 At the same time, the CNM MF systematically increases by $27\pm5$ percentage points.
 We also measure an increase in the warm ionised medium (WIM) by $8\pm6$ percentage points, driven mainly by the rise in the WIM MF in the runs $\mathrm{\Sigma010vFUV}$ and $\mathrm{\Sigma030vFUV}$.
 We explain this behaviour further in Appendix \ref{app:WIMpeak}.

 Furthermore, we detect a significant volume-filling fraction (VFF) of the CNM in the \textsc{AdaptiveG0} models which range from $\mathrm{VFF} = 5 - 12\,\mathrm{per\,cent}$, scaling with the initial $\Sigma_\mathrm{gas}$.
 Overall, with the self-consistent treatment of the FUV radiation field, we achieve a much clearer separation of the thermal gas phases and observe less thermally unstable gas, both in mass and volume.
\end{enumerate}

\section*{Acknowledgements}

The authors thank the referee for their constructive feedback, which significantly improved the quality and clarity of this manuscript.
Funded by the Deutsche Forschungsgemeinschaft (DFG, German Research Foundation) – Project-ID 500700252 – SFB 1601.
TER, SW, and DS acknowledge support by the project ''NRW-Cluster for data-intensive radio astronomy: Big Bang to Big Data (B3D)'' funded through the programme ''Profilbildung 2020'', an initiative of the Ministry of Culture and Science of the State of North Rhine-Westphalia.
The sole responsibility for the content of this publication lies with the authors.
TN acknowledges the support of the Deutsche Forschungsgemeinschaft (DFG, German Research Foundation) under Germany’s Excellence Strategy - EXC-2094 - 390783311 of the DFG Cluster of Excellence ''ORIGINS''.
RW acknowledges the support of the institutional project of the Czech Science Foundation, RVO:67985815.
SCOG acknowledges financial support from the European Research Council via the ERC Synergy Grant ``ECOGAL'' (project ID 855130) and from the Heidelberg Cluster of Excellence (EXC 2181 - 390900948) ``STRUCTURES'', funded by the German Excellence Strategy.
The authors gratefully acknowledge the Gauss Centre for Supercomputing e.V. (www.gauss-centre.eu) for funding this project by providing computing time through the John von Neumann Institute for Computing (NIC) on the GCS Supercomputer JUWELS at J\"ulich Supercomputing Centre (JSC).

\textit{Software:} The software used in this work was in part developed by the DOE NNSA-ASC OASCR Flash Centre at the University of Rochester \citep{Fryxell2000, Dubey2009}.
We used \textsc{Starburst99} \citep{Leitherer1999}.
The simulation results were visualised in part using the \textsc{yt} \citep{Turk2011}, \textsc{numpy} \citep{vanderWalt2011}, \textsc{matplotlib} \citep{Hunter2007}, \textsc{h5py} \citep{Collette2020}, \textsc{IPython} \citep{Perez2007}.

\section*{Data Availability}

The derived data underlying this article will be shared on reasonable request to the corresponding author.
The simulation data will be made available on the \textsc{Silcc Project} data web page: \url{http://silcc.mpa-garching.mpg.de/}.
\bibliographystyle{mnras}
\bibliography{lit}

\appendix\label{sec:appendix}

\section{Tabulated summary}\label{app:tables}

In Table \ref{tab:SFR}, we summarise $\Sigma_\mathrm{SFR}$.
The data is calculated for a time range of $t - t_\mathrm{SF} = [10:100]$~Myr.
We discard the first 10 Myr after the onset of star formation to reduce the artificial impact due to the initial conditions of the simulations.
In the first data column, we quote the median value with the $\nth{75}$ percentile and $\nth{25}$ percentile as upper and lower bounds.
in the second and third data columns, we quote the time-averaged mean and the standard deviation of the mean, respectively.
All values are given in units $10^{-3}\,\mathrm{M_\odot\,yr^{-1}\,kpc^{-2}}$.
We present the long-term evolution of the solar neighbourhood models in Appendix \ref{app:long}.

\begin{table}
 \centering
 \caption{Star formation rate surface density, $\Sigma_\mathrm{SFR}$, and burstiness parameter, $\beta_\mathrm{SF}$, in our models.
 	FUV radiation from young massive stars seems to be negligible for self-regulating $\Sigma_\mathrm{SFR}$ although it decreases the burstiness slightly.}
 \setlength{\tabcolsep}{5pt}
 \begin{tabular}{lcccc}
 \hline
	& median & mean & 1$\sigma$ & $\beta_\mathrm{SF}$\\
	& & [$\times10^{-3}\,\mathrm{M_\odot\,yr^{-1}\,kpc^{-2}}$] & &\\
	\hline \hline
	$\mathrm{\Sigma010vFUV}$ & $3.1_{0.6}^{10}$ & 5.0 & 4.9 & 3.2\footnotemark\\
	$\mathrm{\Sigma030vFUV}$ & $43_{27}^{62}$ & 44 & 20 & 0.8 \\
	$\mathrm{\Sigma100vFUV}$ & $240_{182}^{294}$ & 240 & 59 & 0.5 \\
	\hline
	$\mathrm{\Sigma010}$ & $3.4_{1.6}^{7.2}$ & 4.3 & 3.1 & 1.6 \\
	$\mathrm{\Sigma030}$ & $21_{12}^{36}$ & 30 & 28 & 1.0 \\
	$\mathrm{\Sigma100}$ & $340_{110}^{540}$ & 340 & 220 & 1.3 \\
	\hline
	$\mathrm{\Sigma010vFUV\dagger}$ & $34_{10}^{57}$ & 35 & 24 & 1.5 \\
	$\mathrm{\Sigma010\dagger}$ & $34_{14}^{50}$ & 33 & 20 & 1.1 \\
	$\mathrm{\Sigma010vFUVnoSN\dagger}$ & $93_{70}^{110}$ & 90 & 26 & 0.5\\
	\hline
	\end{tabular}
 \label{tab:SFR}
\end{table}
\footnotetext{We discuss the cause of this stark outlier in Appendix \ref{app:long}.}
\setlength{\tabcolsep}{3pt}
\begin{table*}
 \centering
 \caption{Midplane ($|z| \leq 500\,\mathrm{pc}$) MFs of the individual thermal gas phases during $t = [t_\mathrm{SF} + 10, t_\mathrm{end}]\,\mathrm{Myr}$.}
 \begin{tabular}{l|c|c|c|c|c|c|c|c}
 \hline
	& \multicolumn{2}{|c|}{HIM} & \multicolumn{2}{|c|}{WIM} & \multicolumn{2}{|c|}{WNM} & \multicolumn{2}{|c|}{CNM} \\
	& median$^\mathrm{75\%ile}_\mathrm{25\%ile}$ & mean($\pm1\sigma$) & median$^\mathrm{75\%ile}_\mathrm{25\%ile}$ & mean($\pm1\sigma$) & median$^\mathrm{75\%ile}_\mathrm{25\%ile}$ & mean($\pm1\sigma$) & median$^\mathrm{75\%ile}_\mathrm{25\%ile}$ & mean($\pm1\sigma$) \\
	\hline
	\hline
	$\mathrm{\Sigma010vFUV}$ & 0.1$_{0.1}^{0.2}$ & 0.1($\pm0.1$) & 10$_{3}^{49}$ & 23($\pm25$) & 28$_{14}^{31}$ & 25($\pm10$) & 64$_{34}^{68}$ & 52($\pm19$) \\
	$\mathrm{\Sigma030vFUV}$ & 0.2$_{0.1}^{0.2}$ & 0.2($\pm0.1$) & 24$_{18}^{35}$ & 29($\pm16$) & 22$_{16}^{30}$ & 23($\pm9$) & 51$_{46}^{55}$ & 49($\pm12$) \\
	$\mathrm{\Sigma100vFUV}$ & 0.2$_{0.1}^{0.2}$ & 0.2($\pm0.1$) & 23$_{20}^{18}$ & 25($\pm7$) & 23$_{19}^{29}$ & 23($\pm6$) & 49$_{46}^{56}$ & 52($\pm8$) \\
	\hline
	$\mathrm{\Sigma010}$ & 0.1$_{0.1}^{0.1}$ & 0.1($\pm0.1$) & 7$_{5}^{18}$ & 13($\pm12$) & 59$_{52}^{65}$ & 58($\pm12$) & 34$_{21}^{37}$ & 29($\pm8$) \\
	$\mathrm{\Sigma030}$ & 0.2$_{0.1}^{0.3}$ & 0.2($\pm0.1$) & 9$_{7}^{49}$ & 11($\pm6$) & 69$_{67}^{73}$ & 69($\pm9$) & 20$_{14}^{23}$ & 19($\pm6$) \\
	$\mathrm{\Sigma100}$ & 0.7$_{0.4}^{0.9}$ & 0.7($\pm0.3$) & 22$_{16}^{35}$ & 22($\pm7$) & 48$_{42}^{60}$ & 51($\pm12$) & 25$_{20}^{35}$ & 26($\pm8$) \\
	\hline
	$\mathrm{\Sigma010vFUV\dagger}$ & 0.4$_{0.3}^{0.6}$ & 0.4($\pm0.1$) & 0.9$_{0.7}^{1.2}$ & 0.9($\pm0.3$) & 26$_{20}^{32}$ & 26($\pm7$) & 73$_{67}^{79}$ & 73($\pm7$) \\
	$\mathrm{\Sigma010\dagger}$ & 0.6$_{0.5}^{0.9}$ & 0.7($\pm0.3$) & 1.1$_{1.0}^{1.3}$ & 1.1($\pm0.2$) & 33$_{32}^{35}$ & 34($\pm4$) & 65$_{63}^{67}$ & 65($\pm4$) \\
	$\mathrm{\Sigma010vFUVnoSN\dagger}$ & 0.1$_{0.1}^{0.1}$ & 0.1($\pm0.0$) & 0.0$_{0.0}^{0.0}$ & 0.0($\pm0.0$) & 33$_{26}^{35}$ & 31($\pm5$) & 67$_{64}^{74}$ & 69($\pm5$) \\
	\hline
 \end{tabular}
 \label{tab:mf}
\end{table*}

\begin{table*}
 \centering
 \caption{Midplane ($|z| \leq 500\,\mathrm{pc}$) VFFs of the individual thermal gas phases during $t = [t_\mathrm{SF} + 10, t_\mathrm{end}]\,\mathrm{Myr}$.}
 \begin{tabular}{l|c|c|c|c|c|c|c|c}
 \hline
	& \multicolumn{2}{|c|}{HIM} & \multicolumn{2}{|c|}{WIM} & \multicolumn{2}{|c|}{WNM} & \multicolumn{2}{|c|}{CNM} \\
	& median$^\mathrm{75\%ile}_\mathrm{25\%ile}$ & mean($\pm1\sigma$) & median$^\mathrm{75\%ile}_\mathrm{25\%ile}$ & mean($\pm1\sigma$) & median$^\mathrm{75\%ile}_\mathrm{25\%ile}$ & mean($\pm1\sigma$) & median$^\mathrm{75\%ile}_\mathrm{25\%ile}$ & mean($\pm1\sigma$) \\
	\hline
	\hline
	$\mathrm{\Sigma010vFUV}$ & 51$_{35}^{58}$ & 45($\pm19$) & 17$_{10}^{35}$ & 23($\pm17$) & 25$_{7}^{33}$ & 26($\pm21$) & 6$_{2}^{8}$ & 5($\pm3$) \\
	$\mathrm{\Sigma030vFUV}$ & 43$_{39}^{47}$ & 44($\pm9$) & 24$_{20}^{30}$ & 26($\pm11$) & 26$_{14}^{33}$ & 24($\pm12$) & 7$_{5}^{7}$ & 6($\pm2$) \\
	$\mathrm{\Sigma100vFUV}$ & 34$_{27}^{39}$ & 33($\pm8$) & 22$_{19}^{28}$ & 24($\pm9$) & 31$_{23}^{38}$ & 31($\pm9$) & 12$_{10}^{15}$ & 12($\pm3$) \\
	\hline
	$\mathrm{\Sigma010}$ & 43$_{25}^{52}$ & 41($\pm19$) & 15$_{12}^{17}$ & 18($\pm13$) & 41$_{27}^{50}$ & 40($\pm19$) & 0.5$_{0.3}^{0.7}$ & 0.5($\pm0.2$) \\
	$\mathrm{\Sigma030}$ & 55$_{33}^{60}$ & 48($\pm14$) & 9$_{8}^{12}$ & 10($\pm4$) & 37$_{29}^{55}$ & 41($\pm14$) & 0.4$_{0.4}^{0.5}$ & 0.4($\pm0.1$) \\
	$\mathrm{\Sigma100}$ & 73$_{62}^{77}$ & 70($\pm12$) & 11$_{8}^{17}$ & 12($\pm6$) & 14$_{13}^{19}$ & 17($\pm8$) & 0.6$_{0.4}^{0.9}$ & 0.7($\pm0.4$) \\
	\hline
	$\mathrm{\Sigma010vFUV\dagger}$ & 89$_{83}^{92}$ & 88($\pm5$) & 3$_{2}^{4}$ & 3($\pm1$) & 7$_{4}^{11}$ & 8($\pm3$) & 1$_{1}^{2}$ & 1($\pm1$) \\
	$\mathrm{\Sigma010\dagger}$ & 95$_{92}^{96}$ & 93($\pm4$ ) & 1$_{1}^{2}$ & 2($\pm1$) & 4$_{3}^{5}$ & 5($\pm3$) & 0.4$_{0.3}^{0.6}$ & 0.6($\pm0.3$) \\
	$\mathrm{\Sigma010vFUVnoSN\dagger}$ & 86$_{86}^{86}$ & 86($\pm0.0$) & 0.0$_{0.0}^{0.0}$ & 0.0($\pm0.0$) & 10$_{10}^{10}$ & ($\pm1$) & 4$_{4}^{4}$ & 4($\pm1$) \\
	\hline
	\end{tabular}
 \label{tab:vff}
\end{table*}

We give a quantitative overview of the midplane MF and VFF of the different gas phases in Table \ref{tab:mf} and \ref{tab:vff}, respectively.
Models without hydrogen-ionising radiation do not manage to generate a WIM.
The CNM in \textsc{AdaptiveG0} models is elevated due to less artificial H$_2$ dissociation and PE heating far away from star cluster (sink particles)
Remarkably, we can detect a non-zero midplane VFF of the CNM which scales with the initial $\Sigma_\mathrm{gas}$ in the \text{AdaptiveG0} models.

\section{Choice for the maximum distance of the FUV field calculation}\label{App:distance}

The maximum distance, $d$, for which a star cluster can affect a cell with its FUV radiation is a free parameter of our model.
This distance is equal to the maximum distance for which we calculate the 3D-averaged hydrogen column densities, $N_\mathrm{H,\,tot}$, using the \textsc{TreeRay/OpticalDepth} routine \citep{Wunsch2018}, which is based on \textsc{TreeCol} \citep{Clark2012}.
We argue that the same distance limits should be used for calculating the column densities and for the cells affected by a star's ISRF because the effective strength of the ISRF, $G_\mathrm{eff}$, is related to the amount of extinction, $G_\mathrm{eff} \propto \exp(-2.5\,A_\mathrm{V})$.

Our fiducial choice for the maximum distance for which we calculate 3D-averaged column densities and the ISRF field is $d = 50\,\mathrm{pc}$.
This value is motivated by the approximate average distance of O- and B-type stellar associations in the local ISM.
This model limitation is appropriate for radiation propagation within the ISM midplane.
However, it implies less accuracy for the ISRF along the vertical direction.
Consequently, gas located more than $50\,\mathrm{pc}$ above or below the star-forming disc is always exposed to a maximum ISRF strength of $G_\mathrm{bg} = 0.0948$.
Nonetheless, as discussed in Sect. \ref{sec:theory}, this maximum distance of $d = 50\,\mathrm{pc}$ is reasonable for gas experiencing attenuation from star clusters with maximum masses present in our models.

\begin{figure}
	\centering
	\includegraphics[width=.95\linewidth]{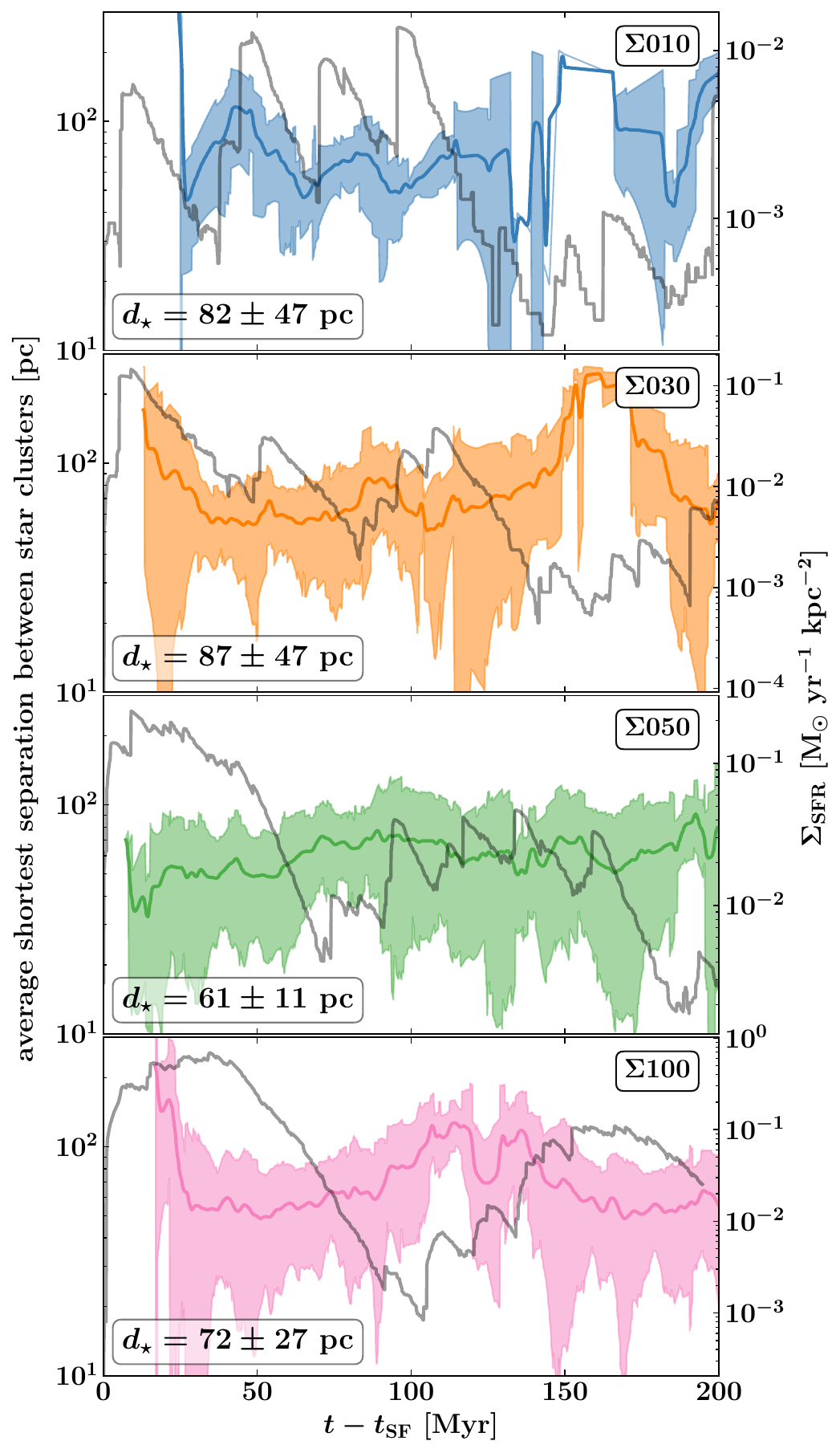}
	\caption{Average shortest separation between two star clusters, $d_\star$, as a function of time.
The colour-coded solid lines show $d_\star$ with 1$\sigma$ scatter as the shaded area.
We quote the global average shortest separation between two star clusters with 1$\sigma$ uncertainty in the top-right boxes of each panel.
For reference, we indicate $\Sigma_\mathrm{SFR}$ as a grey line (right-hand $y$-axis).}
	\label{fig:avg_dist}
\end{figure}

We explore the validity of the assumption of the average distance of O- and B-type stellar associations by examining the average shortest separation between star clusters, $d_\star$, for the static $G_0$ models (Fig. \ref{fig:avg_dist}).
We plot $d_\star$ as a function of $t-t_\mathrm{SF}$ with a 1$\sigma$ scatter as a shaded area.
We also over plot the respective $\Sigma_\mathrm{SFR}$ in grey with the corresponding $y$-axis on the right-hand side.
The average shortest separation between star clusters is weakly anti-correlated with $\Sigma_\mathrm{SFR}$, with a Pearson correlation coefficient of $\rho = -0.38 \pm 0.18$, averaged over the four simulations with varying initial $\Sigma_\mathrm{gas}$.
The weighted average separation is $\overline{d_\star} = 64 \pm 10\,\mathrm{pc}$.
Our fiducial parameter $d = 50\,\mathrm{pc}$ is smaller than the average separation between star cluster sink particles.
This choice of $d$ is sensible, as we aim to prevent the local attenuation of a cell from being influenced by multiple nearby star clusters.
The rationale is to avoid accounting for sightlines that pass through other star clusters, given that star cluster sink particles are treated as point sources.
It's important to note that the local column densities shielding each cell are calculated as 3D averages along multiple lines of sight to the cell.

\begin{figure}
 \centering
 \includegraphics[width=.95\linewidth]{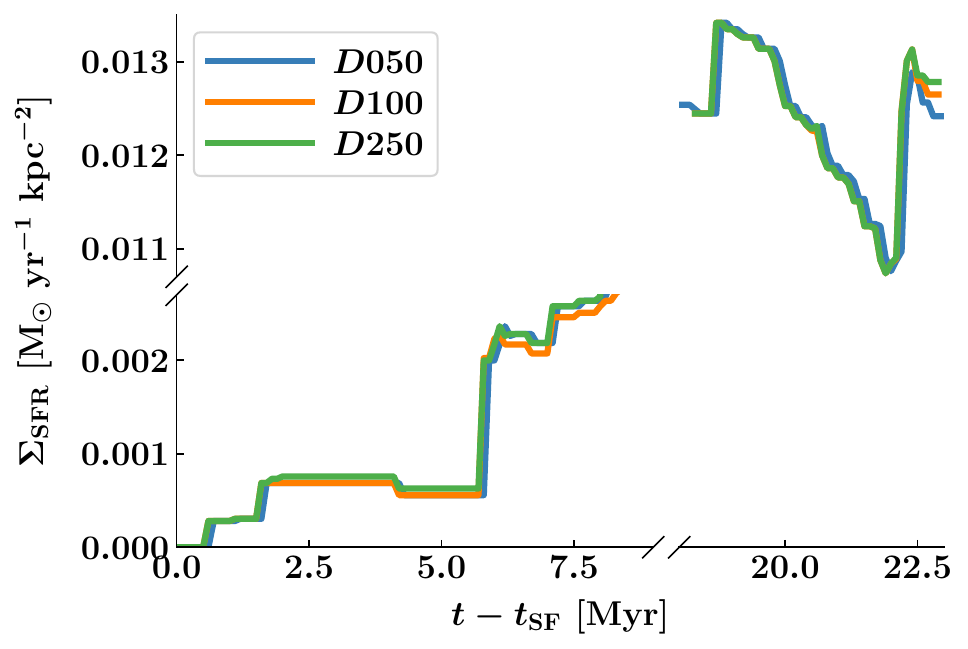}
 \caption{$\Sigma_\mathrm{SFR}$ for varying maximum distance of the FUV field calculation, $d$, ranges between the fiducial 50 pc (\textit{D050}, identical to $\mathrm{\Sigma010vFUV}$) to 250 pc (\textit{D250}, the maximum possible distance given the geometry of our setup).
 All models are based on $\mathrm{\Sigma010vFUV}$.
 The free parameter $d$ affects $\Sigma_\mathrm{SFR}$ only minimally, at most.}
 \label{fig:md_SFR}
\end{figure}

We test the impact of different maximum FUV propagation distances, $d$, on $\Sigma_\mathrm{SFR}$ by running two additional models similar to $\mathrm{\Sigma010vFUV}$, but with $d = 100\,\mathrm{pc}$ and $d = 250\,\mathrm{pc}$, respectively.
A radial distance of $d = 250\,\mathrm{pc}$ is the longest realisable distance in our periodic simulation domain with a side length of $L_\mathrm{x} = L_\mathrm{y} = 500\,\mathrm{pc}$.
The results of the $\Sigma_\mathrm{SFR}$ test are presented in Fig. \ref{fig:md_SFR}.
Please note the linear scale on the ordinate.
We simulate and show the behaviour at the start of star formation and around the first peak of star formation.
The total integrated stellar mass in those time bins is $M_\star = [1.81, 1.69, 1.76] \times 10^4\,\mathrm{M_\odot}$, respectively.
The total stellar mass shows no monotonic trend with varying distances.
We conclude that the free model parameter $d$ minimally affects the SFR, and our fiducial choice of $d = 50\,\mathrm{pc}$ is adequate for all simulations.

\begin{figure}
 \centering
 \includegraphics[width=.95\linewidth]{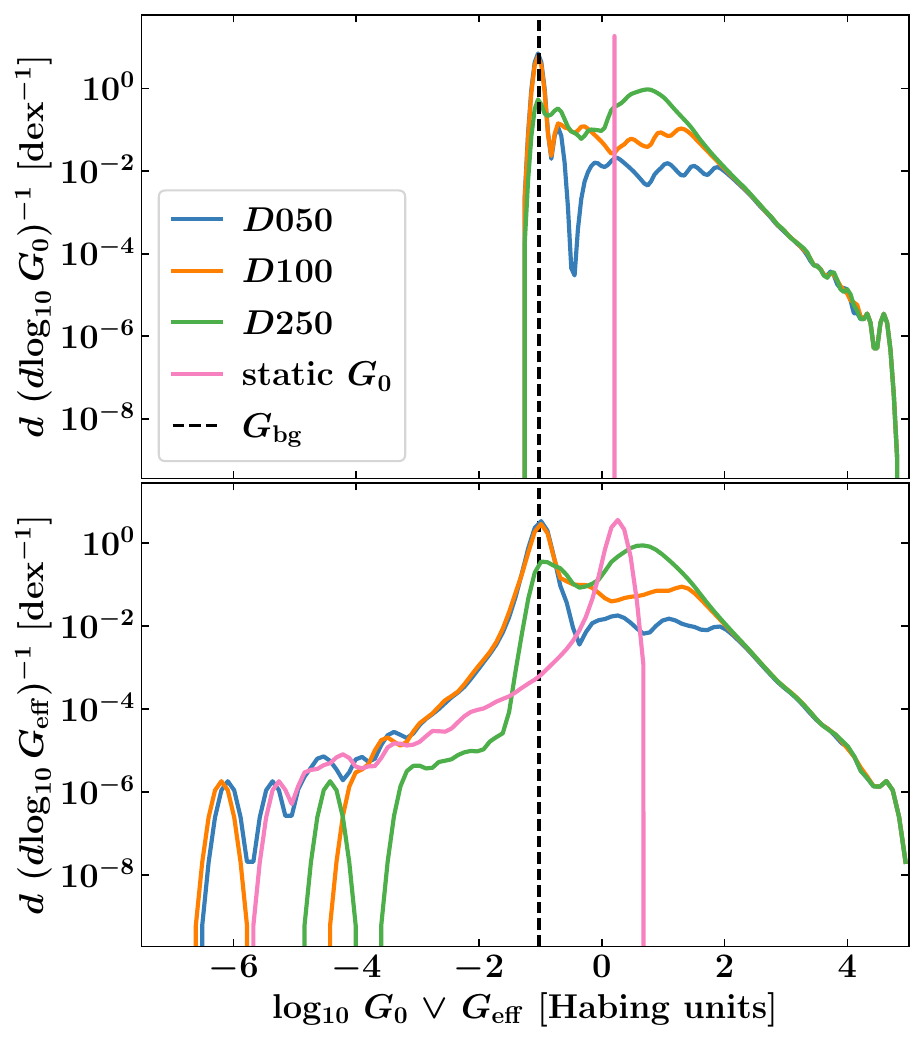}
 \caption{Distribution of $G_0$ (\textit{top}) and $G_\mathrm{eff}$ (\textit{bot}) for one snapshot of the $\mathrm{\Sigma010vFUV}$ simulation with varying maximum FUV propagation distance, $d$.
 The high-end of the distribution (i.e. regions close to star clusters) is identical for \textsc{AdaptiveG0} models.}
 \label{fig:md_G0GeffDist}
\end{figure}

\begin{figure}
 \centering
 \includegraphics[width=.95\linewidth]{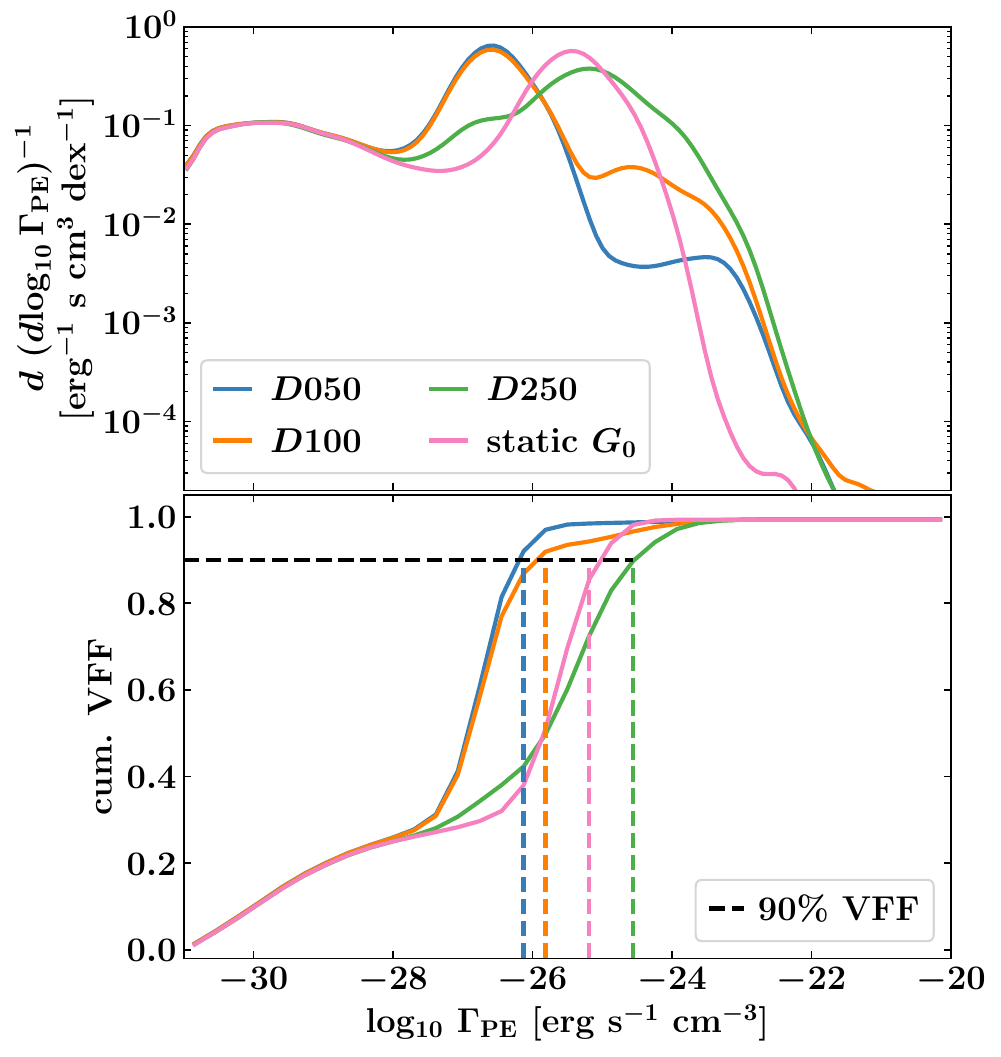}
 \caption{Distribution of $\Gamma_\mathrm{PE}$ for the test models with different $d$, similar to Fig. \ref{fig:heatingDist} and Fig. \ref{fig:vffHeat}.}
 \label{fig:md_VffheatingDist}
\end{figure}

The choice of parameter $d$ influences both the 3D averaged $A_\mathrm{V, 3D}$ and the extent of FUV radiation exposure in the computational domain.
A larger $d$ results in higher $A_\mathrm{V, 3D}$ values due to integration over extended sightlines, while simultaneously increasing the number of cells exposed to FUV radiation from each cluster.
We investigate the effect of varying $d$ on the FUV ISRF distribution in Fig. \ref{fig:md_G0GeffDist}.
This analysis involves rerunning the $\mathrm{\Sigma010vFUV}$ simulation at peak star formation with different $d$ values, following a similar approach to that used in Fig. \ref{fig:heatingDist} and \ref{fig:vffHeat}.
Our results show that the \textsc{AdaptiveG0} models with different FUV propagation distances exhibit similar behaviour at the high end of the FUV ISRF distribution, corresponding to regions in close proximity to star clusters.
In the $D050$ and $D100$ models, the majority of gas cells experience an effective FUV field strength of $G_\mathrm{eff} = G_\mathrm{bg}$.
However, the $D250$ model displays a bimodal $G_\mathrm{eff}$ distribution, with peaks at $G_\mathrm{eff} = 4.8$ and $G_\mathrm{eff} = G_\mathrm{bg}$.
The primary distinction between the $G_\mathrm{eff}$ distributions in the $D050$/$D100$ models and the $D250$ model lies in the cells located farther from the star-forming midplane ISM.

The impact of varying $d$ is also evident in the $\Gamma_\mathrm{PE}$ distribution, as shown in Fig. \ref{fig:md_VffheatingDist}.
The $D050$ and $D100$ realisations (represented by blue and orange lines, respectively) exhibit similar distributions, with overlapping characteristic peaks at low to intermediate heating rates.
Extending the FUV propagation to 100 pc results in a $\sim0.5$ dex higher probability density at moderate $\Gamma_\mathrm{PE} \approx 10^{-25}\,\mathrm{erg\,s^{-1}\,cm^{-3}}$.
However, this leads to only a marginal difference in the cumulative VFF of $\Gamma_\mathrm{PE}$, as depicted in the bottom panel of Fig. \ref{fig:md_VffheatingDist}.
The heating rate to which 95 per cent of the volume is exposed differs by $\sim0.3$ dex between $D050$ and $D100$, and by 0.6 dex when compared to the static $G_0$ model.
The situation is qualitatively different for the $D250$ test model.
The $\Gamma_\mathrm{PE}$ to which 95 per cent of the volume is exposed is a full order of magnitude higher in $D250$ compared to the static $G_0$ model and $\sim1.5$ dex higher compared to $D050$ and $D100$.
It's important to note that while this leads to enhanced heating rates, massive star formation remains largely unaffected due to the dominance of HII region-generating EUV radiation and stellar winds in the immediate vicinity of young massive stars.

Our current model for the FUV ISRF is limited to accessing the 3D-averaged visual extinction calculated for each gas cell at each timestep.
The computation of the local column densities and hence the visual extinction needs a reasonably small value of $d$ to capture only the local gas structure instead of averaging out the properties of the midplane ISM.
Averaging the column density over radial lines of sight with a 250 pc length (i.e. half the size of our computational domain in the horizontal plane) will result in a single plane-averaged visual extinction in the midplane ISM.
Local gas and dust structures would be fully ignored and averaged out over in the midplane ISM

\begin{figure}
 \centering
 \includegraphics[width=.95\linewidth]{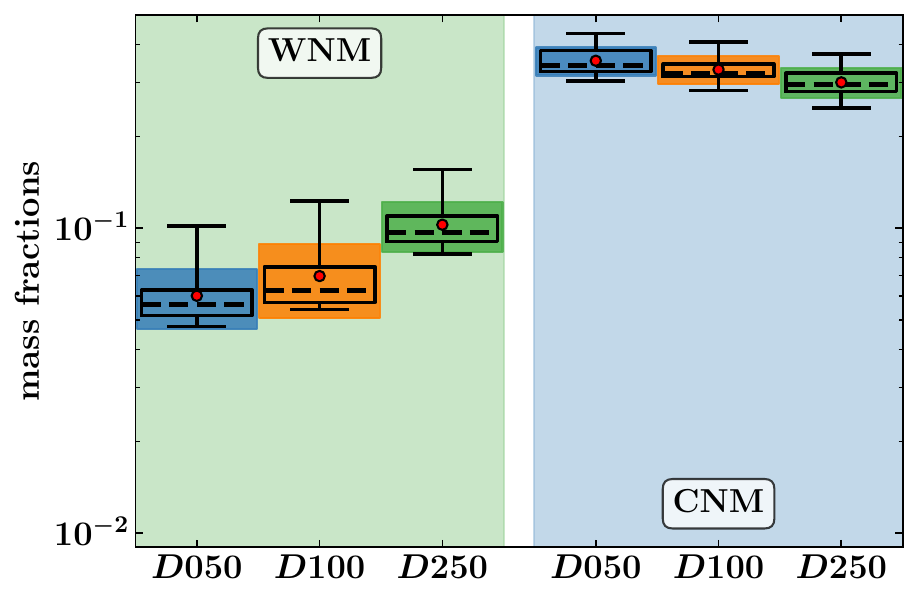}
 \caption{WNM and CNM MF for different values of $d$ measured in a $\sim5\,\mathrm{Myr}$ window around the peak of star formation.
 The plot has the same layout as Fig. \ref{fig:chemistry}.
 The models $D050$ and $D100$ are in reasonable agreement with each other.
 Especially for the WNM MF, $D250$ breaks out of this agreement.
 The slight increase in the WNM can be explained by increased $\Gamma_\mathrm{PE}$ VFF.}
 \label{fig:md_MF}
\end{figure}

Similar to Fig. \ref{fig:chemistry}, we show the WNM and CNM MF of the $D050$, $D100$, and $D250$ test models around the peak of star formation in Fig. \ref{fig:md_MF}.
The averaged ($\pm1\sigma$) values of the WNM MF are 6($\pm1$), 7($\pm2$), and 10($\pm2$) per cent, respectively.
For the CNM MF, we find 35($\pm4$), 33($\pm3$), and 30($\pm3$) per cent during that 5 Myr period around peak star formation.
The CNM MF only decreases slightly between the fiducial model $D050$ and $D250$, while the impact on the WNM mass fraction is more pronounced.

Combining the above tests demonstrates a reasonable choice for $d$ should be below 100 pc.
In the context of massive star formation, the free parameter $d$ choice does not influence the outcome, as expected from theoretical considerations.
The heating rates and gas phase structure near star cluster sink particles do not experience any change with a change in $d$.
For regions further away from active star clusters, the choice of $d$ has an impact and must be chosen reasonably.
In our current model limitations of the 3D averaged column density calculations, a sensible choice, based on the average separation of star clusters (see Fig. \ref{fig:avg_dist}), is $d = 50\,\mathrm{pc}$.
Increasing $d$ to scales larger than 100 pc leads to problems with the consistent computation of local visual extinctions and column densities, based on the \textsc{TreeRay/OpticalDepth} algorithm.

\section{How does the far-ultraviolet interstellar radiation field affect the chemistry?}\label{app:chemistry}

\begin{figure}
 \centering
 \includegraphics[width=.95\linewidth]{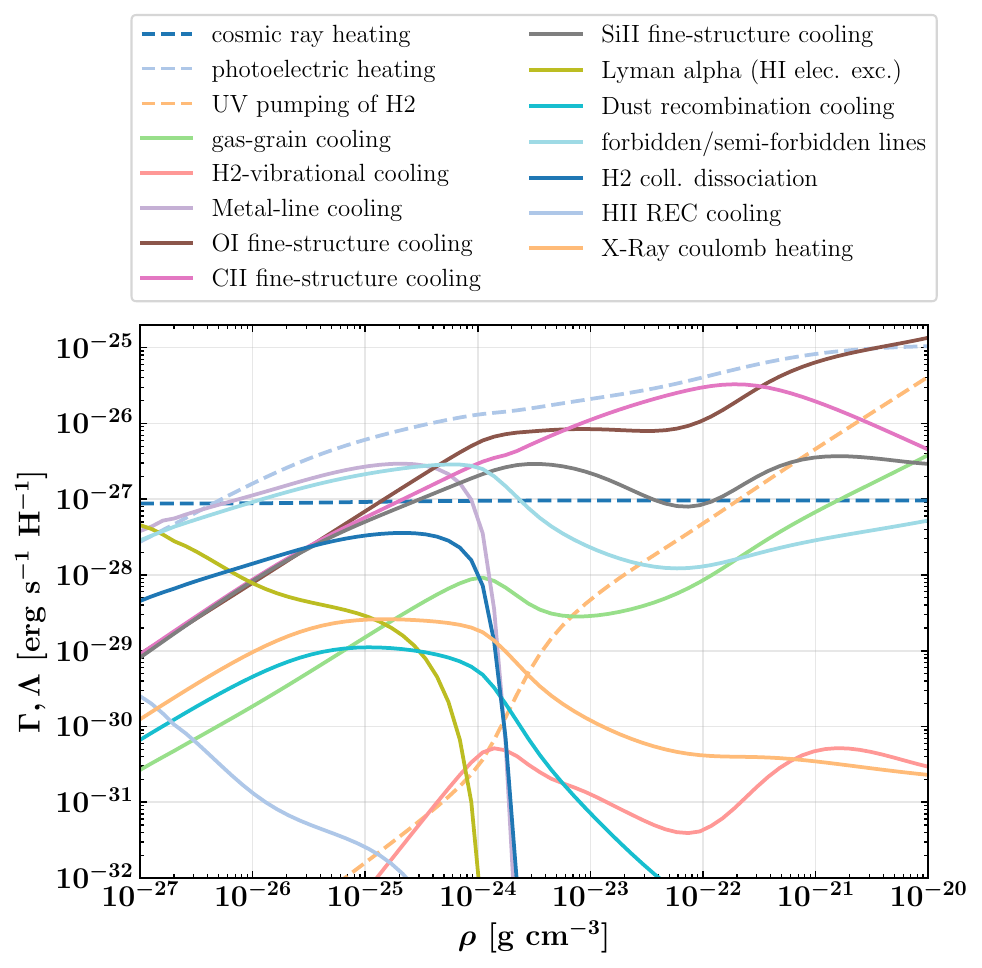}
 \caption{Comparison of the main heating ($\Gamma$) and cooling ($\Lambda$) processes in our chemical network for a given density range in the absence of photoionisation.
 For better comparison, we keep other input parameters, i.e., external shielding factors, radiation fields, dust temperature, and specific internal energy constant for each density bin.
 Fine-structure cooling of neutral oxygen, singly-ionised carbon, and doubly-ionised silicon are the main cooling processes in the warm ISM.}
 \label{fig:HeatCool}
\end{figure}

We list all processes in which the FUV ISRF directly affects our chemical network based on \citet{Nelson1997} and \citet{Glover2007}, and show a comparison of the main heating and cooling processes employed in our simulation for a given gas density in Fig. \ref{fig:HeatCool}.
In addition to changing the chemical composition of the gas, we also account for the changes in thermal energy due to the chemical processes \citep[see][for details]{Glover2007}.\newline

\noindent\textit{Photoelectric heating}\newline
\noindent\citep[based on][]{Bakes1994, Bergin2004}
\begin{align}
 \Gamma_\mathrm{PE} &= 1.3 \times 10^{-24} \cdot \epsilon \cdot G_\mathrm{eff} \cdot n_\mathrm{H, \text{tot}} \\
 \epsilon &= 0.049 \cdot \left(1 + \left(\Psi / 963\right)^{0.73}\right)^{-1} \notag\\
 &+ \left(0.037 \cdot \left(T / 10^{4}\right)^{0.7}\right) \cdot \left(1 + 4 \times 10^{-4} \cdot \Psi\right) \\
 \Psi &= G_\mathrm{eff} \cdot \sqrt{T} \cdot n_\mathrm{e^-}^{-1},
\end{align}with the PE heating efficiency, $\epsilon$ \citep{Wolfire2003}, electron number density, $n_\mathrm{e^-}$, and total hydrogen number density, $n_\mathrm{H, tot}$.\newline

\noindent\textit{Photodissociation of H$_2$}
\begin{align}
 &R_\mathrm{pd,\,H_2} = R_\mathrm{pd,\,H_2\,thin} \,f_\mathrm{dust,\,H_2} \,f_\mathrm{shield,\,H_2},\\
 &R_\mathrm{pd,\,H_2\,thin} = 3.34\times10^{-11} G_0\,\mathrm{s^{-1}},
\end{align}with the H$_2$-photodissociation rate in optically thin gas, $R_\mathrm{pd,\,H_2\,thin}$, the dust shielding of H$_2$ factor, $f_\mathrm{dust,\,H_2} = \exp (-4.18\,A_\mathrm{V})$ \citep{Heays2017} with visual extinction, $A_\mathrm{V, 3D}$, and the H$_2$ self-shielding factor, $f_\mathrm{shield,\,H_2}$ provided by \citet{Draine1996}.\newline

\noindent\textit{Photodissociation of CO}
\begin{align}
 &R_\mathrm{pd,\,CO} = R_\mathrm{pd,\,CO\,thin} \,f_\mathrm{dust,\,CO} \,f_\mathrm{shield,\,CO},\\
 &R_\mathrm{pd,\,CO\,thin} = 1.43\times10^{-10}\, (G_0 / 1.7)\,\mathrm{s^{-1}},
\end{align}with the CO-photodissociation rate in optically thin gas, $R_\mathrm{pd,\,CO\,thin}$ \citep{Heays2017}, the dust shielding of CO factor, $f_\mathrm{dust,\,CO} = \exp (-2.5\,A_\mathrm{V})$ \citep{Dishoeck1988}, with visual extinction, $A_\mathrm{V, 3D}$, and the CO self-shielding factor, $f_\mathrm{shield,\,CO}$ provided by \citet{Lee1996}.\newline

\noindent\textit{Other metals and ions}\newline
In the chemical network based on \citet{Nelson1997}, it is assumed that any carbon not bound in CO exists as C$^+$, due to carbon's low ionisation potential of $\sim11.3\,\mathrm{eV}$.
This assumption is reasonable for static $G_0$ implementations.
However, in \textsc{AdaptiveG0} models where much of the ISM experiences a weaker ISRF, C$^+$ might potentially recombine into atomic C.
We do not track higher ionisation states of carbon, such as C$^{++}$.
Similarly, we do not account for the ionisation states of oxygen and assume all oxygen not bound in CO to exist in its ground state, O.
Lastly, we do not track the evolution of silicon but adopt a silicon abundance of $\chi_\mathrm{Si} = 1.5 \times 10^{-5}$. For the calculation of the free electron abundance, we assume silicon to be always singly ionised, resulting in $\chi_\mathrm{e} = \chi_\mathrm{H^+} + \chi_\mathrm{C^+} + \chi_\mathrm{Si^+}$.
A more detailed treatment of higher ionisation states of metals, such as carbon and oxygen, may influence the overall cooling rates of the ISM \citep[see e.g.][]{Wolfire2003}.
However, we do not expect this potential inaccuracy to significantly affect our results regarding molecular hydrogen gas or star formation properties.
As discussed in \citet{Rathjen2021} and informed by \citet{Seifried2017, Seifried2020} and \citet{Joshi2019}, the CO abundance in our simulation framework is likely underresolved at our maximum spatial resolution of $dx\approx3.9\,\mathrm{pc}$.
Consequently, we refrain from making claims about CO abundance or its influence on our models' chemical dynamics.\newline

\noindent\textit{Dust temperature, $T_\mathrm{dust}$}\newline
\noindent We model the equilibrium dust temperature $T_\mathrm{dust}$ by balancing the ISRF heating and collisions with the gas with the thermal emission of the dust \citep{Glover2010, Glover2012},
\begin{align}
 &\Gamma_\mathrm{ISRF} - \Lambda_\mathrm{dust} + \Gamma_\mathrm{gd} + \Gamma_\mathrm{H_2} = 0,\\
 &\Gamma_\mathrm{ISRF} = \exp (-2.5\,A_\mathrm{V}) \Gamma_\mathrm{thin},\\
 &\Gamma_\mathrm{thin} = 5.6\times10^{-24}\, n_\mathrm{H,\,tot}\,(G_0 / 1.7)\,\mathrm{erg s^{-1} cm^{-3}}
\end{align}with the dust heating rate through absorption of FUV photons, $\Gamma_\mathrm{ISRF}$, the optically thin dust heating rate, $\Gamma_\mathrm{thin}$ \citep{Goldsmith2001}, the dust radiative cooling rate, $\Lambda_\mathrm{dust}$, the energy transfer rate through collisions from gas to dust, $\Gamma_\mathrm{gd}$, and the dust heating rate through H$_2$ formation on dust grains, $\Gamma_\mathrm{H_2}$.

\section{Cosmic ray ionisation heating vs. photoelectric heating}\label{app:PECR}

Another important heating mechanism in the ISM is the ionisation of molecular and atomic hydrogen through CRs.
Assuming each ionisation process deposits $20\,\mathrm{eV}$ of heat energy in to the ISM \citep{Goldsmith1978}, we compute the CR heating rate as
\begin{align}
 \Gamma_\mathrm{CR} = 3.2\times10^{-11}\zeta_\mathrm{CR}(n_\mathrm{H} + n_\mathrm{H2}) [\mathrm{erg} \mathrm{s}^{-1} \mathrm{cm}^{-3}],
\end{align}
with the CR ionisation rate, $\zeta_\mathrm{CR}$. The value of $\zeta_\mathrm{CR}$ scales with the initial $\Sigma_\mathrm{gas}$ of each simulation between $\zeta_\mathrm{CR} = [3\times10^{-17} : 3\times10^{-16}]$.

\begin{figure}
 \centering
 \includegraphics[width=.95\linewidth]{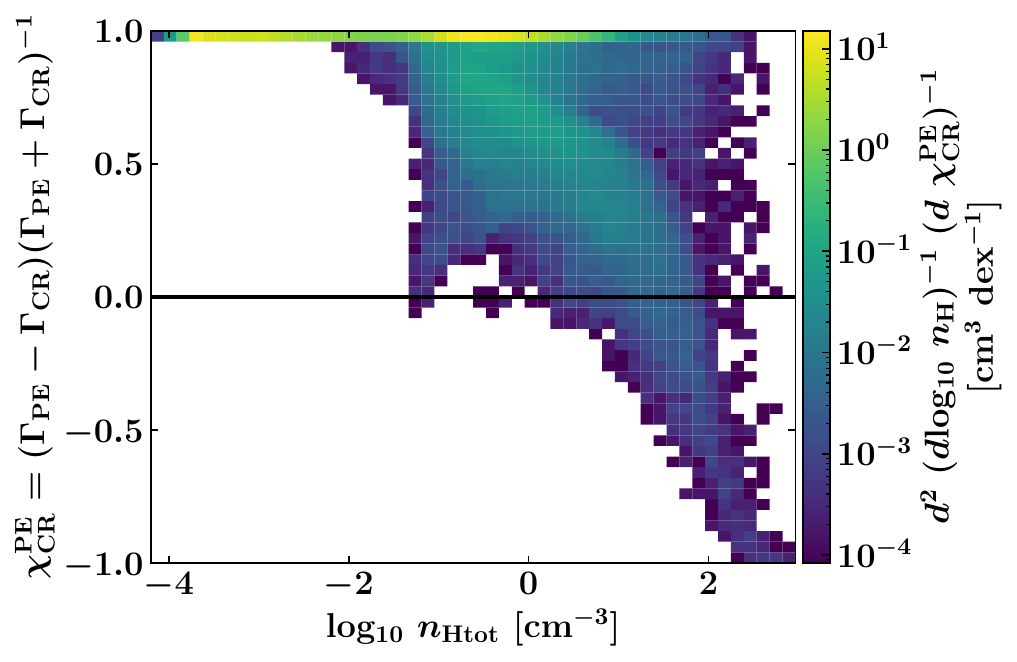}
 \caption{Joint PDF of the relative strength of the CR and PE heating mechanisms, $\chi^\mathrm{PE}_\mathrm{CR}$, and the total hydrogen number density, $n_\mathrm{Htot}$ of the same model and snapshot as in the heating rate analysis in Sect. \ref{sec:heating}.
 PE heating is the dominant heating mechanism of the two.}
 \label{fig:PECR}
\end{figure}

We introduce the relative strength of $\Gamma_\mathrm{CR}$ compared to $\Gamma_\mathrm{CR}$ as $\chi^\mathrm{PE}_\mathrm{CR} = (\Gamma_\mathrm{PE} - \Gamma_\mathrm{CR})(\Gamma_\mathrm{PE} + \Gamma_\mathrm{CR})^{-1}$ and show the joint PDF of $\chi^\mathrm{PE}_\mathrm{CR}$ and $n_\mathrm{Htot}$ in Fig. \ref{fig:PECR}.
A value of $\chi^\mathrm{PE}_\mathrm{CR} = 1$ indicates that no CR ionisation heating is present (e.g. in an already fully ionised region) while a value of $\chi^\mathrm{PE}_\mathrm{CR} = -1$ would indicate a region fully shielded from the FUV ISRF and only penetrable by CRs.
We therefore see a moderate anti-correlation of $\chi^\mathrm{PE}_\mathrm{CR}$ and $n_\mathrm{Htot}$ with a Pearson correlation coeffiction of $\rho = -0.42$.
However, most of the midplane gas, $88\,\mathrm{per\,cent}$ by mass and $98\,\mathrm{per\,cent}$ by volume, is exposed to stronger $\Gamma_\mathrm{PE}$ than $\Gamma_\mathrm{CR}$.

We want to note that in this iteration of the \textsc{Silcc Project}, $\zeta_\mathrm{CR}$ is to be assumed constant throughout an individual simulation and it does not scale with local gas properties or CR energy densities.
A more detailed analysis of the impact of CR ionisation heating and its interplay with PE heating carried out with an updated and more self-consistent model for it will be presented in \citet{Brugaletta2025}.

\section{Star formation rate long-term evolution}\label{app:long}

\begin{figure}
 \centering
 \includegraphics[width=.95\linewidth]{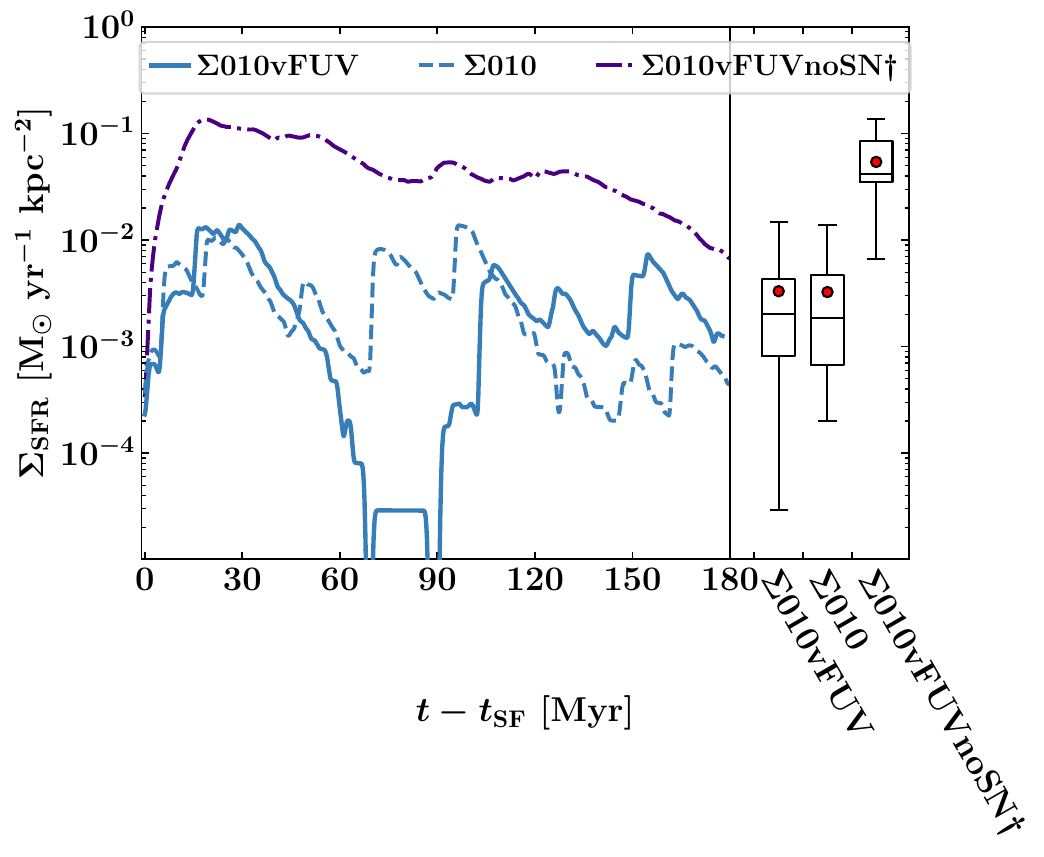}
 \caption{Long-term evolution of $\Sigma_\mathrm{SFR}$ as a function of simulated time after the onset of star formation, $t - t_\mathrm{SF}$ for the solar neighbourhood models, $\mathrm{\Sigma010vFUV}$, $\mathrm{\Sigma010}$ and $\mathrm{\Sigma010vFUVnoSN\dagger}$.
 We also include the box plots for the $\Sigma_\mathrm{SFR}$ distribution, similar to Fig. \ref{fig:sfr_sum}.}
 \label{fig:long_sfr}
\end{figure}

\begin{table}
 \centering
 \caption{Star formation rate surface density, $\Sigma_\mathrm{SFR}$, and burstiness parameter, $\beta_\mathrm{SF}$, in our $\mathrm{\Sigma010vFUV}$, $\mathrm{\Sigma010}$, and $\mathrm{\Sigma010vFUVnoSN\dagger}$ runs.
 The data is calculated for a time range of $t-t_\mathrm{SF} = [10, 180]\,\mathrm{Myr}$.
 We discard the first 10 Myr after the onset of star formation to reduce the artificial impact due to the initial conditions of the simulations.
 We quote the median value in the first data column with the $\nth{75}$ percentile and $\nth{25}$ percentile as upper and lower bounds.
 In the second and third data columns, we quote the time-averaged mean and the standard deviation of the mean, respectively.
 All values are given in units $10^{-3}\,\mathrm{M_\odot\,yr^{-1}\,kpc^{-2}}$.
 The burstiness parameter is defined as $\beta_\mathrm{SF} \equiv \mathrm{IQR(\Sigma_\mathrm{SFR})}~\mathrm{Median^{-1}(\Sigma_\mathrm{SFR})}$.}
 \setlength{\tabcolsep}{5pt}
 \begin{tabular}{lrccc}
 \hline
 & median & mean & 1$\sigma$ & $\beta_\mathrm{SF}$\\
 & & [$\times10^{-3}\,\mathrm{M_\odot\,yr^{-1}\,kpc^{-2}}$] & & \\
 \hline \hline
 $\mathrm{\Sigma010vFUV}$ & $1.9_{0.3}^{4.2}$ & 3.2 & 3.6 & 2.1 \\
 $\mathrm{\Sigma010}$ & $1.9_{0.7}^{5.7}$ & 3.3 & 3.4 & 2.6 \\
 $\mathrm{\Sigma010vFUVnoSN\dagger}$ & $42_{35}^{86}$ & 54 & 34 & 1.2\\
 \hline
 \end{tabular}
 \label{tab:longSFR}
\end{table}

We show the long-term evolution of $\Sigma_\mathrm{SFR}$ in the solar neighbourhood models $\mathrm{\Sigma010vFUV}$ and $\mathrm{\Sigma010}$, as well as $\mathrm{\Sigma010vFUVnoSN\dagger}$ in Fig. \ref{fig:long_sfr}.
We further evolve only those models with a lower initial $\Sigma_\mathrm{gas}$, balancing computational costs against necessity.
As shown in Sect. \ref{sec:sfr}, the self-consistent treatment of the FUV ISRF generated by stellar clusters has a negligible impact on $\Sigma_\mathrm{SFR}$.
We detect an unexpectedly strong dip in $\Sigma_\mathrm{SFR}$ for $\mathrm{\Sigma010vFUV}$ during the time-frame $t-t_\mathrm{SF} = [67.4 : 92.4]\,\mathrm{Myr}$.
However, this decrease in $\Sigma_\mathrm{SFR}$ does not stem from self-regulation through stellar feedback but from large-scale gas dynamical effects.
A rather weak galactic outflow establishes itself before the quiescence phase, partially depleting the star-forming gas reservoir.
This extinguishes further star formation, which in hand allows the gas to fall back and accumulate onto the midplane ISM\footnote{See \citet{Rathjen2023} for more discussion about the cyclical nature of the SFR governed by galactic outflow and subsequent inflow and the dynamical impact of CNMs in launching and supporting galactic outflows.}.
While we would expect that the infall and accumulation of gas onto the midplane ISM would increase the gas density and trigger star formation events, our simulations reveal a more complex behaviour.
In the $\mathrm{\Sigma010}$ model, supernovae-driven gas compression in the midplane ISM triggers subsequent star formation.
However, the $\mathrm{\Sigma010vFUV}$ model exhibits more spatially dispersed initial star formation, preventing the formation of quiescent regions where stellar feedback could initiate new star formation events.
Additionally, the absence of support by the not yet fully established CR pressure gradient leads to the previously outflowing gas falling back onto the midplane ISM. This can trigger further star formation in $\mathrm{\Sigma010}$, but this effect is suppressed in $\mathrm{\Sigma010vFUV}$.

Our subgrid sink particle star cluster formation prescription has a set of checks in place before star cluster particles can be formed to ensure the physical plausibility of the method.
We require the gas to be above a threshold density of $\rho_\mathrm{thr} = 2\times10^{-21}\,\mathrm{g\,cm^{-3}}$.
Also, we do not form new sink particles within 6 cells of an existing sink particle due to numerical stability (\textit{overlap criterion}).
We further demand (\textit{i}) that the gas in the surrounding cells is in a converging flow ($v_\mathrm{rad} \leq 10^{-5}\,c_\mathrm{s}$, with the radial gas velocity, $v_\mathrm{rad}$, and the local sound speed of the cell, $c_\mathrm{s}$, \textit{converging flow criterion}); (\textit{ii}) that the gas sits in a gravitational potential minimum (\textit{potential criterion}); (\textit{iii}) that the gas is Jeans unstable (\textit{Jeans criterion}).

During the quiescent phase, while gas is accumulating in the midplane ISM, we indeed observe simulation cells with overall densities getting as high as $\rho\sim 10\times\rho_\mathrm{thr}$ but the additional checks prevent the formation of more star cluster particles.
We check for the star cluster sink particle formation and accretion conditions at every hydrodynamical timestep, $dt$, for each cell in the domain.
During the 25 Myr ($t-t_\mathrm{SF} = [67.4, 92.4]\,\mathrm{Myr}$), the simulation evaluated a total of 6090 $dt$ and prevented sink particle formation a total of $\sim14800$ times.
That is on average $\sim2.4$ prohibited sink particles per hydrodynamical timestep.
Out of those, $\sim88\,\mathrm{per\,cent}$ have been prohibited due to the overlap criterion, another $\sim10\,\mathrm{per\,cent}$ due to the potential criterion and finally $\sim3\,\mathrm{per\,cent}$ due to the converging flow criterion.
Please note that no further checks are made once one prohibiting criterion is triggered.
This means that even though $\sim88\,\mathrm{per\,cent}$ of the preventive checks are due to the overlap criterion, the parcel of gas in question could also have been Jeans stable, in a diverging flow or outside a gravitational potential minimum.
This behaviour is indeed to some extent peculiar and not very common.
However, we want to stress that it is not nonphysical and fully results from the overall gas dynamics in the simulation during this period.
The physical reasoning behind preventing sink particles from forming within another sink particle's accretion radius is the idea that the gas within a sink particle's radius is dynamically heated by interacting with the stars.
When a parcel of gas is in conditions that it would be star-forming but it is also in the vicinity of another sink particle we let the gas be accreted onto that existing sink particle instead.

Moreover, as soon as the dynamical state of the gas is again favourable for star cluster sink particle formation, $\Sigma_\mathrm{SFR}$ picks up again and levels in at the same magnitude as $\Sigma_\mathrm{SFR}$ in $\mathrm{\Sigma010}$, indicating again the process of self-regulation through primarily hydrogen-ionising radiation, as well as stellar winds and SNe.
An animation\footnote{Which will be hosted at the \textsc{Silcc Project} website (\url{https://hera.ph1.uni-koeln.de/~silcc/}).} of the overall evolution of the simulation illustrates this explanation further.
We quote the averaged $\Sigma_\mathrm{SFR}$ for the long-term evolution runs in Table \ref{tab:longSFR}.

\section{The peaks in the warm ionised medium mass fractions}\label{app:WIMpeak}

In Sect. \ref{sec:chemistry} Fig. \ref{fig:mf_time}, we see two outlier peaks in the WIM MF and corresponding troughs in the CNM MF, once in $\mathrm{\Sigma010vFUV}$, and once in $\mathrm{\Sigma030vFUV}$.
Those outliers can also be seen in the MF distribution in Fig. \ref{fig:chemistry}.
FUV radiation is not able to photoionise the gas and can hence be excluded as the direct source of this momentary increase in warm ionised gas.
We explore whether secondary effects of the self-consistent FUV treatment or other factors are responsible for the observed MFs.
We focus on the run $\mathrm{\Sigma030vFUV}$ but our conclusions are transferable to $\mathrm{\Sigma010vFUV}$.

\begin{figure}
 \centering
 \includegraphics[width=.95\linewidth]{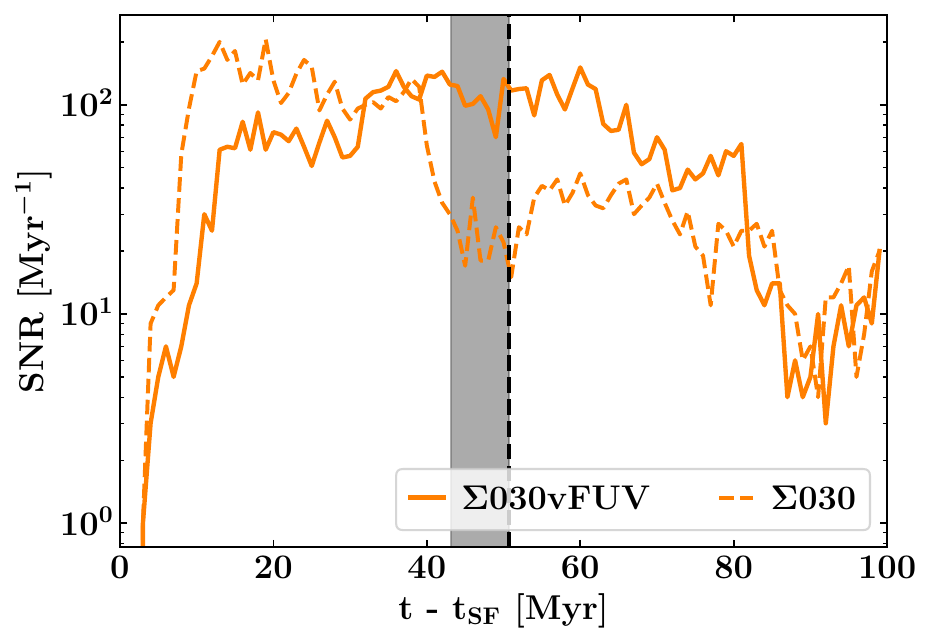}
 \caption{SNe rate (SNR) for the models with initial $\Sigma_\mathrm{gas} = 30\,\mathrm{M_\odot\,pc^{-2}}$.
 The grey shaded area indicates the period in which the WIM MF in $\mathrm{\Sigma030vFUV}$ peaks.}
 \label{fig:SNR030}
\end{figure}

\begin{figure}
 \centering
 \includegraphics[width=.95\linewidth]{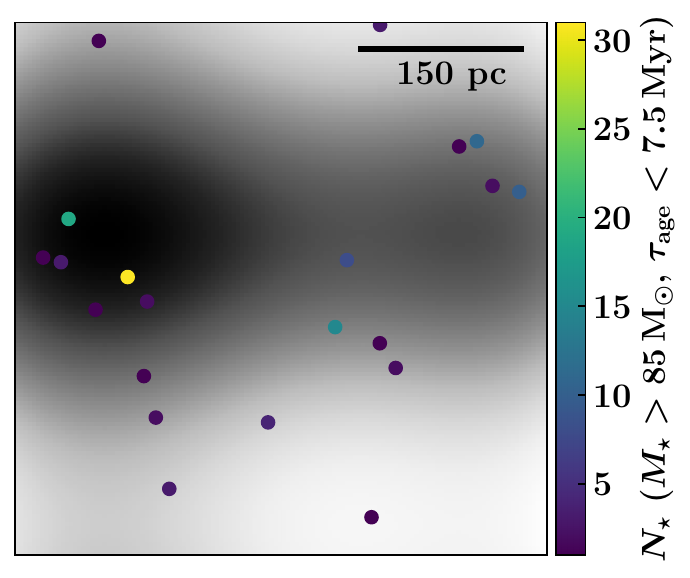}
 \caption{Spatial distribution in the $x-y-$plane (i.e. disc seen face-on) of star cluster sink particles which host super-massive($M_\star > 85\,\mathrm{M_\odot}$) and young ($\tau_\mathrm{age} < 7.5\,\mathrm{Myr}$) stars.
 The colour bar encodes the number of those stars, $N_\star$.
 A slice through the midplane ($z = 0\,\mathrm{pc}$) of a 3D Gaussian kernel density estimate of the $N_\star$-weighted spatial distribution is shown in greyscale.}
 \label{fig:Nstars}
\end{figure}

The WIM MF of $\mathrm{\Sigma030vFUV}$ peaks at $t - t_\mathrm{SF} \approx 49.2\,\mathrm{Myr}$ and starts to build-up during the preceding $\sim7.5\,\mathrm{Myr}$.
As seen in Fig. \ref{fig:sfr}, $\Sigma_\mathrm{SFR}$ slightly increases during that period by a factor of $\sim2$.
That increase in magnitude of $\Sigma_\mathrm{SFR}$ alone is not sufficient to explain the peak in the WIM MF, as similar behaviour in $\Sigma_\mathrm{SFR}$ seen in all simulations does not lead to outliers of that extent.
However, $\Sigma_\mathrm{SFR}$ only measures the total mass transformed into stars, which tells only half the truth.
It is not only important to know how many stars are formed but also with what masses.
The relative impact of stellar feedback of 10 stars with $M_\star = 10\,\mathrm{M_\odot}$ is lesser than the impact of stellar feedback of a single $M_\star = 100\,\mathrm{M_\odot}$ star (see e.g. Fig. \ref{fig:FUV-EUV_Mstar}).
In our star formation model, stars are formed stochastically by sampling from a given IMF, whenever enough gas mass is accreted.
We use the same random seed for each simulation but the spatial and temporal distribution of star formation changes of course between our simulations.
This leads to a non-predictable variation of the stellar mass distribution in models with varying gas dynamics and initial conditions.
But not only the masses of the individual stars are important but also their clustering.
A star embedded in a dense environment has its feedback efficiency greatly reduced through radiative losses.
When multiple stars exist clustered in space and time, their feedback efficiency is boosted and the feedback can penetrate deeper into the surrounding ISM \citep[see e.g.][]{Rathjen2021, Smith2021, Andersson2024}.
First, we can inspect the supernova rate (SNR) over time of $\mathrm{\Sigma030vFUV}$ and $\mathrm{\Sigma030}$ in Fig. \ref{fig:SNR030}.
We indicate the time of the WIM MF peak with a vertical dashed line and the build-up period before as a grey-shaded area.
The average SNRs prior the WIM peak period are $\mathrm{SNR = 93\pm31\,\mathrm{Myr^{-1}}}$ and $\mathrm{SNR = 111\pm35\,\mathrm{Myr^{-1}}}$ for $\mathrm{\Sigma030vFUV}$ and $\mathrm{\Sigma030}$, respectively.
During the WIM peak period, the SNR in $\mathrm{\Sigma030vFUV}$ reaches a quasi-steady state while the SNR in $\mathrm{\Sigma030}$ drops drastically, resulting in $\mathrm{SNR = 32\pm8\,\mathrm{Myr^{-1}}}$ and $\mathrm{SNR = 95\pm31\,\mathrm{Myr^{-1}}}$, respectively.
This shows, that the magnitude of star formation and stellar feedback is not responsible for the peak in the WIM MF of $\mathrm{\Sigma030vFUV}$.

In Fig. \ref{fig:Nstars}, we show the $x-y-$plane spatial distribution of star cluster sink particles which host a stellar population with young ($\tau_\mathrm{age} < 7.5\,\mathrm{Myr}$) and super-massive ($M_\star > 85\,\mathrm{M_\odot}$) stars existing at $t - t_\mathrm{SF} \approx 49.2\,\mathrm{Myr}$ in $\mathrm{\Sigma030vFUV}$, colour-coded by the number of those stars, $N_\star$, in each cluster.
It is evident, that star formation during this period is highly clustered.
We further highlight the clustering by showing a $N_\star$-weighted 3D Gaussian kernel density estimate slice ($z = 0\,\mathrm{pc}$) in greyscale.
The amount of young, super-massive stars within a small region ($L < 150\,\mathrm{pc}$) exceeds the total number of similar stars outside this region.
This strong clustering of massive stars leads to highly efficient photoionisation and stellar wind feedback, which is responsible for the temporary increase in the WIM MF.
Those very massive stars, however, only live for a short time before they explode as core-collapse SNe and the warm ionised gas of the ISM decreases again towards a quasi-equilibrium state while also replenishing the CNM.

\bsp
\label{lastpage}
\end{document}